\documentclass[manuscript]{aastex}

\usepackage{subfigure}
\usepackage{natbib}%
\usepackage{graphicx}
\usepackage{multirow}
\usepackage{epstopdf}
\usepackage{amsmath}
\usepackage{url}
\usepackage{hyperref}
\usepackage{tikz}
\usetikzlibrary{calc,shapes}
\usepackage{xspace}
\usepackage{refcount}
\usepackage{lineno}
\linenumbers

\newcommand{\band}{{\sc{Band}}\xspace}
\newcommand{\comp}{{\sc Comp}\xspace}
\newcommand{\sbpl}{{\sc{Sbpl}}\xspace}
\newcommand{\pwrlw}{{\sc{Pl}}\xspace}
\newcommand{\fermi}{\emph{Fermi}\xspace}
\newcommand{\epeak}{$E_{\rm peak}$\xspace}
\newcommand{\eprest}{$E_{\rm peak}^{\rm rest}$\xspace}
\newcommand{\eiso}{$E_{\rm iso}$\xspace}
\newcommand{\Liso}{$L_{\rm iso}$\xspace}
\newcommand{\ebreak}{$E_{\rm b}$\xspace}
\newcommand{\ebrest}{$E_{\rm break}^{\rm rest}$\xspace}

\newcommand{\gbm}{GBM\xspace}

\shorttitle{GBM Spectral Catalog}
\shortauthors{Gruber, et al.}

\begin{document}

\title{The \emph{Fermi} GBM Gamma-Ray Burst Spectral Catalog: \\
         Four Years Of Data}

\author{
David~Gruber\altaffilmark{1,14}, 
Adam~Goldstein\altaffilmark{2},
Victoria~Weller~von~Ahlefeld\altaffilmark{1,3}, 
P.~Narayana~Bhat\altaffilmark{2}, 
Elisabetta~Bissaldi\altaffilmark{4,5},
Michael~S.~Briggs\altaffilmark{2},  
Dave~Byrne\altaffilmark{6},
William~H.~Cleveland\altaffilmark{7},
Valerie Connaughton\altaffilmark{2}, 
Roland~Diehl\altaffilmark{1}, 
Gerald~J.~Fishman\altaffilmark{8},
Gerard~Fitzpatrick\altaffilmark{6}, 
Suzanne~Foley\altaffilmark{6}, 
Melissa~Gibby\altaffilmark{9}, 
Misty~M.~Giles\altaffilmark{9}, 
Jochen~Greiner\altaffilmark{1}, 
Sylvain~Guiriec\altaffilmark{10},
Alexander J.~van der Horst\altaffilmark{11},
Andreas~von~Kienlin\altaffilmark{1}, 
Chryssa~Kouveliotou\altaffilmark{8}, 
Emily~Layden\altaffilmark{2},
Lin~Lin\altaffilmark{2,13},
Charles A.~Meegan\altaffilmark{2}, 
Sin\'ead~McGlynn\altaffilmark{6}
William S.~Paciesas\altaffilmark{2}, 
V\`{e}ronique~Pelassa\altaffilmark{2},
Robert~D.~Preece\altaffilmark{2},
Arne~Rau\altaffilmark{1},
Colleen A. Wilson-Hodge\altaffilmark{8},
Shaolin~Xiong\altaffilmark{2},
George~Younes\altaffilmark{7} and
Hoi-Fung~Yu\altaffilmark{1}
}
\altaffiltext{1}{Max-Planck-Institut f$\rm \ddot{u}$r extraterrestrische Physik, Giessenbachstrasse 1, 85748 Garching, Germany} 
\altaffiltext{2}{University of Alabama in Huntsville, 320 Sparkman Drive, Huntsville, AL 35805, USA}
\altaffiltext{3}{School of Physics and Astronomy, University of Edinburgh, James Clerk Maxwell Building, Mayfield Road, EH9 3JZ Edinburgh, United Kingdom}
\altaffiltext{4}{Istituto Nazionale di Fisica Nucleare, Sezione di Trieste, I-34127 Trieste, Italy}
\altaffiltext{5}{Dipartimento di Fisica, Universita' di Trieste, I-34127 Trieste, Italy}
\altaffiltext{6}{School of Physics, University College Dublin, Belfield, Stillorgan Road, Dublin 4, Ireland}
\altaffiltext{7}{Universities Space Research Association, 320 Sparkman Drive, Huntsville, AL 35805, USA}
\altaffiltext{8}{Space Science Office, VP62, NASA/Marshall Space Flight Center, Huntsville, AL 35812, USA}
\altaffiltext{9}{Jacobs Technology, Inc., Huntsville, Alabama}
\altaffiltext{10}{NASA Goddard Space Flight Center, Greenbelt, MD 20771, USA}
\altaffiltext{11}{Astronomical Institute, University of Amsterdam, Science Park 904, 1098 XH Amsterdam, The Netherlands}
\altaffiltext{12}{Los Alamos National Laboratory, PO Box 1663, Los Alamos, NM 87545, USA}
\altaffiltext{13}{Sabanc\i~University, Faculty of Engineering and Natural Sciences, Orhanl\i$-$ Tuzla, \.{I}stanbul 34956, Turkey}
\altaffiltext{14}{Planetarium S\"{u}dtirol, Gummer 5, 39053 Karneid, Italy}

\begin{abstract}
In this catalog we present the updated set of spectral analyses of GRBs detected by the \emph{Fermi} Gamma-Ray Burst Monitor (GBM) during its first four years of operation. It contains two types of spectra, time--integrated spectral fits and spectral fits at the brightest time bin, from 943 triggered GRBs. Four different spectral models were fitted to the data, resulting in a compendium of more than 7500 spectra.  The analysis was performed similarly, but not identically to \citet{goldstein12}. All 487 GRBs from the first two years have been re-fitted using the same methodology as that of the 456 GRBs in years three and four. We describe, in detail, our procedure and criteria for the analysis, and present the results in the form of parameter distributions both for the observer-frame and rest-frame quantities.  The data files containing the complete results are available from the High-Energy Astrophysics Science Archive Research Center 
(HEASARC).

\end{abstract}
\keywords{gamma rays: bursts --- methods: data analysis}

\section{Introduction}
During its first four years of operation, the \emph{Fermi} Gamma-Ray Burst Monitor \citep[GBM,][]{Meegan} provided the scientific community with an enormous sample of Gamma-Ray Burst (GRBs) data, significantly expanding our understanding of the physical properties and characteristics of GRBs. In addition, discoveries of new and intriguing phenomena were associated with many individual GRBs \citep[e.g.][]{ackermann10, guiriec11, ackermann11, axelsson12, guiriec13}. 

Here, we present the second GBM GRB spectral catalog which will provide the most comprehensive resource of GRB spectral properties up to date. In order to be as complete and uniform as possible, our methodology follows closely, but is not identical to the procedures employed in the GRB Catalog from the Burst And Transient Source Experiment (BATSE) \citep{Kaneko06} and the first GBM GRB spectral Catalog \citep{goldstein12}. We include representative spectral fits for all GBM bursts from the first four years of operation (July 14, 2008 to July 13, 2012).   

For each GRB, we show two types of spectra: Time-integrated spectra (henceforth labelled \emph{F} for \emph{fluence}) and spectra at the brightest time bin (henceforth labelled \emph{P} for \emph{peak-flux}). 
A set of four empirical models was applied to the data in both cases. The selection of these model functions is based on tradition \citep{Band93, Kaneko06, goldstein12} and mathematical complexity. The signal-to-noise ratios in the data of \gbm bursts rarely support models with more than four free parameters, which is why we resort to models with two, three or four free fit parameters and only in exceptional cases additive terms \citep[e.g. a blackbody component][]{axelsson12} can be added. Making use of the Castor statistics \citep{ackermann11}, we derive the best model for each GRB and present the distribution and characteristics of the model parameters.

This catalog is organized as follows: In Section~2 we present a short overview of the GBM, in Section~3 we describe the  methodology used in the production of this catalog, including  detector selection, data types, energy selection and background fitting, and the source selection. We then offer a description of the spectral models used in this catalog in Section~4, present the spectral analysis methods in Section~5 and the results in Section~6. Finally, in Section~7 we conclude with a summary and a discussion.  

\section{\fermi GBM}
The \fermi\footnote{Formerly known as the \emph{Gamma-Ray Large Area Space Telescope} or GLAST} Gamma-Ray Space Telescope was successfully launched on 2008 June 11 into a Low Earth orbit (LEO) of $\sim565$~km altitude at an 25.6 degree inclination. Its payload comprises two instruments, the GBM and the Large Area Telescope \citep[][LAT]{Atwood}.
The goal of \gbm is to augment the science return from \fermi with its prime objective being joint spectral and timing analyses of GRBs seen in common with the LAT. In addition, \gbm provides near real-time burst locations which permit (i) the \fermi spacecraft to repoint the LAT towards the observed GRB and (ii) to perform follow-up observations with ground-based facilities.
Compared to other high-energy spacecraft, the great advantage of \gbm is its capability to observe the whole unocculted sky at any given time with a Field of View (FoV) of $\ge 8$~sr and its very broad energy coverage. Therefore, along with GRBs, \gbm offers great capabilities to observe all kinds of high-energy astrophysical phenomena, such as e.g., Solar Flares \citep[e.g.][]{grubersfl11, sfl12}, Soft Gamma Repeaters \citep[SGRs, e.g.][]{Lin, kienlin12} and Terrestrial Gamma-Ray Flashes \citep[TGFs, e.g.][]{briggs11}.

Designed to study the gamma-ray sky in the energy band of $\sim$8 keV--40 MeV, GBM is composed of twelve sodium iodide (NaI) and two bismuth germanate (BGO) scintillation detectors. With a thickness of 1.27~cm  and a diameter of 12.7~cm, the NaI crystals cover an energy range from 8 keV--1 MeV. They are oriented around the spacecraft such that the position of the GRB can be determined.

The two BGO crystals have a diameter and thickness of 12.7~cm, covering an energy range of 200 keV--40 MeV,  and are located on opposite sides of the spacecraft so that at least one is illuminated from any direction. A source location is calculated in spacecraft coordinates and used in the production of the detector response matrices (see Section~\ref{sec:datana}).
 
For more details on the GBM detectors and their calibration, refer to \citet{Meegan, Bissaldi, paciesas12}.

\section{Method}
During the first four years of operation, GBM triggered on a total of 954 GRBs \citep{kienlin13}, 943 of which are presented in this catalog.  The remaining bursts are excluded due to a low accumulation of counts or a lack of spectral/temporal coverage.  In order to deliver the 
most useful analysis to the community, we have attempted to make the method as systematic and uniform as possible; circumstances under which deviations were employed are clearly indicated. Details of the detector and data selection as well as the 
process used to fit the data are described in this section. Many of the criteria are adopted from the GBM Burst Catalog \citep{paciesas12} and we have attempted to maintain this in all aspects. However, due to the nature of spectral analysis we demand stricter criteria to 
ensure that we have adequate signal in all energy channels. This  effectively reduces the GRB sample from that used in the 
burst catalog.  

We highlight that this catalog only presents the analysis of GRBs that \emph{triggered} the GBM. There is a non-negligible amount of GRBs that did not trigger GBM whose temporal and spectral properties are presented elsewhere \citep{untrig}. These GRBs do not have different properties compared to the triggered GRB sample but simply occurred during times when the \gbm triggering algorithm was switched off (e.g. when the spacecraft was at latitudes of high geomagnetic activity). 

\subsection{Detector Selection}

The detector selection is consistent with \citet{goldstein12}, i.e. a maximum of 3 NaI detectors together with one BGO detector were used for the spectral analysis.
Since the effective area (i.e. detection efficiency) of the NaI detectors decreases rapidly for high incidence angles \citep{Bissaldi} only detectors with source angles $\le 60^{\circ}$ are  used for the spectral analysis. In addition, it has been verified that the detectors were neither obstructed by the spacecraft nor by the solar panels of \emph{Fermi}. However, due to small inaccuracies in the spacecraft mass model or location uncertainties, the blockage code does not always return a subset of detectors that is free from blockage. This is evident when the low-energy data deviate strongly from the fit model \citep{goldstein12}.   When this occurs we remove these detectors from the selected sample.  If more than 3 NaI detectors are qualified for the spectral fitting, the NaI detectors with the smallest source angles were used to avoid a fitting bias toward lower energies. 
\subsection{Data Types}
\gbm persistently records two different types of science data, called CTIME (fine time resolution, coarse spectral resolution of 8 energy channels) and CSPEC (coarse time resolution, full spectral resolution of 128 energy channels). CTIME (CSPEC) data have a nominal time resolution of 0.256~s (4.096~s) which is increased to 64~ms (1.024~s) whenever GBM triggers on an event. After 600~s in triggered mode, both data types return to their non-triggered time resolution. The third and primary data type used in this catalog is the ``Time Tagged Events'' (TTE) which consist of individual events, each tagged with arrival time (2~$\mu$s precision), energy (128 channels) and detector number. The TTE data are generated and stored on-board in a continuously recycling buffer. When \gbm enters trigger mode, the buffered pre-trigger TTE are transmitted as science data along with $\sim 300$~s of post-trigger TTE.

For the purpose of this catalog, we choose a standard time binning of 1024 ms for bursts longer than 2~s in duration as defined by the burst $\rm T_{90}$ \citep{Kouveliotou} presented in \citet{kienlin13} and 64~ms for bursts of duration 2 s and shorter.  The time history of TTE typically starts at $\sim$30 s before trigger and extends to $\sim$300 s after trigger.  This TTE data time span is adequate for the analysis of most GRBs.  For GRBs that have evident precursors or emissions that last more than 300 s after trigger, we use the CSPEC data, which extend $\sim$4000 s before and after the 
burst for triggered events. CSPEC data were used for 76 GRBs in this catalog.

\subsection{Energy Selection and Background Fitting}
With the optimum subset of detectors selected, the best time and energy selections are chosen to fit the data.  The available 
and reliable energy channels in the NaI detectors lie between $\sim$8 keV and $\sim$1 MeV. This selection excludes the overflow 
channel at high energies and those channels $< 8$~keV where the transmission of gamma-rays is poor due to the silicone pad in front of the NaI crystal and the Multi Layer Insulation (MLI) around the detectors \citep{Bissaldi}.
We perform a similar selection to the BGO detector for each burst, selecting channels between $\sim$300 keV 
and $\sim$38 MeV.  We select enough pre- and post-burst data to sufficiently model the 
background and fit a single energy dependent polynomial (choosing up to $4^{\rm th}$ order) to the background. For each 
detector the time selection and polynomial order are varied until the $\chi^2$ statistic map over all energy channels is 
minimized, resulting in an adequate background fit. This approach is rather subjective in that it is dependent on the observer's choice of the background intervals. In the future, it may be advantageous to implement more objective background selection methods such as the ``direction dependent background fitting (DBBD)'' method presented in  \citet{szecsi}.

\subsection{Source Selection}
Knowing the background model, the background-subtracted count rates are summed over all NaI detectors for a given burst to produce a single GRB count light curve.
Using this light curve, only time bins (1.024~s for long burst and 64~ms for short bursts) with a signal-to-noise ratio greater or equal to 3.5 were selected, in agreement with \citet{goldstein12}. This time selection is then applied to all detectors for a given burst. 

This criterion ensures that there is adequate signal to successfully perform a spectral fit and constrain the parameters of the fit. This does however eliminate some faint bursts from the catalog sample (i.e., those with no time bins with signal above 3.5 sigma). While the possibility remains that not all signal from the GRB was selected, this method nevertheless provides the most objective way to obtain a selection of intrinsic GRB counts as including less significant bins would only increase the uncertainty in the measurements.

This selection is what we refer to as the \emph{F} selection, since it is a time-integrated selection and the derived photon and energy fluences are representative of but not equal to the fluence over the total duration of the burst. We draw attention to the fact that time bins with a signal-to-noise ratio less than 3.5, which were not included in the fitting process, also contribute to the photon and energy fluence. For more than 80~\% of the GRBs in the catalogue the ratio of count fluence (without the intervals with SNR$<3.5$) vs the total counts (with the intervals with SNR$<3.5$) is larger than 0.8. So while there are some bursts for which the fluence can be considerably underestimated, the other option would be to overestimate the fluence of those bursts by including background that contaminates the spectral fits. 

The other selection performed is a 1.024~s peak photon flux selection for long bursts ($\rm T_{90}>2$~s) and 64 ms peak count rate flux selection for short bursts ($\rm T_{90} \le 2$~s).  This selection is made by adding the count rates in the NaI detectors again and selecting the single bin of signal with the highest background-subtracted count rate.  This selection is a snapshot of the energetics at the most intense part of the burst and is depicted as the \emph{P} selection.  

\begin{figure}
	\begin{center}
		\begin{minipage}[t]{0.5\textwidth}
		\centering
		\subfigure[]{\label{acumtime}\includegraphics[scale=0.45]{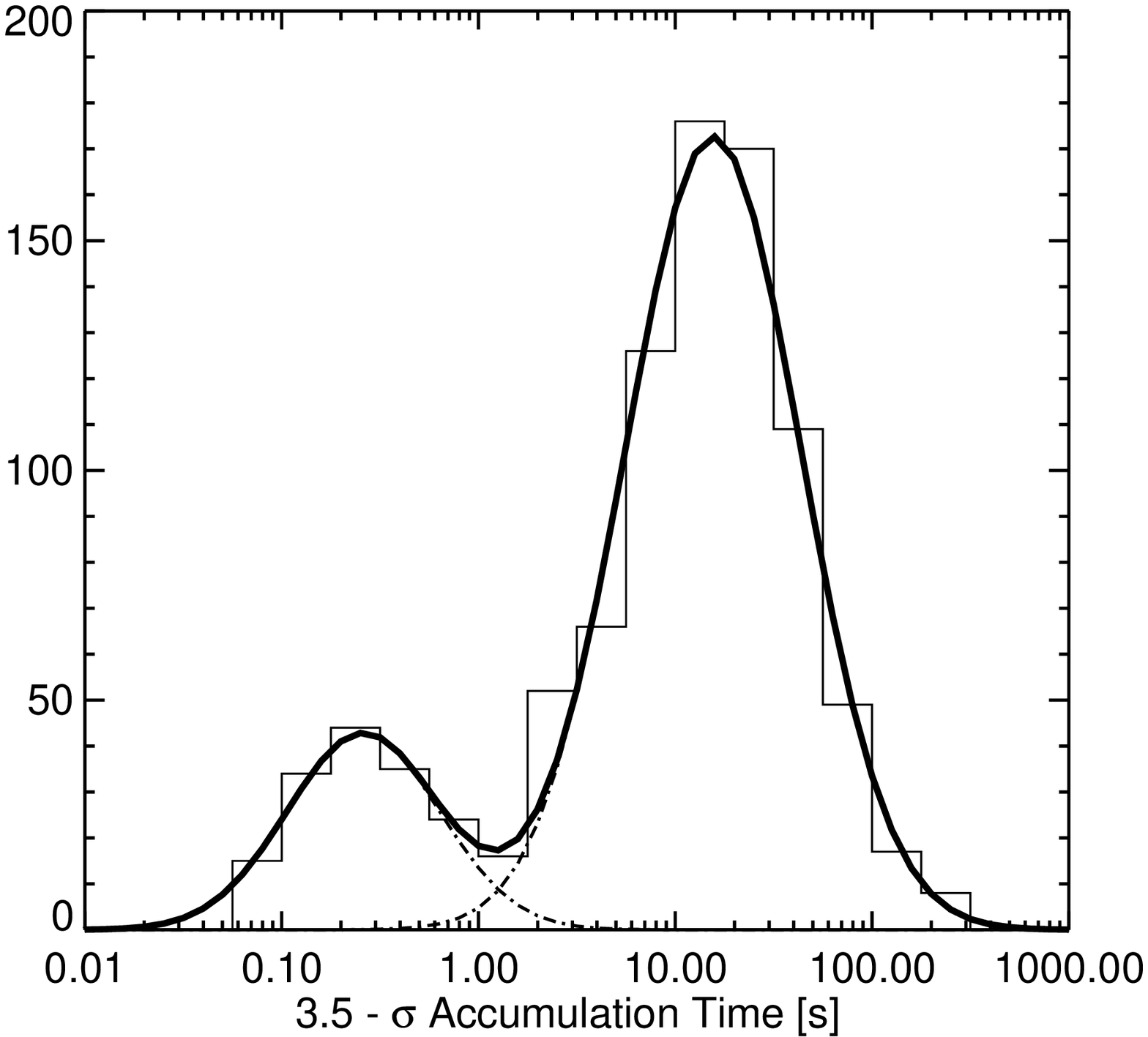}}\\
		\end{minipage}
		\begin{minipage}[b]{1\textwidth}
		\subfigure[]{\label{timepfluence}\includegraphics[scale=0.45]{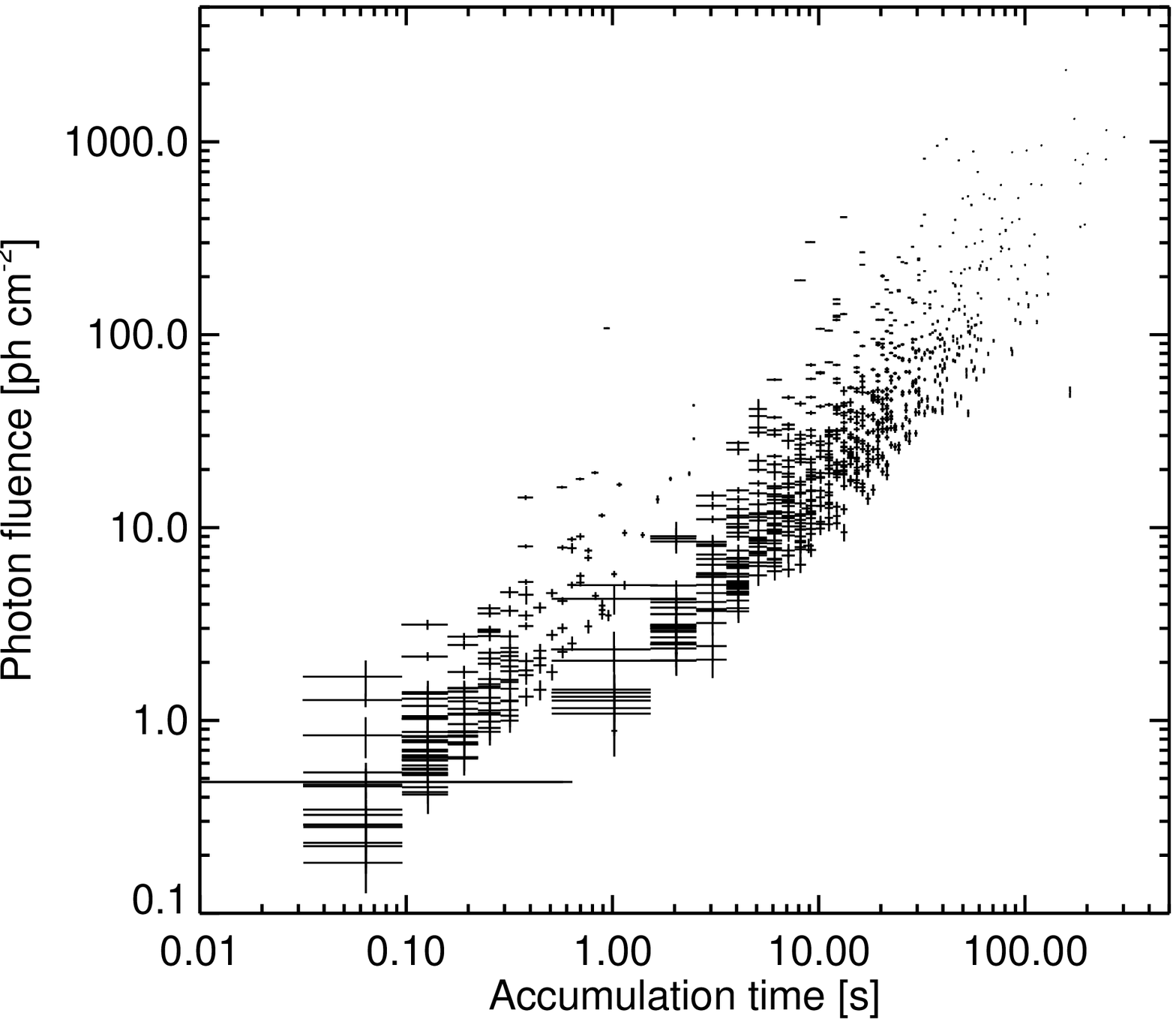}}
		\subfigure[]{\label{timepflux}\includegraphics[scale=0.45]{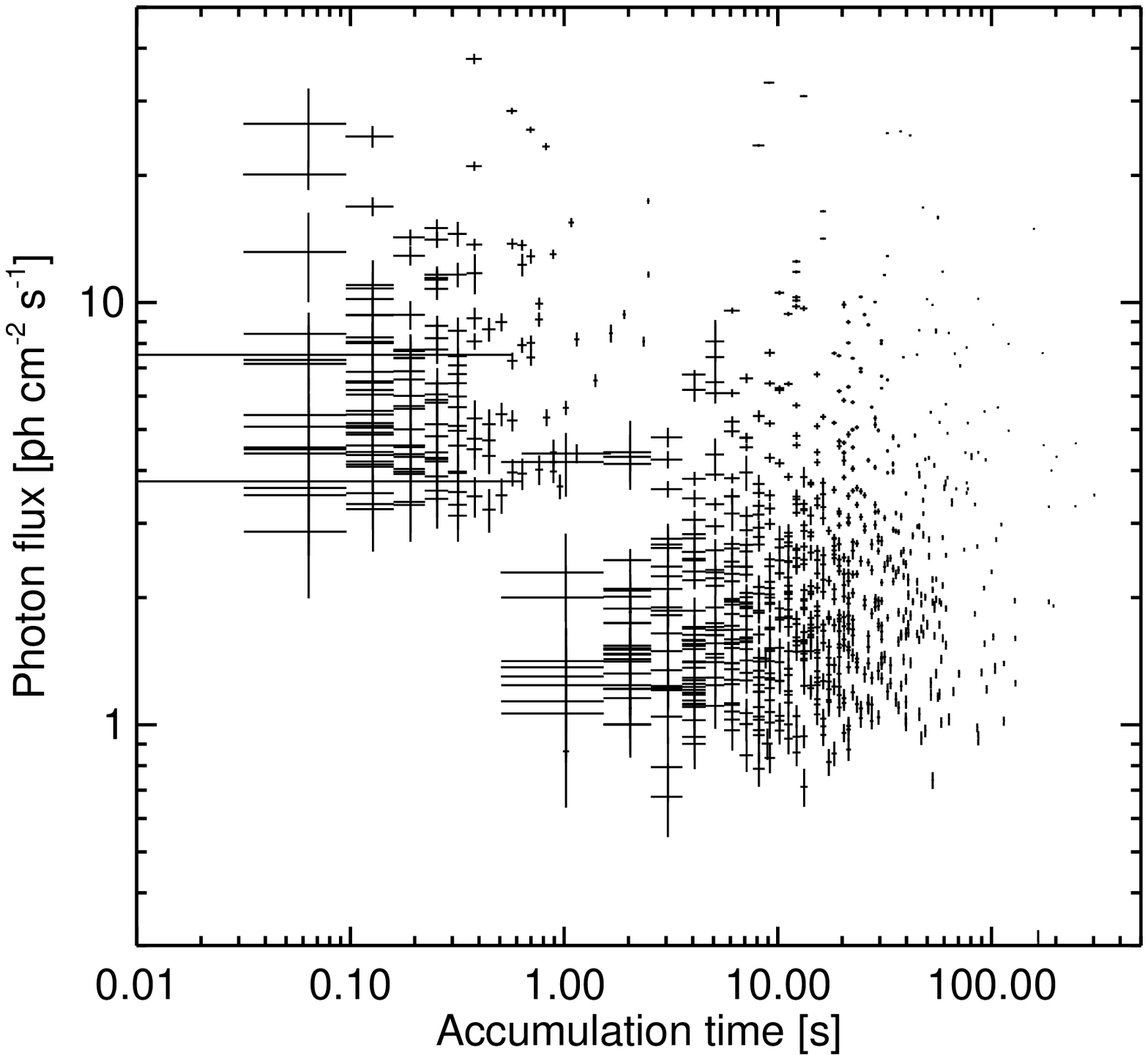}}
		\end{minipage}
	\end{center}
\caption{\subref{acumtime} is the distribution of the accumulation times based on the 3.5$\sigma$ signal-to-noise selections.  Note 
the similarity to the traditional $\rm t_{90}$ distribution, with the minimum near 1 second.  No other estimation of the duration was 
factored into the production of the accumulation time. \subref{timepfluence} and \subref{timepflux} show the comparison of the model photon fluence and model photon flux to the accumulation time respectively.  The photon fluences and fluxes shown in these figures are from the estimated BEST model fits. \label{allTime}}
\end{figure}

Figure~\ref{allTime} shows the distribution of accumulation times used for the fitting process based on the signal-to-noise selection criteria.  The distribution of accumulation times reported is similar to the observed emission time of the burst, excluding quiescent periods, \citep[e.g.][]{Mitrofanov} and peaks at $~0.26$~s and $\sim 15$~s for short and long GRBs, respectively. The dividing duration time scale between the two classes of GRBs is $\sim 1.27$~s and is, as expected from the employed source selection methodology, somewhat smaller than the canonical $2$~s \citep{Kouveliotou}. Figure \ref{allTime} also includes the comparisons of the model photon fluence and model photon flux compared to the accumulation time. In both cases two specific regions are visible for long and short GRBs. In addition, there is a clear correlation between the photon fluence (flux) and accumulation time in Figure~\ref{timepfluence} and Figure~\ref{timepflux} shows a relationship indicating the existence of two different burst groups, similar to the ones delineated by the hardness--duration relationship found by \citet{Kouveliotou}.

\section{Models}
We chose four spectral models to fit the spectra of GRBs in our selection sample. These models include a single power law 
(\pwrlw), Band's GRB function (\band), an exponential cut-off power-law (\comp), and a smoothly broken power law (\sbpl). All 
models are formulated in units of photon flux with energy (\emph{E}) in keV and multiplied  by a normalization constant \emph
{A} ($ \rm ph \ s^{-1} \ cm^{-2} \ keV^{-1}$). Below we describe each model and its features in detail.

\subsection{Power-Law Model}
The single power law with two free parameters is defined as
\begin{equation}
f_{ \rm{PL} } ( E ) = A \left( \frac{ E }{ E_{piv} } \right )^{ \lambda }
\end{equation}
where \emph{A} is the amplitude and $\lambda$ is the spectral index. The pivot energy ($E_{piv}$) normalizes the model to the 
energy range under consideration and helps reduce cross-correlation of other parameters.  In all cases, $E_{piv}$ is 
held fixed at 100 keV.  While most GRBs exhibit a spectral break in the GBM passband, some GRBs are too weak to 
adequately constrain this break in the fits. These bursts are well fit by the single power--law.

\subsection{Band's GRB function}
Band's GRB function \citep{Band93} has become the standard for fitting GRB spectra, and therefore we include it in our analysis:

\begin{equation}
\begin{split}
&  f_{\rm{ BAND} } ( E ) = \\
& A \begin{cases} 
\biggl(\frac{E}{100 \ \rm keV }\biggr)^{\alpha} \exp \biggl[- \frac{ (\alpha +2) E}{ E_{\rm peak} } \biggr], \ E  < \frac{ (\alpha - \beta) \ 
E_{\rm peak} } { \alpha +2} \\
\biggl( \frac{E}{ 100 \ \rm keV } \biggr)^{ \beta } \exp (\beta -\alpha) \biggl[ \frac{(\alpha-\beta ) E_{\rm peak}}{100 \ \rm keV \ (\alpha 
+2)} \biggr]^{\alpha-\beta }, \\ E \geq \frac{(\alpha -\beta ) \ E_{\rm peak}}{\alpha +2}
\end{cases}
\end{split}
\end{equation}
The four free parameters are the amplitude, \emph{A}, the low and high energy spectral indices, $\alpha$ and $\beta$, respectively, and the $\nu F_{\nu}$ peak energy, \epeak. This function is essentially a smoothly broken power law with a 
curvature defined by its spectral indices. The low-energy index spectrum is a power law only asymptotically. 

\subsection{Comptonized Model}
The Comptonized model is an exponentially cutoff power-law, which is a subset of the Band function in the limit 
that $\beta \to -\infty$:
\begin{equation}
f_{\rm{\sc{COMP}}}(E) = A \ \Bigl(\frac{E}{E_{piv}}\Bigr) ^{\alpha} \exp \Biggl[ -\frac{(\alpha+2) \ E}{E_{\rm peak}}  \Biggr]
\end{equation}	  
The three free parameters are the amplitude \emph{A}, the low energy spectral index $\alpha$ and \epeak. $E_{piv}$ is 
again fixed to 100 keV, as for the power law model. 

\subsection{Smoothly Broken Power-Law}
The final model that we consider in this catalog is a broken power-law characterized by one break with flexible curvature and is able 
to fit spectra with sharp or smooth transitions between the low and high energy power laws.  This model, first published in 
\citet{Ryde}, where the logarithmic derivative of the photon flux is a continuous hyperbolic tangent, has been re-parametrized 
\citep{Kaneko06} as given below:
\begin{equation}
f_{\rm{\sc{SBPL}}}(E)=A \biggl(\frac{E}{E_{piv}} \biggr)^b  \ 10^{(a - a_{piv})}
\end{equation}
where
\begin{equation}
	\begin{split}
&a=m\Delta \ln \biggl(\frac{e^q+e^{-q}}{2}\biggr), \\
&\\
&a_{piv}=m\Delta \ln \biggl(\frac{e^{q_{piv}}+e^{-q_{piv}}}{2} \bigg), \\
&\\
&q=\frac{\log (E/E_b)}{\Delta}, \quad q_{piv}=\frac{\log(E_{piv}/E_b)}{\Delta},\\
&\\
&m=\frac{\lambda_2-\lambda_1}{2}, \quad b=\frac{\lambda_1+\lambda_2}{2}.
	\end{split}
\end{equation}
In the above relations, the low- and high-energy power law indices are $\lambda_1$ and $\lambda_2$ respectively, \ebreak is 
the break energy in keV, and $\Delta$ is the break scale in decades of energy.  The break scale is independent and not 
coupled to the power law indices as for the Band function, and represents an additional degree of freedom.  
However, \citet{Kaneko06} found that an appropriate value for $\Delta$ for GRB spectra is 0.3, therefore we fix $\Delta$ at this 
value.  
In addition, we tested the behavior of $\Delta$ for some bright GRBs by letting it vary during the fit process. The results of this study are presented in Section~6.

We choose to fit these four different functions because the measurable spectrum of GRBs is dependent on 
intensity. Less intense bursts (in the observer frame) provide less data to support a large number of 
parameters.  This may appear obvious, but it allows us to determine why in many situations a particular empirical function 
provides a poor fit, while in other cases it provides an accurate fit.  For example, the energy spectra of GRBs are normally well fit 
by two smoothly joined power laws.  For particularly bright GRBs, the \band and \sbpl functions are typically an accurate 
description of the spectrum, while for weaker bursts the \comp function is most appropriate.  Bursts that have signal significance 
on the order of the background fluctuations do not have a detectable distinctive break in their spectrum and so the power law is 
the most appropriate function.  

\section{Data Analysis}\label{sec:datana}
The spectral analysis of all bursts was performed using RMfit, version 4.0rc1.  RMfit employs a modified, forward-folding 
Levenberg-Marquardt algorithm for spectral fitting which means that the aforementioned models are used as trial source spectra which are  converted to predicted detector count histograms. These histograms, in turn, are then compared to the actual, observed data.

In order to work properly, a method must be established to associate the energy deposited in the 
detectors to the energy of the detected photons, which depends on effective area and the angle of the detector 
to the incoming photons. We use detector response matrices (DRMs) to convert the photon energies into detector 
channel energies. DRMs contain information about the effective area of the detector, effects of the angular dependence of the detector efficiency, partial energy deposition in the detector, energy dispersion and nonlinearity of the detector and, finally, atmospheric and spacecraft scattering of photons into the detector. Therefore, DRMs are functions of photon energy, measured energy, the direction to the source (with respect to the spacecraft) and the orientation of the latter with respect to the Earth. The response matrices for all GRBs in the 
catalog were made using GBMRSP v2.0 of the response generator and version 2 of the GBM DRM database, and all responses 
employ atmospheric response modeling to correct for atmospheric scattering.  In particular, we use RSP2 files, which 
contain multiple DRMs based on the amount of slew the spacecraft experiences during the burst.  A new DRM is calculated for 
every $2^\circ$ of slew, changing the effective area of each detector based on its angle to the source.  These DRMs are then all 
stored in a single RSP2 file for each detector.  During the fitting process, each DRM is weighted by the counts fluence through 
the detector during each $2^\circ$ slew segment.

At all times, the Castor C-Statistic (C-Stat), which is a modified log likelihood statistic based on 
the Cash parametrization \citep{Cash} is used in the model-fitting process as a figure of merit to be minimized.  This statistic is 
preferable over the more traditional $\chi^2$ statistic minimization because of the non-Gaussian counting statistics present 
when dividing the energy spectra of GBM GRBs into 128 channels. The drawback of this
statistic is that it does not provide an estimation of the goodness-of-fit, since no standard probability distribution 
for likelihood statistics exists. 
Normally, the goodness-of-fit must be estimated for the model in use. This is usually done by simulating the model many times in order to calculate a test statistic, C-Stat in this case, for each fake model of each dataset. The observed value of the test statistic is then compared to the constructed fake test statistic distribution. This would allow to reject a model with a certain level of confidence. However, performing these kind of simulations is unfeasible for such a large sample of GRBs. In addition, any goodness-of-fit method can never provide a probability that any given model is the adequate representation of the burst emission, i.e. the correct model. Therefore, we refrain from performing extensive goodness-of-fit tests and present the fit parameters for each model fit, independent of the goodness-of-fit. As we only apply four different models, it can be instructive to study the fit parameters even if the model is not a perfect representation of the data. We do, however, apply additional selection criteria and employ cuts to define a GOOD and BEST sample of the obtained fit results. These are explained in the following subsections.

 Finally, we caution the reader that the $1\sigma$ errors reported in this catalog are of statistical nature only. Systematic effects were not considered which generally are non-negligible. For example, for weak events a different selection of \emph{on} and \emph{off} intervals for the background fitting can have effects as large as 30\% for the final reported value of \epeak \citep[see also][]{collazzi}.
 
\subsection{The GOOD sample}

We classify fitted burst models as GOOD if the parameter error of \emph{all} model parameters is within certain limits. This is a more conservative approach compared to \citet{Kaneko06} and \citet{goldstein12} who only required the error of the parameter of interest to be within given limits.  The motivation behind this new approach is to show parameter values of models which are \emph{globally} well--constrained, rather than basing interpretations on constrained parameters of overall poorly constrained models. We simultaneously require for the low-energy power law an error less than 0.4,  for high-energy power law indices an error less than 1.0 and for all other parameters we require a relative error of 0.4 or less. These criteria are an arbitrary choice but are in line with other GRB catalogs \citep{Kaneko06, goldstein12}. 
Applying these criteria, the number of bursts that classify as GOOD for each employed model can be seen in Table~\ref{BestTable}. We draw the reader's attention to the fact that for many GRBs there can be several models which qualify as GOOD.

\subsection{The BEST sample}

In addition, we define a BEST sample in order to determine which of the GOOD models is the best representation of the burst emission. Besides the necessity of having constrained parameters -- already required for the GOOD sample -- we compare the difference in C-Stat ($\Delta$C-Stat) per degree of freedom between the various models. In order to assess if a statistically more complex model, i.e. a model with more free fit parameters (hypothesis H1) is preferred over a simpler model (null hypothesis H0) according to its difference in C-Stat, we created a set of $5\times10^4$ synthetic GRB spectra using the fit parameters of H0, obtained from the fit to the real data, as source counts. Similarly, the background counts of the synthesized spectra are estimated from the real data. The input source counts are then folded through the detector response matrix (DRM). Finally, Poisson noise was added to the sum of the source and background counts. 
The synthetic spectra were then fitted with both models, adding Poisson fluctuations to each energy channel of the background spectrum during the fit process.

Integrating the $\Delta$C-Stat distribution from 0 to 99.73\% (i.e. a 3$\sigma$ confidence interval) we identified a critical $\Delta$C-Stat$_{\rm crit}$. If the $\Delta$C-Stat observed in the real data exceeds this critical value, then the null hypothesis is rejected and the statistically more complex model is preferred. 

As the full GRB sample contains nearly 1000 GRBs doing the aforementioned analysis for all bursts is not feasible. We used four typical bursts which are located in four energy fluence classes, separated by one order of magnitude. These bursts are GRB~120608.489, GRB~110227.240, GRB~120129.580 and GRB~120526.303 with an energy fluence [10~keV -- 1~MeV] of $5.2\times 10^{-7}$~erg~cm$^{-2}$, $1.9\times10^{-6}$~erg~cm$^{-2}$, $5.8\times10^{-5}$~erg~cm$^{-2}$, $1.3\times^{-4}$~erg~cm$^{-2}$, respectively. 

For each of the aforementioned GRBs, we created a \pwrlw vs \comp and a \comp vs \band distribution and determined the $\Delta$C-Stat$_{\rm crit}$ for each model comparison. To assess which of the four models qualifies for the BEST sample we used the critical values as a function of fluence. For the comparison between the \band and the \sbpl such simulations were not necessary, as both functions have the same number of degrees of freedom. Therefore, the model with the lower C-Stat value was used.

As there is no observable correlation between $\Delta$C-Stat$_{\rm crit}$ and the energy fluence (see Table~\ref{CstatFluence}) we use the average $\Delta$C-Stat$_{\rm crit}$ value, for the model selection of all bursts, i.e. for \pwrlw vs \comp we use $\Delta$C-Stat$_{\rm crit}=8.58$ and for \comp vs \band we use $\Delta$C-Stat$_{\rm crit}=11.83$

The key idea for the BEST parameter sample is to obtain the best estimate of the observed properties of GRBs.  By using model comparison, the 
preferred model is selected, and the parameters are reviewed for that model.  The models contained herein and in most 
GRB spectral analyses are empirical models, based only on the data received; therefore the data from different GRBs tend to 
support different models.  Perhaps it will be possible to determine the physics of the emission process by investigating the 
tendencies of the data to support a particular model over others.  This is the motivation to provide a 
sample that contains the best picture of the global properties of the data, that prompts the investigation of the BEST sample.

Applying the BEST criteria we are left with 941 GRBs for the fluence spectra and 932 GRBs for the \emph{P} spectra (see again Table\ref{BestTable}). These numbers are smaller than the total number of GRBs in this catalog (943). This is due to the fact that for some bursts the spectral fit did not converge properly and these bursts have been excluded from the BEST sample.

In Table \ref{BestTable} we present the composition of models for the BEST samples.  From this table, it is apparent that the 
\emph{F} spectral fits strongly favors the \comp model over the others in over half of all GRBs.  The \band and \sbpl 
are favored by only few GRBs in the catalog.  It should be noted that the number of GRBs best fit by \pwrlw increases in the 
\emph{P} selection mainly due to the fact that the smaller statistics from the short integration time are unable to support a model 
more complex than the \pwrlw.  

\subsection{The redshift sample}

\begin{figure}
	\begin{center}
		\includegraphics[scale=0.5]{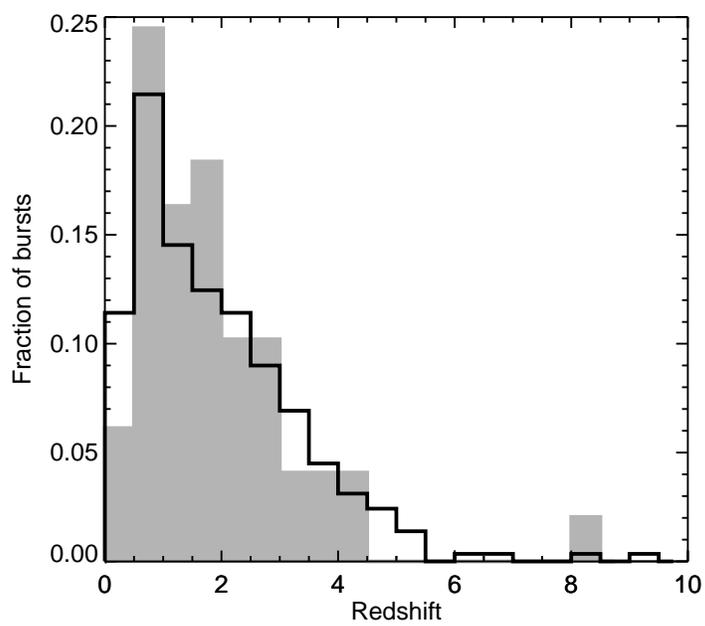}
	\end{center}
\caption{Redshift distribution of all GRBs with known redshift to date (black histogram) and GBM GRBs with known redshift (gray filled histogram) normalized to the area.}\label{fig:redshift}
\end{figure}

In addition to the BEST sample, we form a sample of GRBs with known redshift\footnote{\tt{www.mpe.mpg.de/\textasciitilde jcg/grbgen.html}} (determined either spectroscopically or photometrically) to investigate the rest-frame properties of the GRBs \citep[see e.g.][]{restframe}. The redshift distribution of the GBM GRBs with known redshift to date is shown in Figure~\ref{fig:redshift}. The redshift sample contains 45 triggered GRBs with an additional 3 untriggered GRBs \citep{untrig} for the \emph{F} spectral fits. As apparent from Figure~\ref{fig:redshift}, the sample of GBM GRBs with measured redshift is compatible with the full sample of GRBs with redshift and therefore it can be concluded that the GBM observations do not introduce a new bias for bursts with measured redshift. This is also confirmed by an KS-test which yields a probability of $98$~\% that the two distributions are drawn from the same population.

At all times the cosmological parameters obtained from the \emph{Planck} mission \citep{planck} with $H_0=67.3$~km~s$^{-1}$~Mpc$^{-1}$, $\Omega_m=0.315$ \citep{cosmo} were used for this analysis.

\section{Results}
\subsection{Time-integrated \emph{F} spectral fits}
\begin{figure}
	\begin{center}
		\subfigure[]{\label{allloidx}\includegraphics[scale=0.45]{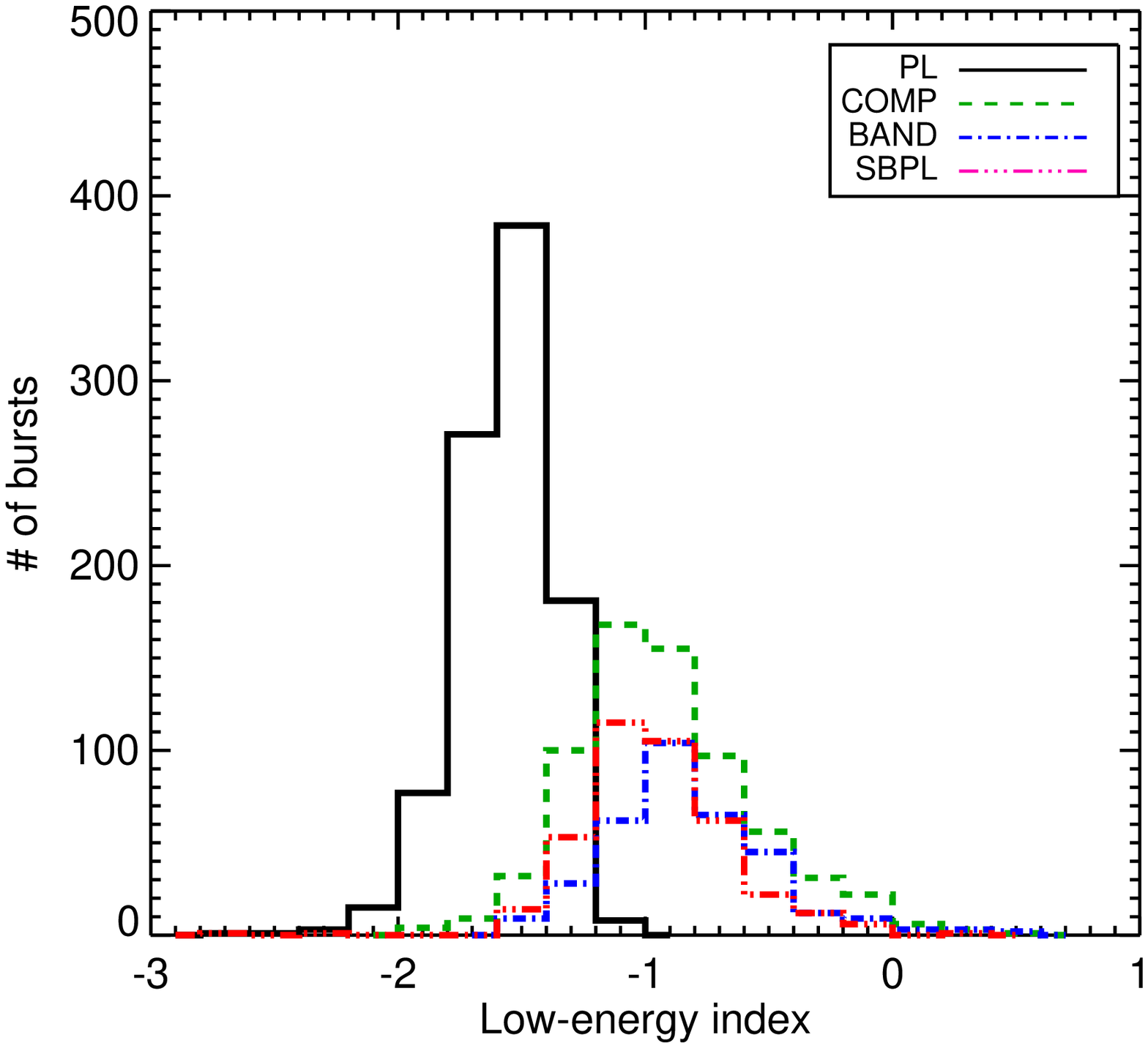}}
		\subfigure[]{\label{allloidxbest}\includegraphics[scale=0.45]{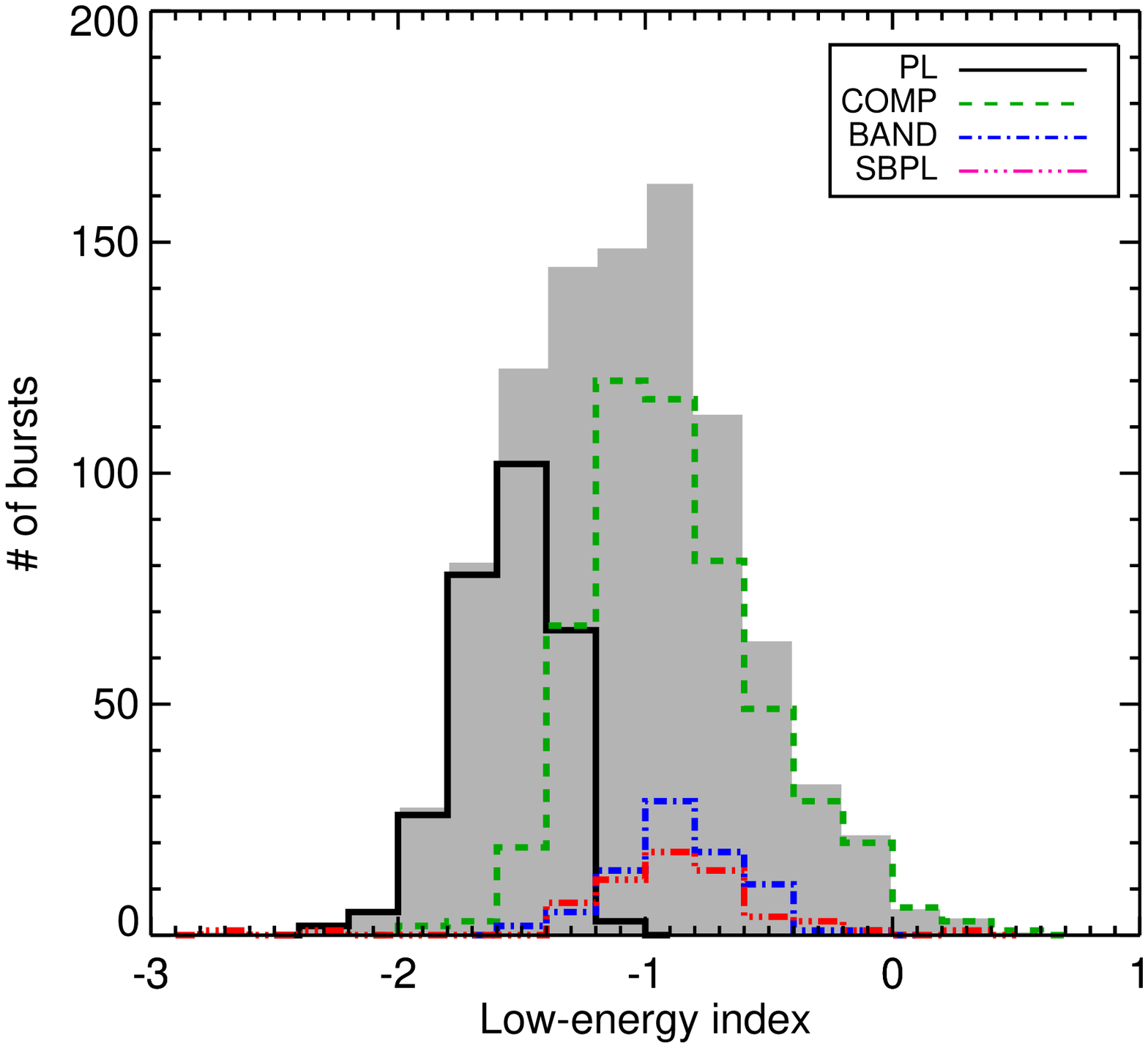}}
	\end{center}
\caption{Distribution of the low-energy indices obtained from the GOOD \emph{F} spectral fits \subref{allloidx}. The BEST parameter distribution (gray filled histogram) and its constituents for the low-energy index is shown in \subref{allloidxbest}. 
}\label{loenindex}
\end{figure}

\begin{figure}
	\begin{center}
		\subfigure[]{\label{allhiidx}\includegraphics[scale=0.45]{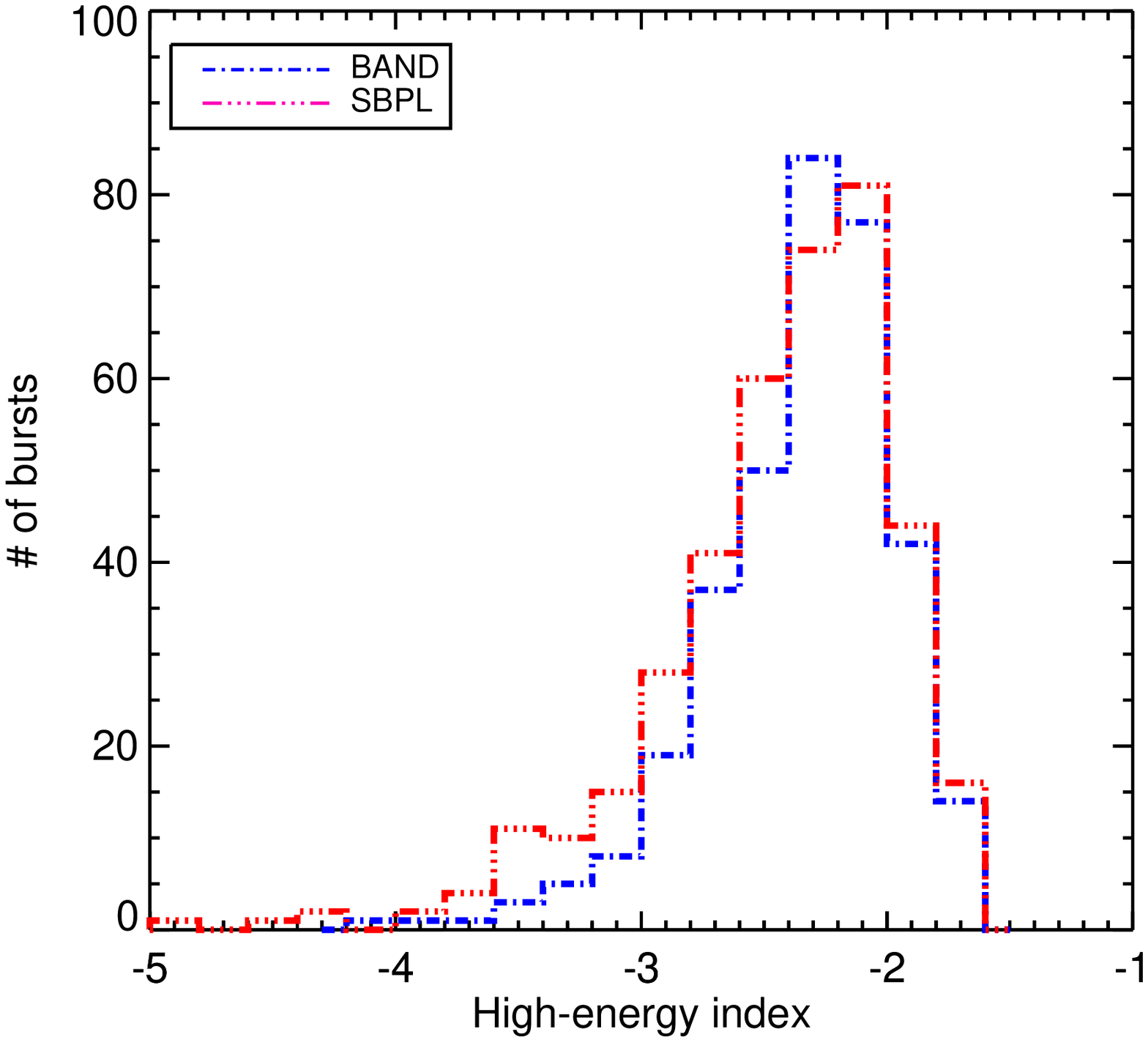}}
		\subfigure[]{\label{allhiidxbest}\includegraphics[scale=0.45]{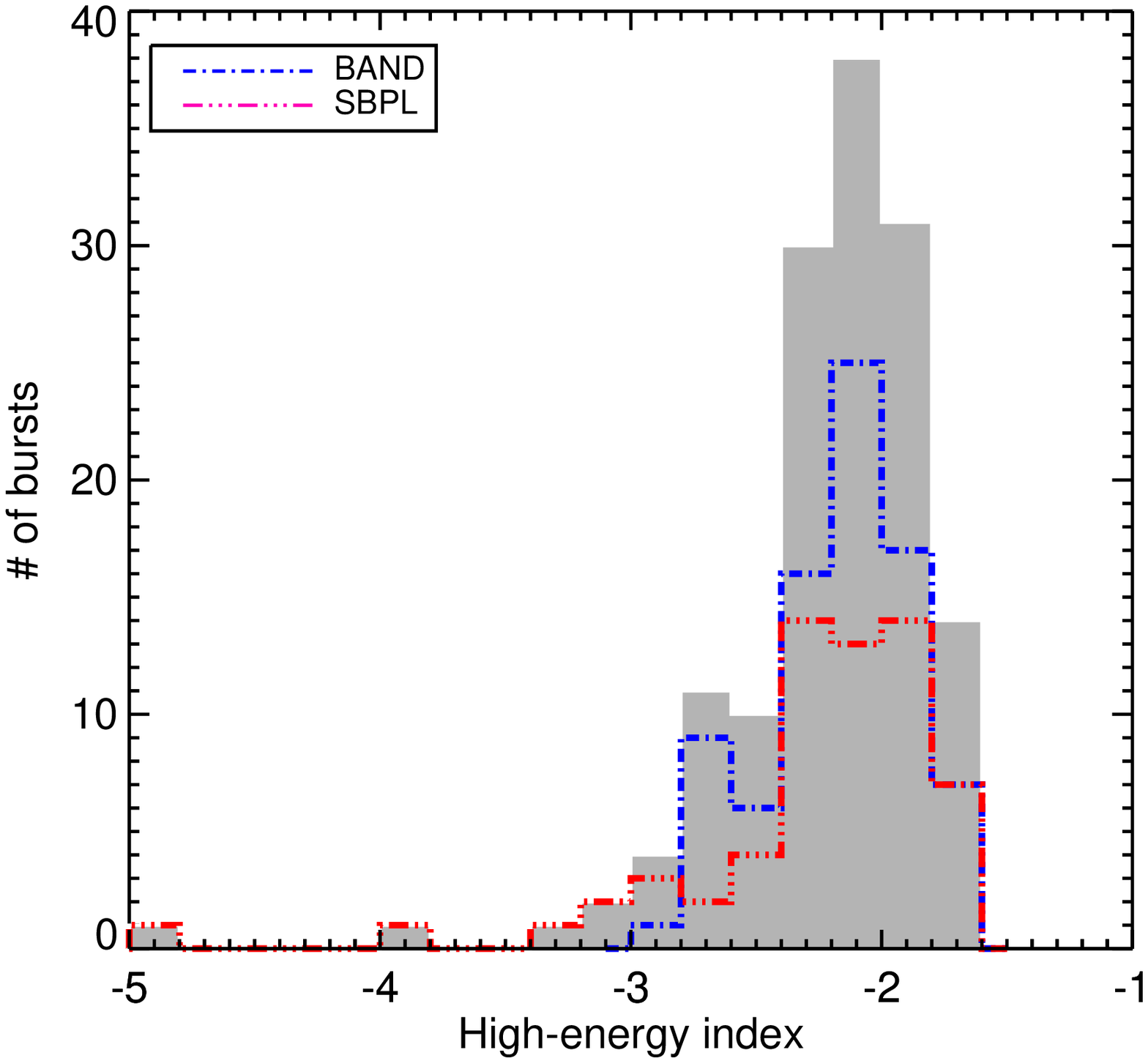}}
	\end{center}
\caption{Distribution of the high-energy indices obtained from the GOOD \emph{F} spectral fits \subref{allhiidx}. The BEST parameter distribution (gray filled histogram) and its constitutents for the high-energy index is shown in \subref{allhiidxbest}. 
}\label{highenindex}
\end{figure}

The time-integrated spectral distributions depict the overall emission properties and do not take into account any spectral evolution.  The low-energy indices, as shown in Figure \ref{loenindex}, are distributed about a $-1.1$ power law typical of most GRBs.  Up to 17\% of the BEST low-energy indices exceed $\alpha > -2/3$, violating the synchrotron ``line-of-death'' \citep{LineOfDeath}, while an additional 18\% of the indices are $\alpha < -3/2$, violating the synchrotron cooling limit.
The high-energy indices in Figure \ref{highenindex} peak at a slope of about $-2.1$ and 
have a long tail toward steeper indices.  Note that very steep high-energy indices in the 
distribution of all high-energy index values indicate that the spectrum of these GRBs mimics closely a \comp model, which is 
equivalent to a \band function with a high-energy index of $- \infty$. 
A significant fraction of \band model fits results in a high-energy power-law index $\beta > -2$ which would indicate a divergent energy flux at high energies. However, it was pointed out by \citet{kocevskibeta} that the inclusion of \fermi/LAT upper limits in the fitting process results in considerably steeper (softer) high-energy power-law indices. The median value decreased from $\beta \sim -2.2$ from the \gbm-only fits to $\beta \sim -2.5$ for the \gbm and LAT joint spectral fits. Indeed, not a single case of their sample had a high-energy power-law index $> -2$ after LAT data had been included. \citet{kocevskibeta} conclude that intrinsic spectral breaks and/or softer-than-measured high-energy spectra must be fairly common in the GRB population in order to explain the lack of LAT-detected GRBs.

The comparison of the single power law index to the low- 
and high-energy indices makes evident that the simple power law index is averaged over the break energy, resulting in an 
index that is on average steeper than the low-energy index, yet shallower than the high-energy index.  In addition, the lack of power-law indices steeper than $-2$ suggest that \gbm does not detect a population of GRBs with low \epeak. If a burst had a Band spectrum with \epeak near or below the \gbm energy threshold a single power-law fit should have an index smaller than -2.

\begin{figure}
	\begin{center}
		\subfigure[]{\includegraphics[scale=0.45]{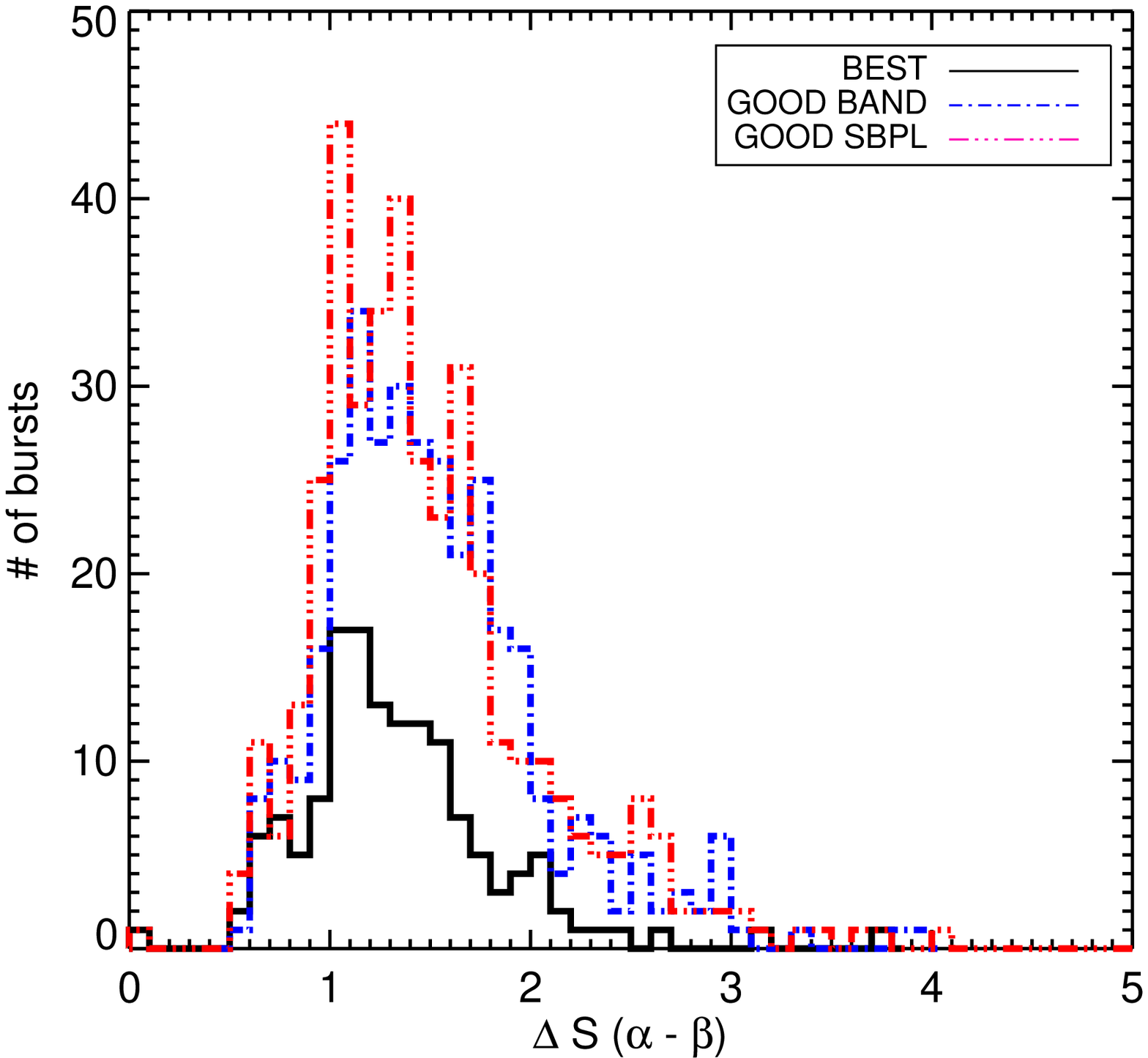}}
	\end{center}
\caption{Distributions of $\Delta S$, the difference between the low- and high-energy spectral indices ($\alpha - \beta$) for the \emph{F} spectral fits. The BEST  (black solid line), the GOOD BAND power-law indices (blue dash-dotted line) and the GOOD \sbpl power-law indices (red dash-dot-dot-dotted line) for the \emph{F} spectral fits are shown.  The first bin contains values less than 0, indicating that the centroid value of $\alpha$ is steeper than the centroid value of $\beta$.  \label{deltasbestfluence}}
\end{figure}

We also show in Figure~\ref{deltasbestfluence} the difference between the time-integrated low- and high-energy spectral indices, $\Delta S =(\alpha-\beta)$.  This quantity is useful since the synchrotron shock model (SSM) makes predictions of this value in a number of cases \citep{Preece} and the power--law index, $p$, of the electron distribution can be inferred from $\Delta S$.  The distributions of $\Delta S$ obtained from the GOOD and BEST \emph{F} spectral fits are consistent with the time-resolved results in \citet{Preece}, as well as the time--integrated results in \citet{Kaneko06}, peaking at $\Delta S \sim 1$ with a median value of $1.25$.

\begin{figure}
	\begin{center}
		\subfigure[]{\label{allebreak}\includegraphics[scale=0.45]{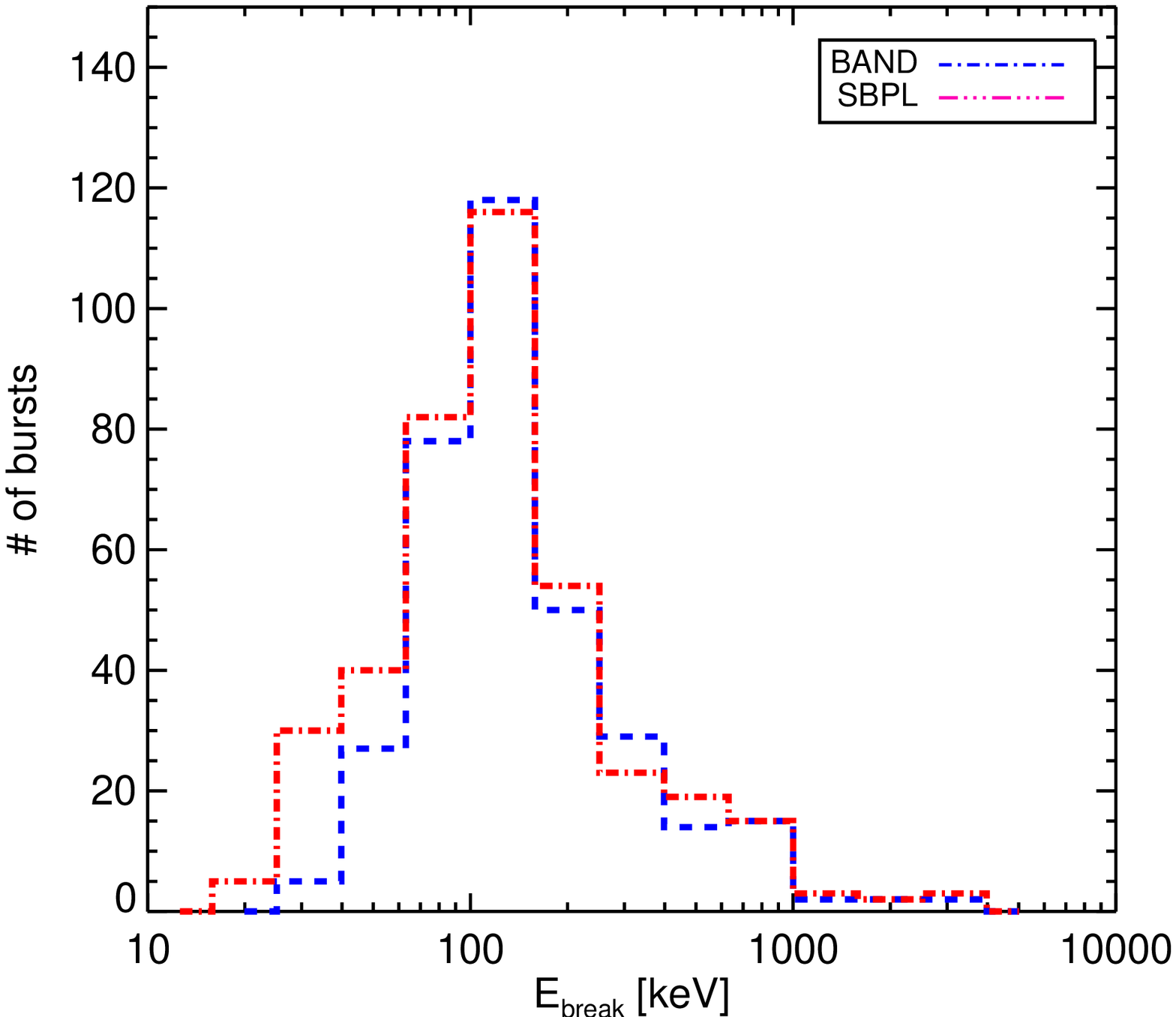}}
		\subfigure[]{\label{allebreakbest}\includegraphics[scale=0.45]{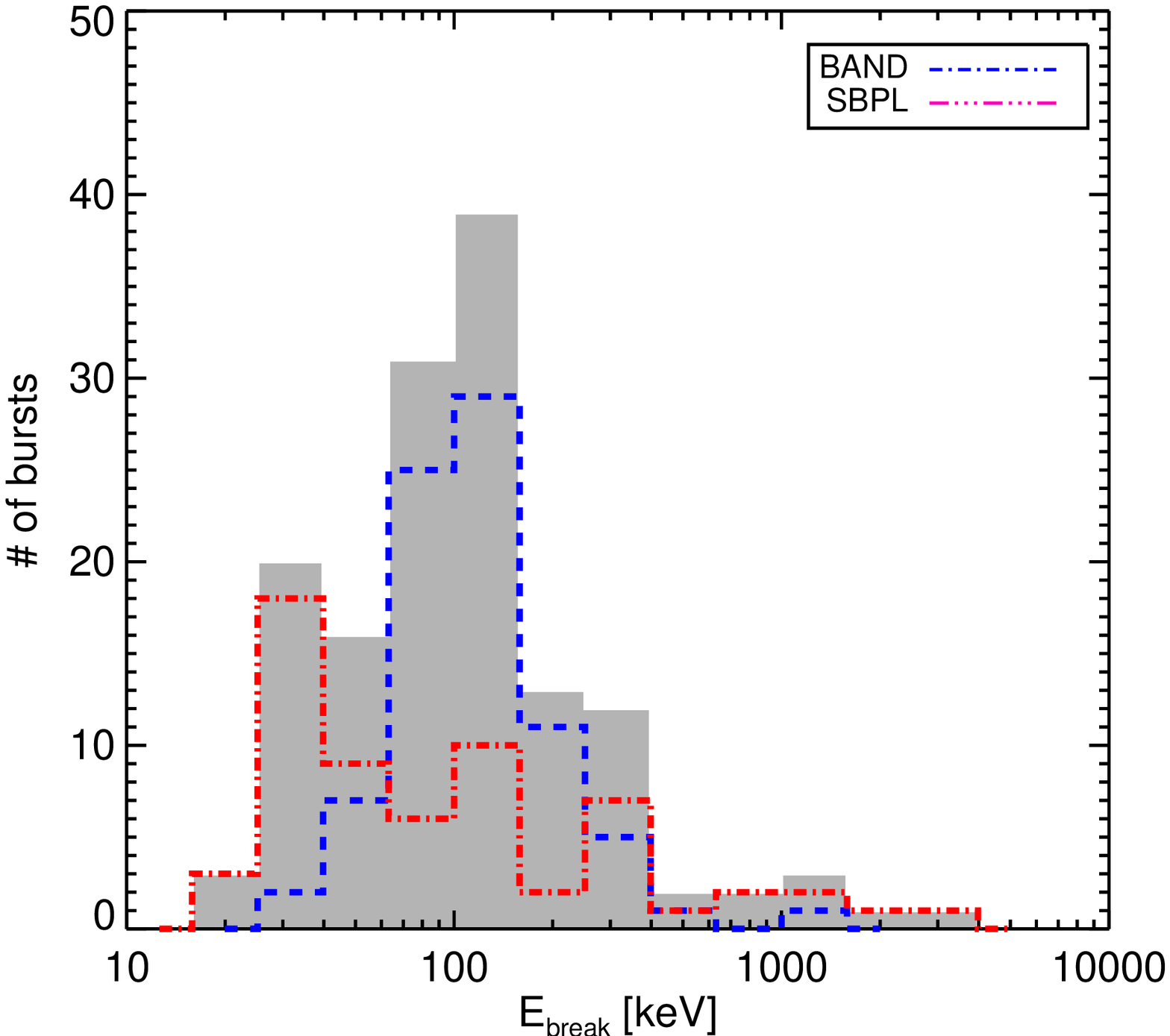}}
	\end{center}
\caption{Distribution of the \ebreak obtained from the GOOD \emph{F} spectral fits \subref{allebreak}. The BEST parameter distribution (gray historgam) and its constituents for \ebreak is shown in \subref{allebreakbest}. The \ebreak of the \band model fits has been derived following \citet{Kaneko06}. 
}\label{ebreakf}
\end{figure}

\begin{figure}
	\begin{center}
		\subfigure[]{\label{allepeak}\includegraphics[scale=0.45]{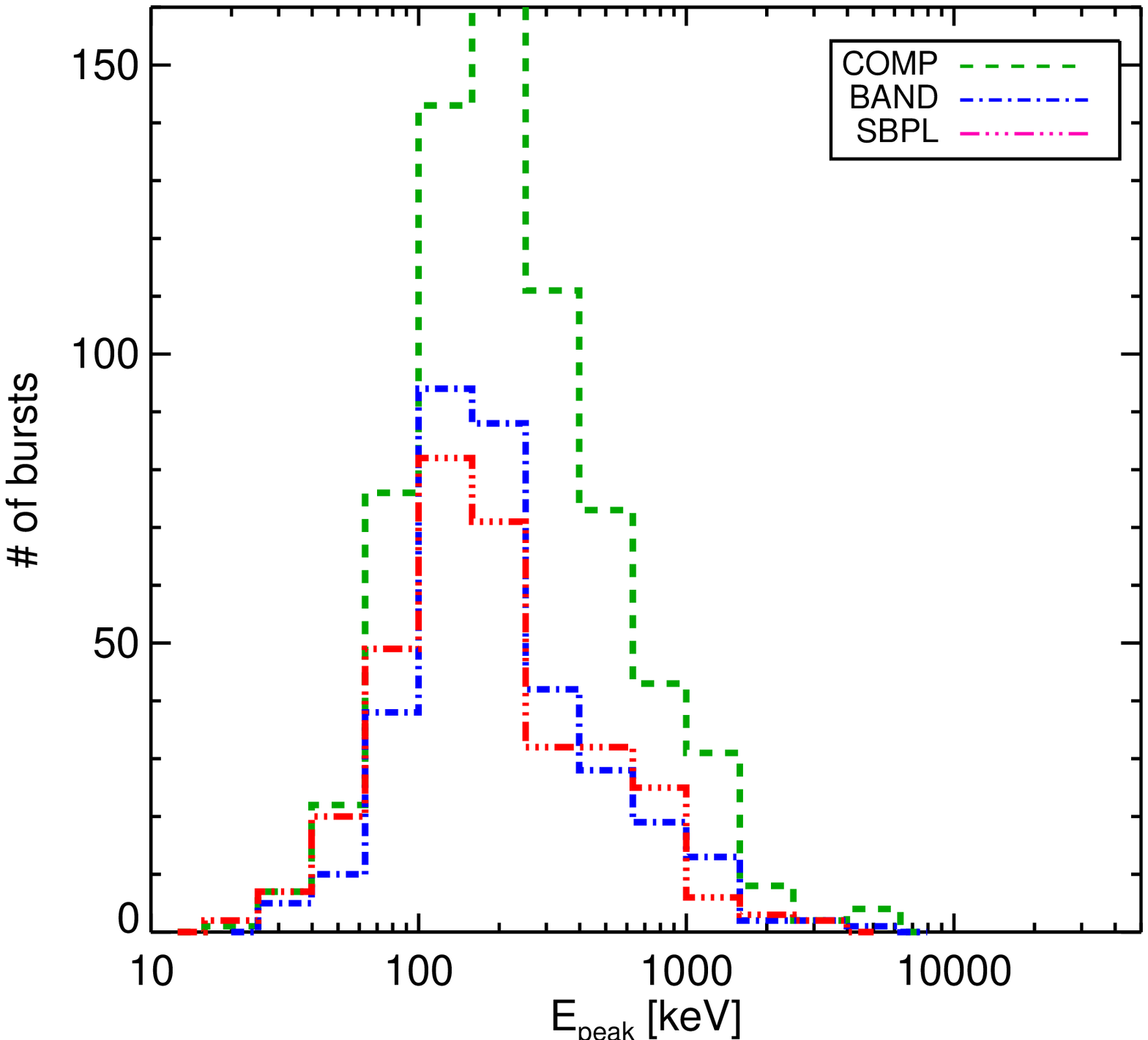}}
		\subfigure[]{\label{allepeakbest}\includegraphics[scale=0.45]{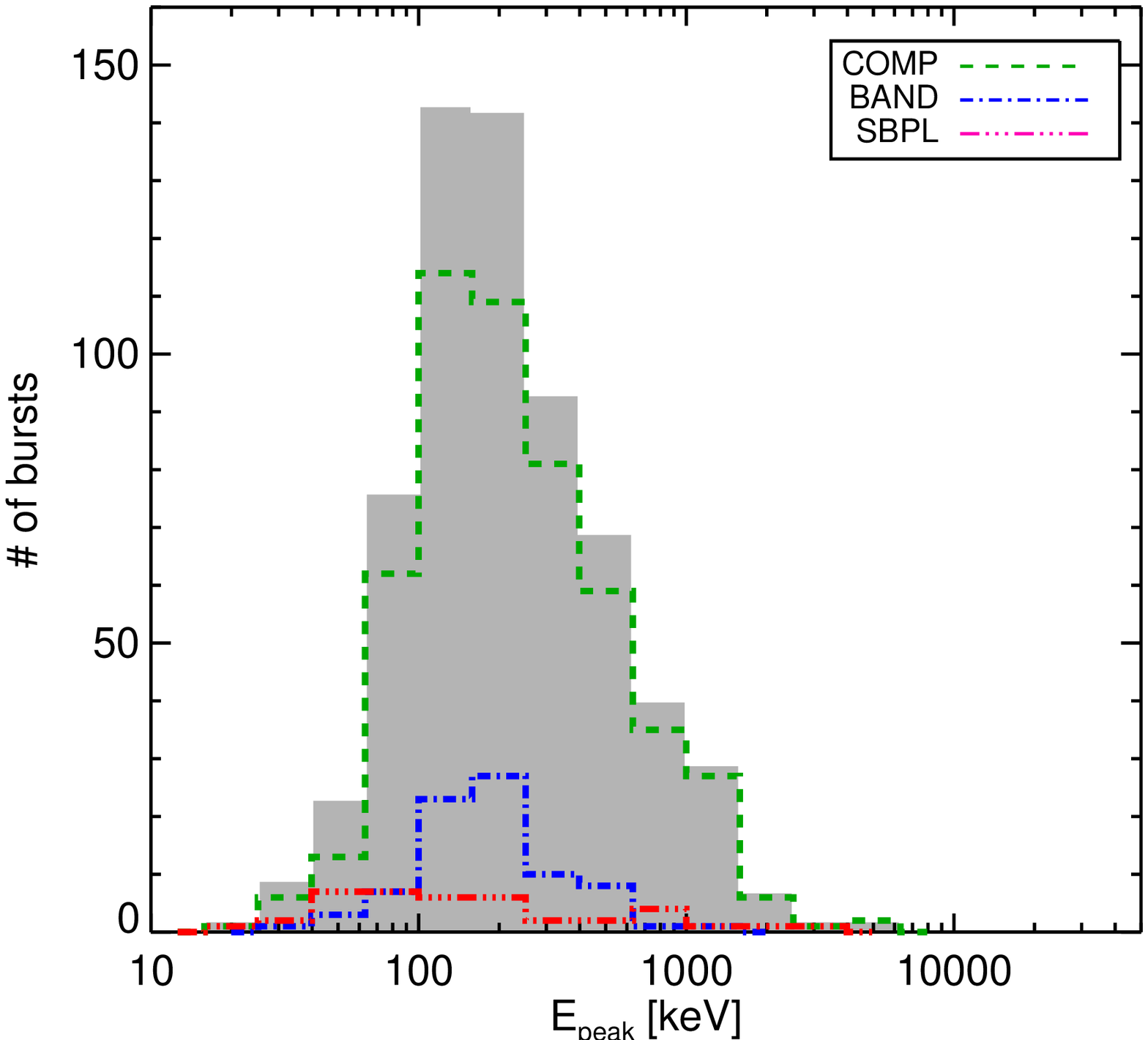}}
	\end{center}
\caption{Distribution of the \epeak obtained from the GOOD \emph{F} spectral fits \subref{allepeak}. The BEST parameter distribution (gray filled histogram) and its constituents for \epeak is shown in \subref{allepeakbest}. 
}\label{epeakf}
\end{figure}

In Figure \ref{ebreakf} and Figure \ref{epeakf}, we show the distributions for the break energy, \ebreak and the peak of the power density 
spectrum, \epeak, respectively.  \ebreak is the energy at which the low- and high-energy power laws are joined, which is not 
necessarily representative of the \epeak.  However, one can easily derive \ebreak from the \band \epeak values following \citet{Kaneko06}. The \ebreak for the GOOD \sbpl and \band fits has a clustering about 120 keV spanning roughly two orders of magnitude. The \ebreak distribution for the BEST sample is skewed to slightly smaller values than the GOOD sample implying that the \sbpl is more likely to be statistically preferred if \ebreak is low.  
The GOOD and BEST \epeak distributions all generally peak around 150 - 200~keV  and 
cover just over two orders of magnitude, which is consistent with previous findings \citep{Mallozzi95, Lloyd00} from BATSE.  As 
discussed in \citet{Kaneko06}, although the \sbpl is parametrized with \ebreak, the \epeak can be derived from the 
functional form.  We have calculated the \epeak for all bursts with low-energy index shallower than -2 and high-energy 
index steeper than -2.  

The overall distribution of \epeak is similar to that found using the BATSE Large Area Detectors, which 
had a much smaller bandwidth and larger collecting area.  This would seem to indicate that it is unlikely a hidden 
population to be undiscovered within the ~10 keV - 40 MeV range which is in line with the results by \citet{harris98}.  

\begin{figure}
	\begin{center}
		\subfigure[]{\label{allepeakvsalpha}\includegraphics[scale=0.45]{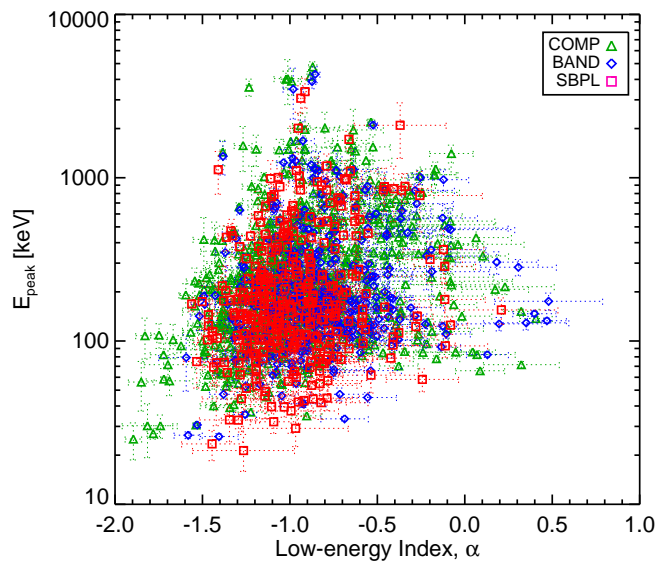}}
		\subfigure[]{\label{allepeakvssigalpha}\includegraphics[scale=0.45]{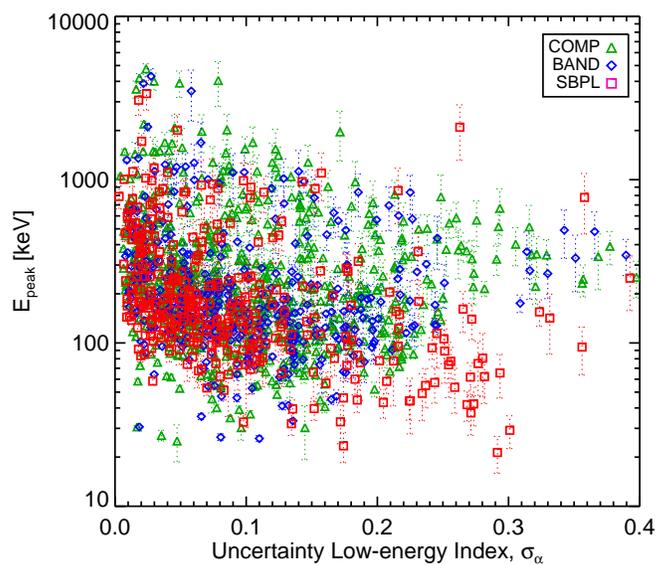}}
	\end{center}
\caption{Comparison of the low-energy index and \epeak for three models from the \emph{F} spectral fits.  This comparison 
reveals a correlation between \subref{allepeakvsalpha}  the \epeak energy and the low-energy power-law index and \subref{allepeakvssigalpha} the \epeak and the uncertainty on the low-energy index. Generally a  lower energy 
\epeak tends to result in a softer and a less constrained low-energy index. \label{alphaepeak}}
\end{figure}

The value of \epeak can 
strongly affect the measurement of the low-energy index of the spectrum, as shown in Figure \ref{alphaepeak}.  A general trend 
appears to show that spectra with smaller \epeak values also have smaller values of the low-energy power-law index. This is due to the fact that when \epeak is close to the instrument's lower energy sensitivity limit, $\alpha$ has not yet reached its asymptotical value and is thus, on average, softer than it is in reality.
In addition, smaller \epeak values tend to increase the uncertainty in the measurement of the low-energy index, 
mostly due to the fact that a spectrum with a low \epeak will exhibit most of its curvature near the lower end of the instrument 
bandpass.  

\begin{figure}
	\begin{center}
		\subfigure[]{\label{alldeltaep}\includegraphics[scale=0.45]{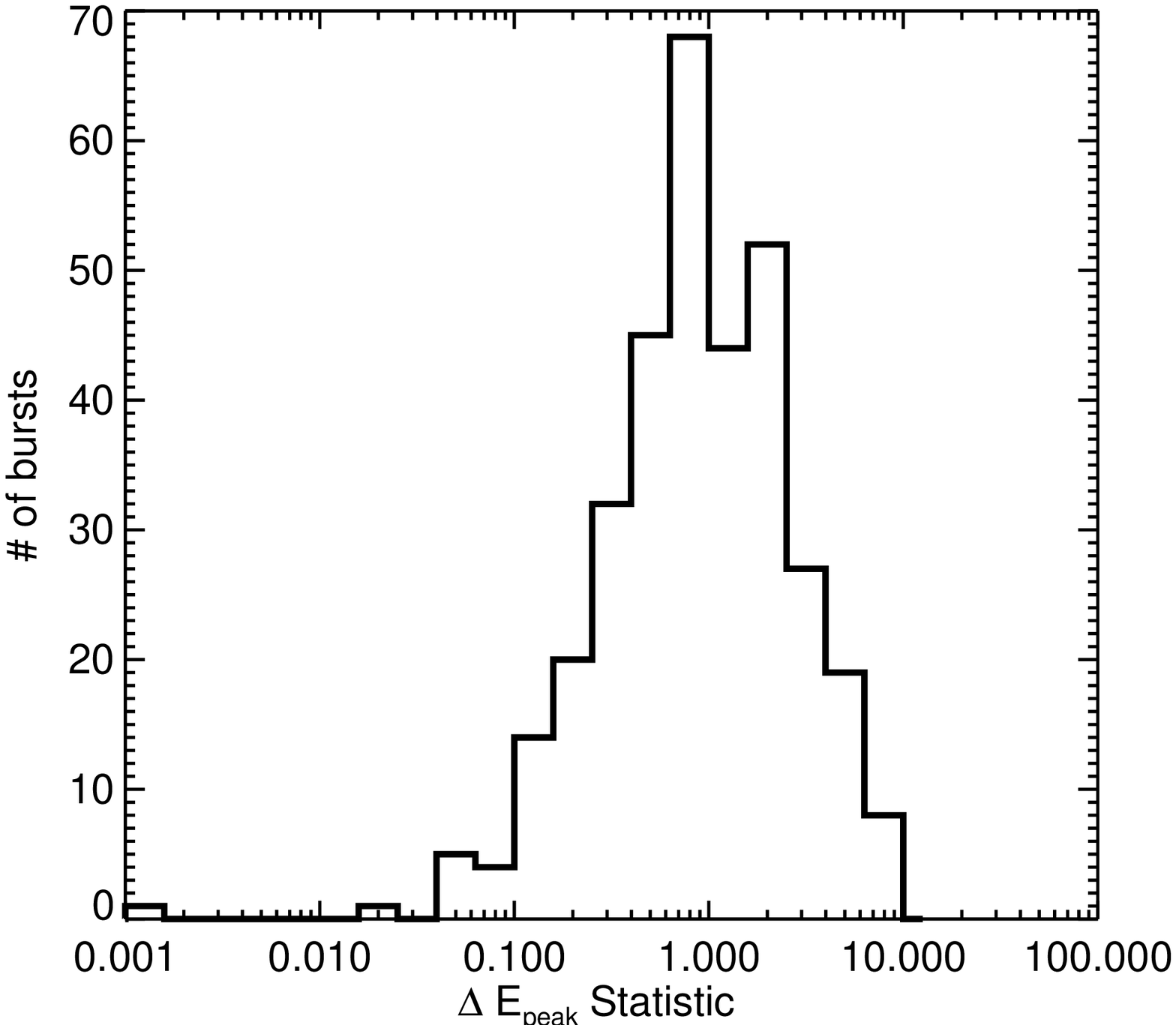}}
	\end{center}
\caption{Distribution of the $\Delta E_{\rm peak}$ statistic for the GOOD \comp and \band models from \emph{F} spectral fits.  A value less than 1 indicates the \epeak values are within errors, while a value larger than 1 indicates the \epeak values are not within errors. \label{deltaepeakf}}
\end{figure}

It is of interest to study the difference in the value of \epeak between the \band and \comp functions since they are the two 
main functions used to study GRB spectra, and \comp is a special case of \band.  To study the relative deviation between the 
two values we calculate a statistic based on the difference between the values and taking into account their 1$\sigma$ errors.  
This statistic can be calculated by
\begin{equation}
	\Delta E_{\rm peak}=\frac{|E_{\rm peak}^{C}-E_{\rm peak}^{B}|}{\sigma_{E_{\rm peak}}^{C}+\sigma_{E_{\rm peak}}^{B}}
\end{equation}
where C and B indicate the \comp and \band values respectively.  This statistic has a value of unity when the deviation 
between the \epeak values exactly matches the sum of the 1$\sigma$ errors.  A value less than one indicates the \epeak values are within errors, and a value greater than one indicates that the \epeak values are not within 1$\sigma$ errors of 
each other.  Figure \ref{deltaepeakf} depicts the distribution of the statistic. Roughly 46\% of the \band and \comp \epeak
values are found to be outside the combined errors.  This indicates that, although \comp is a special case of \band, a 
significant fraction of the \epeak values can vary by more than 1$\sigma$ based on which model is chosen. If the allowed error range is extended to 3$\sigma$ then only 14\% of the \epeak values are not consistent. 

\begin{figure}
	\begin{center}
		\begin{minipage}[t]{1\textwidth}
		\subfigure[]{\label{allefluxf}\includegraphics[scale=0.45]{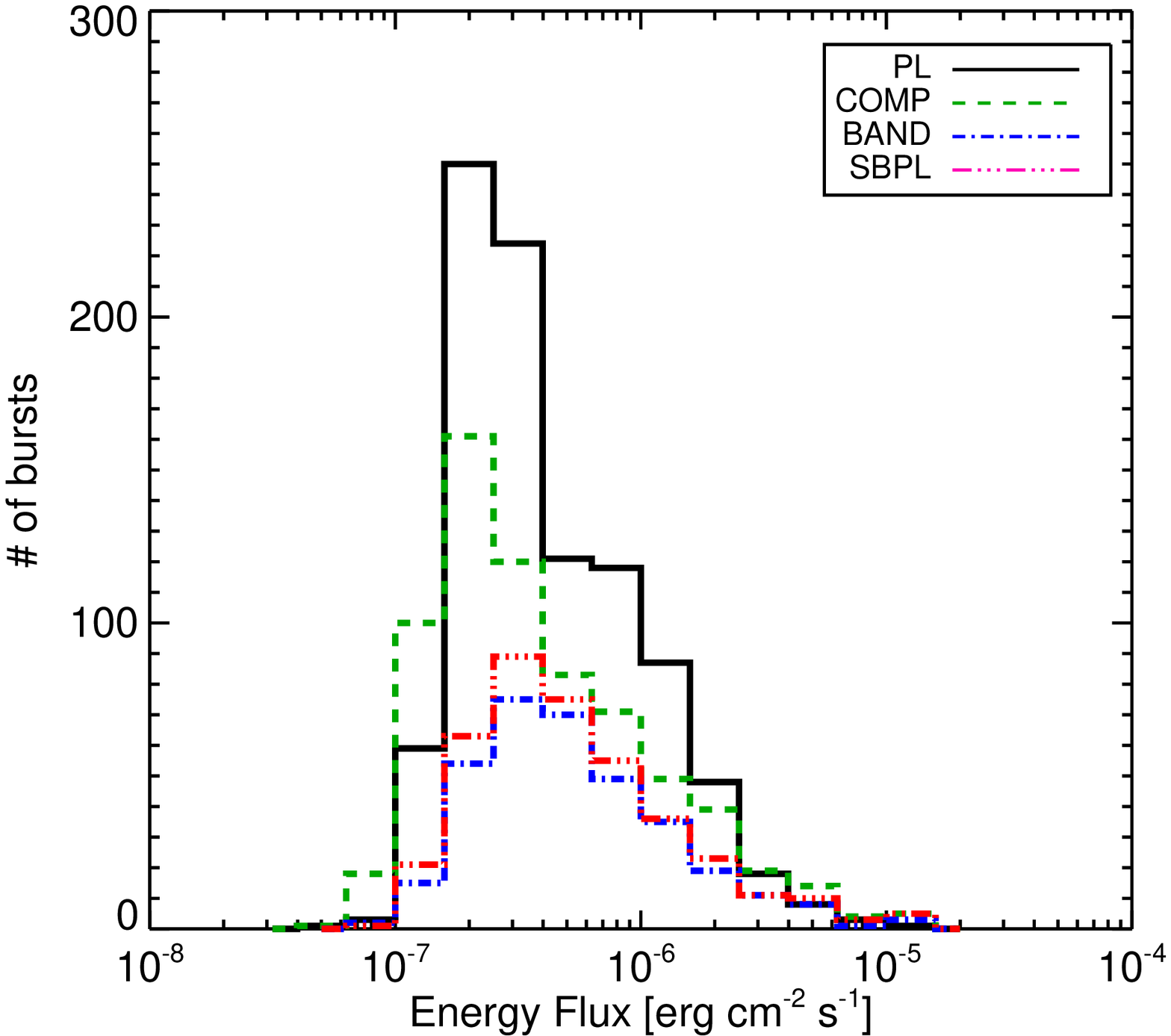}}
		\subfigure[]{\label{allefluxf40MeV}\includegraphics[scale=0.45]{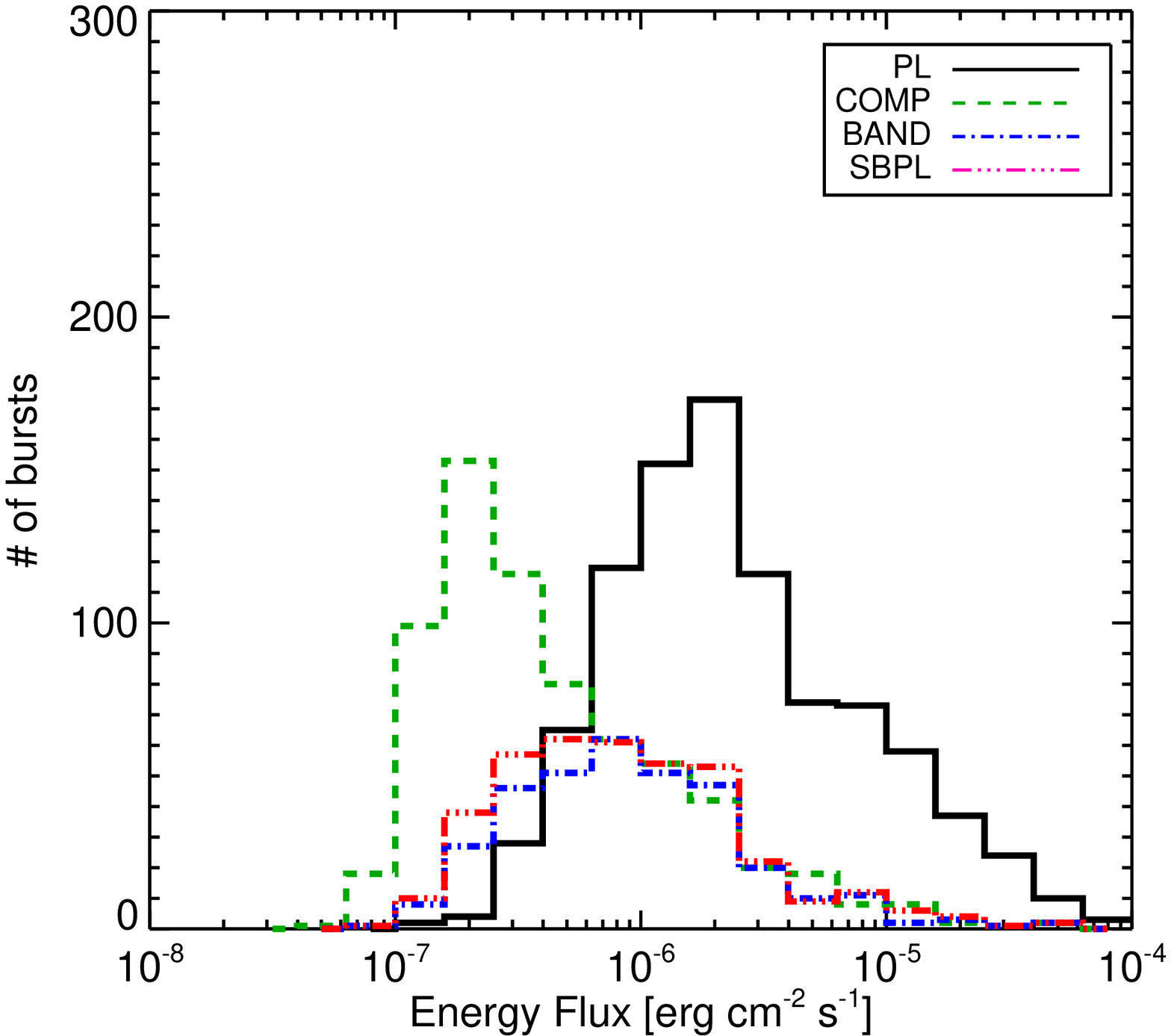}}
		\end{minipage}
		\begin{minipage}[t]{1\textwidth}
		\subfigure[]{\label{allefluxfbest}\includegraphics[scale=0.45]{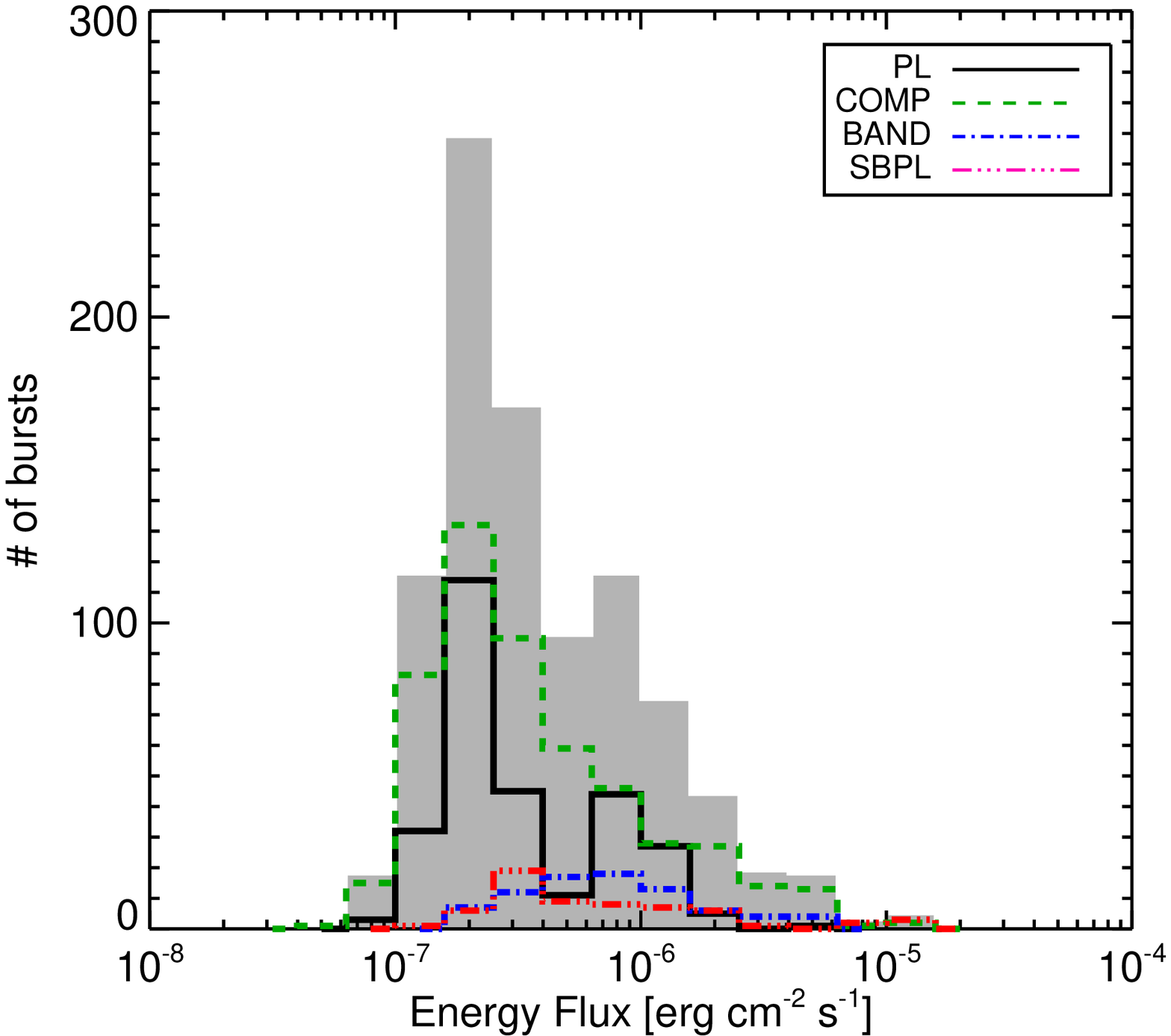}}
		\subfigure[]{\label{allefluxf40MeVbest}\includegraphics[scale=0.45]{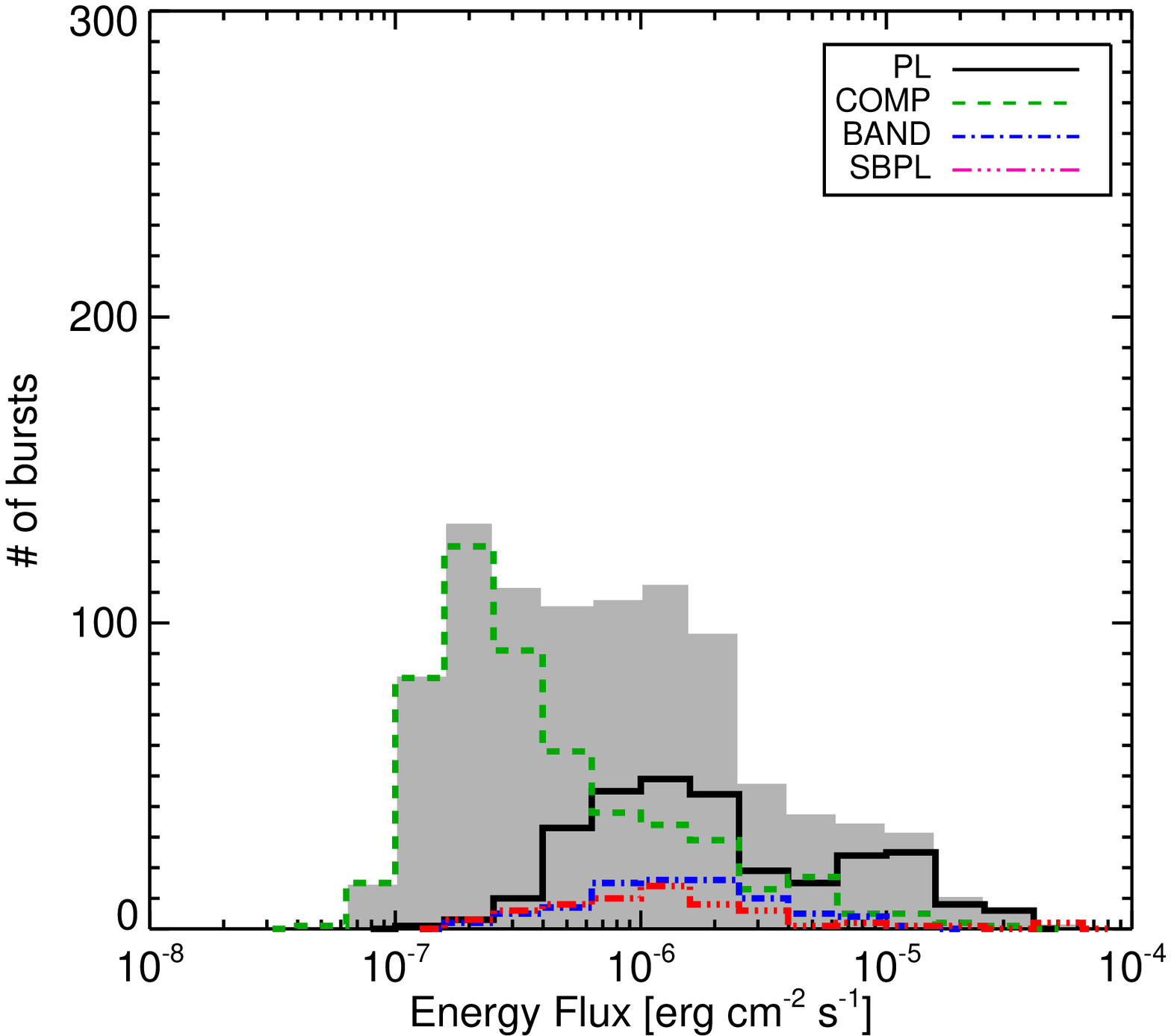}}
		\end{minipage}
	\end{center}
\caption{Distributions of energy flux in the 10~keV--1~MeV \subref{allefluxf} and 10~keV--40~MeV \subref{allefluxf40MeV} band from the GOOD \emph{F} spectral fits. Note that the plotted distributions contain the flux on two different timescales: 1024 ms (long GRBs) and 64 ms (short GRBs). The BEST parameter distribution for the energy flux in both energy ranges is shown in \subref{allefluxfbest} and \subref{allefluxf40MeVbest}. The gray filled histogram shows the total distribution and the constituents are shown in colors.\label{enluxf}}
\end{figure}

\begin{figure}
	\begin{center}
		\begin{minipage}[t]{1\textwidth}
		\subfigure[]{\label{allpfluxf}\includegraphics[scale=0.45]{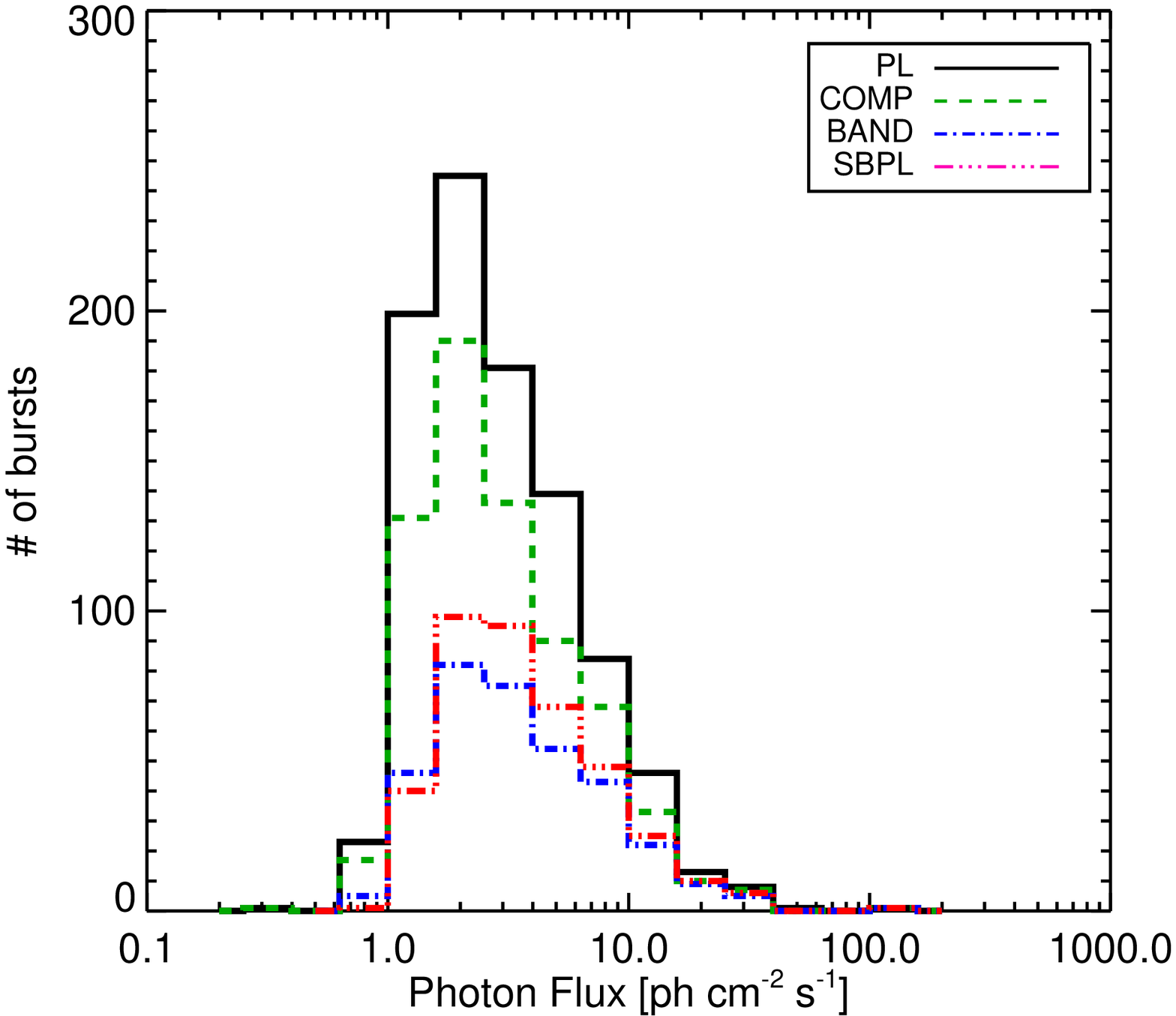}}
		\subfigure[]{\label{allpfluxf40MeV}\includegraphics[scale=0.45]{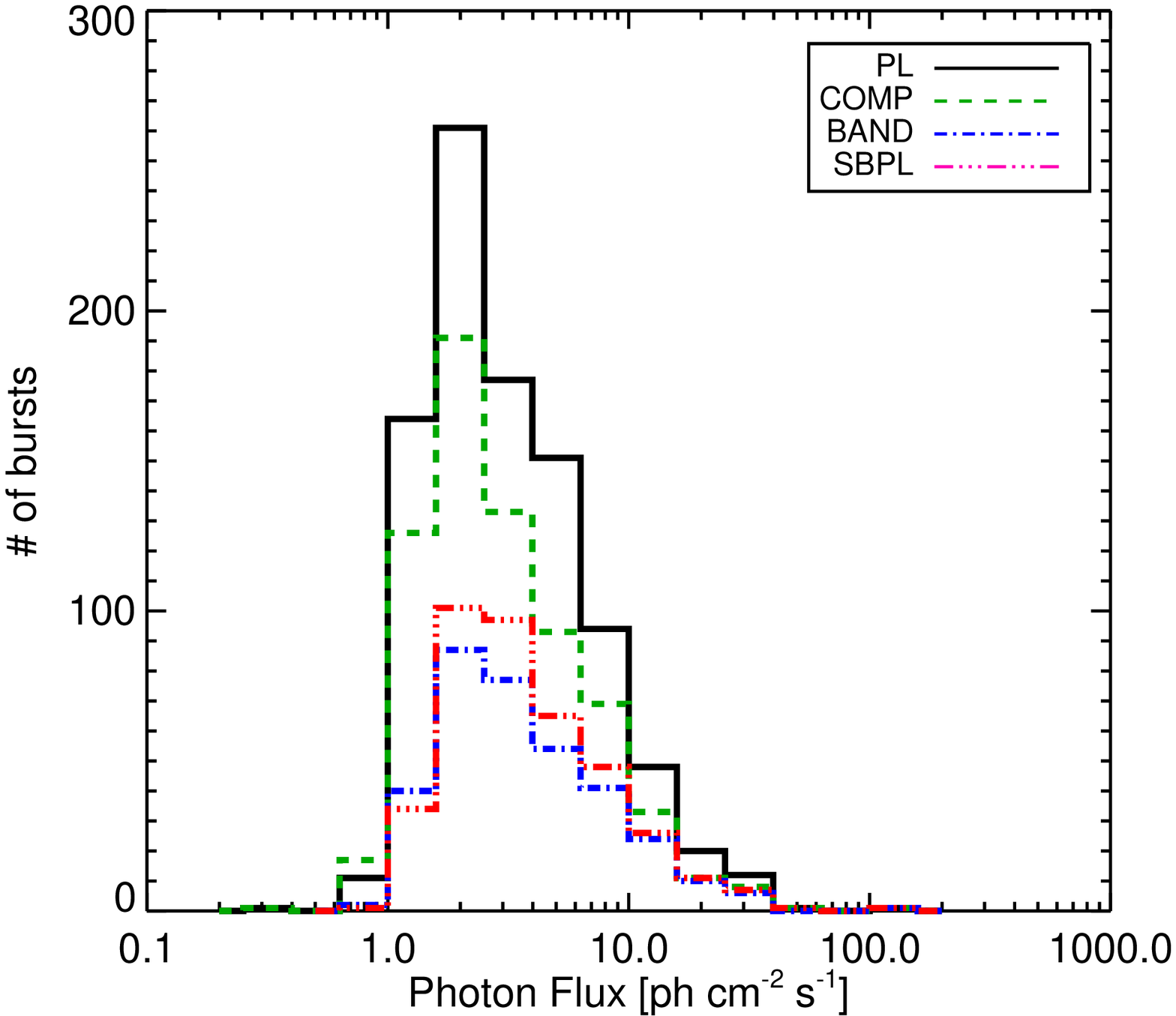}}
		\end{minipage}
		\begin{minipage}[t]{1\textwidth}
		\subfigure[]{\label{allpfluxfbest}\includegraphics[scale=0.45]{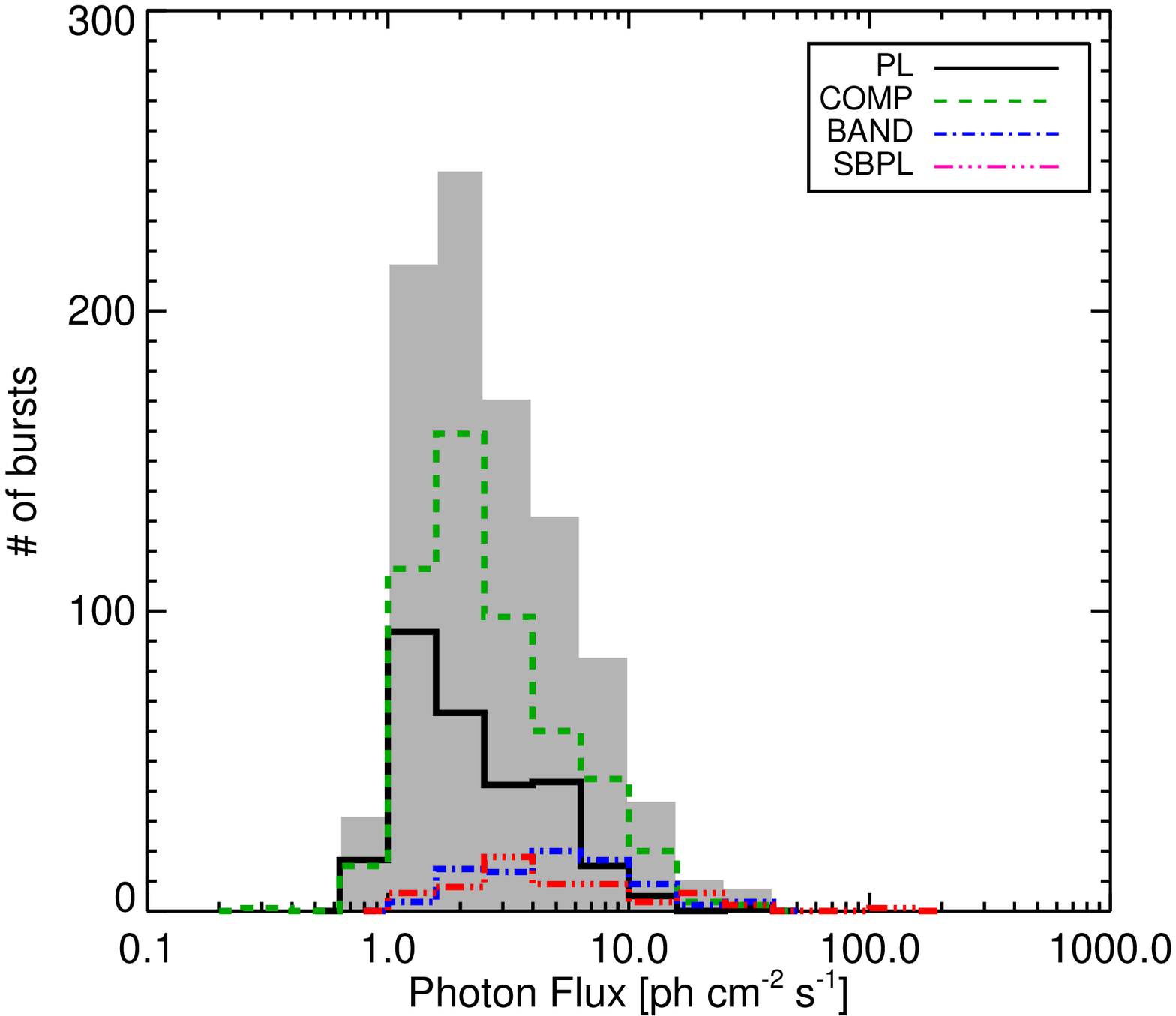}}
		\subfigure[]{\label{allpfluxf40MeVbest}\includegraphics[scale=0.45]{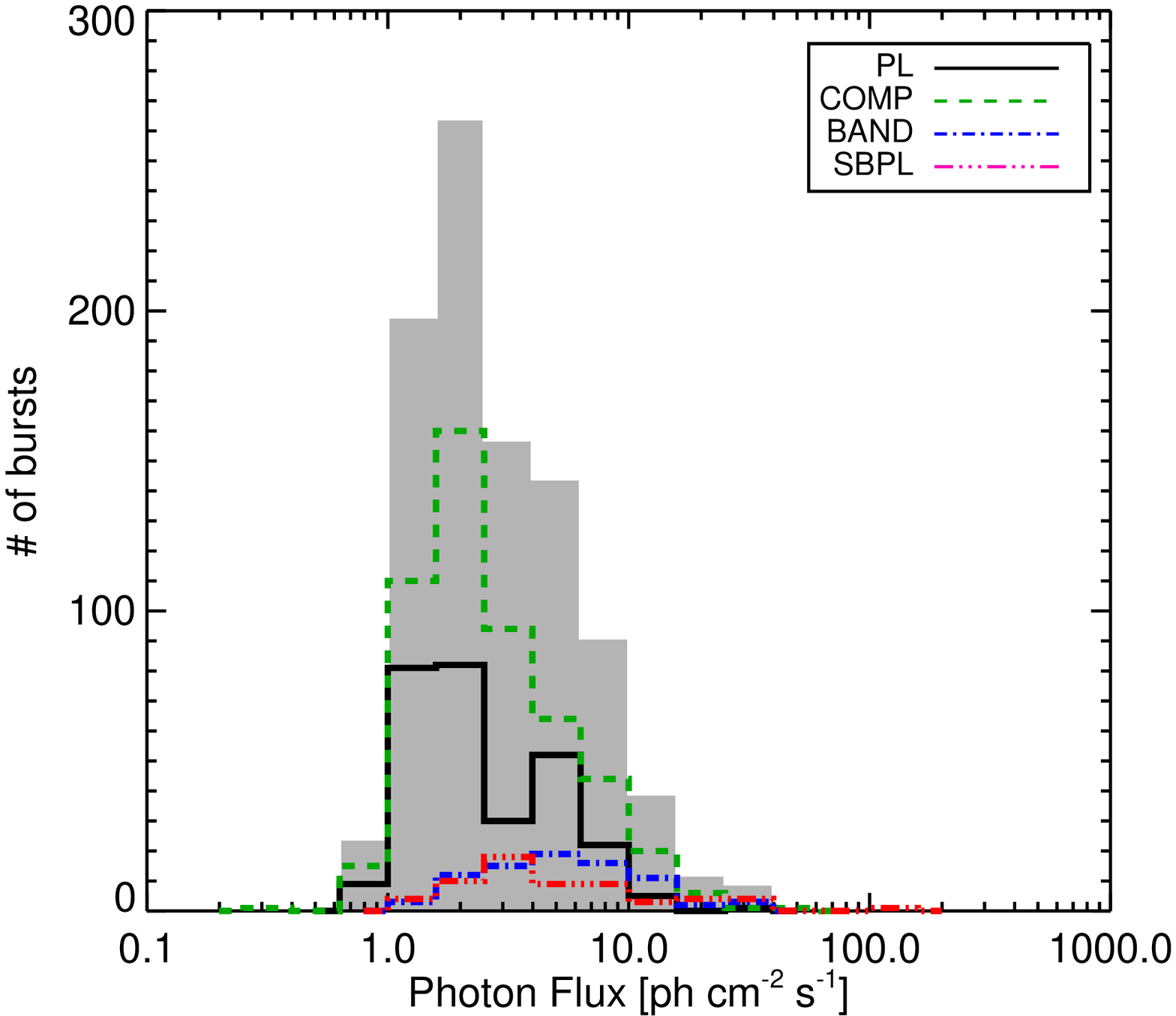}}
		\end{minipage}
	\end{center}
\caption{Distributions of photon flux in the 10~keV--1~MeV \subref{allpfluxf} and 10~keV--40~MeV \subref{allpfluxf40MeV} band from the GOOD \emph{F} spectral fits. Note that the plotted distributions contain the flux on two different timescales: 1024 ms (long GRBs) and 64 ms (short GRBs). The BEST parameter distribution for the photon flux in both energy ranges is shown in \subref{allpfluxfbest} and \subref{allpfluxf40MeVbest}. The gray filled histogram shows the total distribution and the constituents are shown in colors.\label{pfluxf}}
\end{figure}

\begin{figure}
	\begin{center}
		\begin{minipage}[t]{1\textwidth}
		\subfigure[]{\label{fig:enfluence}\includegraphics[scale=0.45]{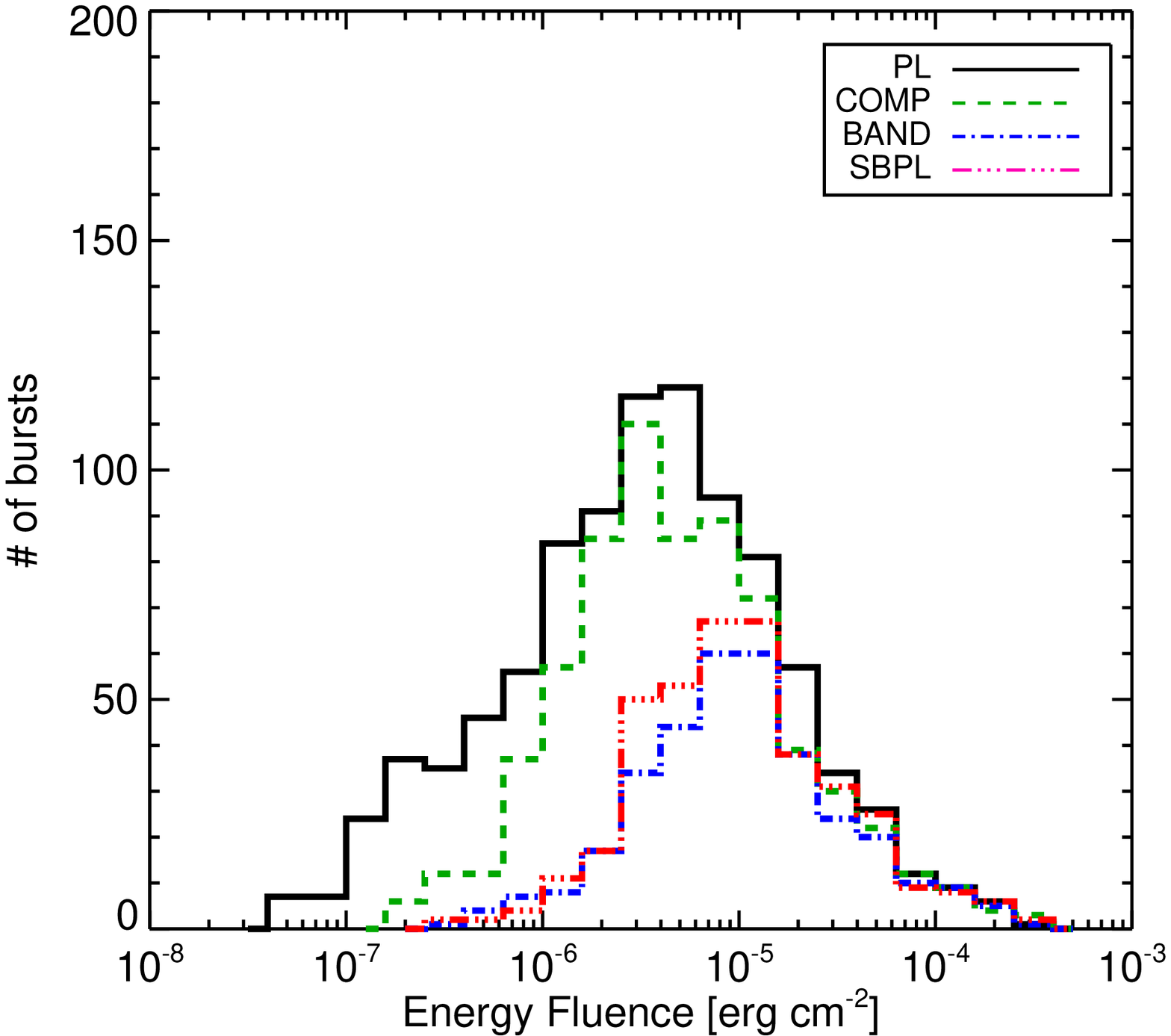}}
		\subfigure[]{\label{fig:enfluence40}\includegraphics[scale=0.45]{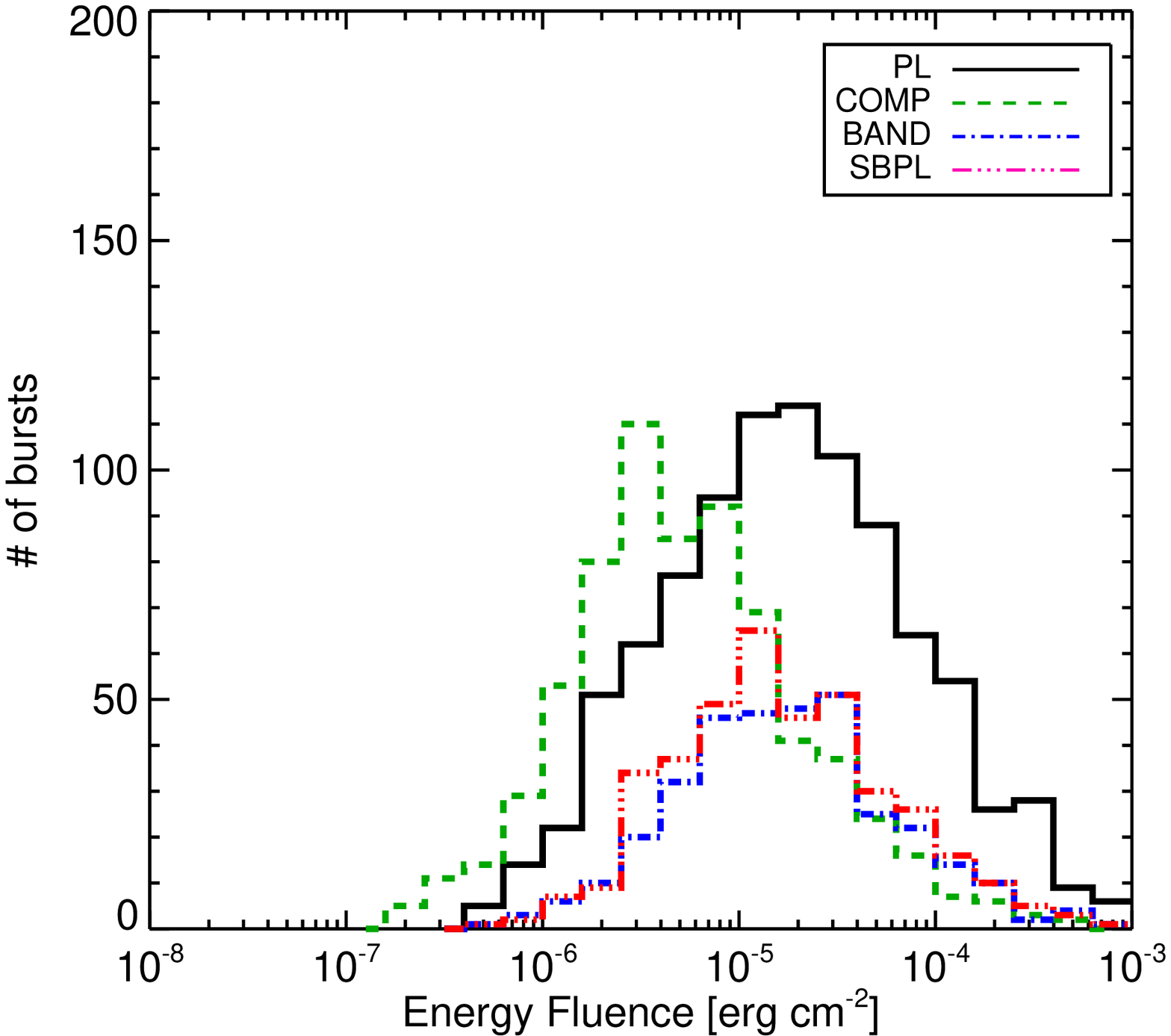}}
		\end{minipage}
		\begin{minipage}[t]{1\textwidth}
		\subfigure[]{\label{fig:enfluencebest}\includegraphics[scale=0.45]{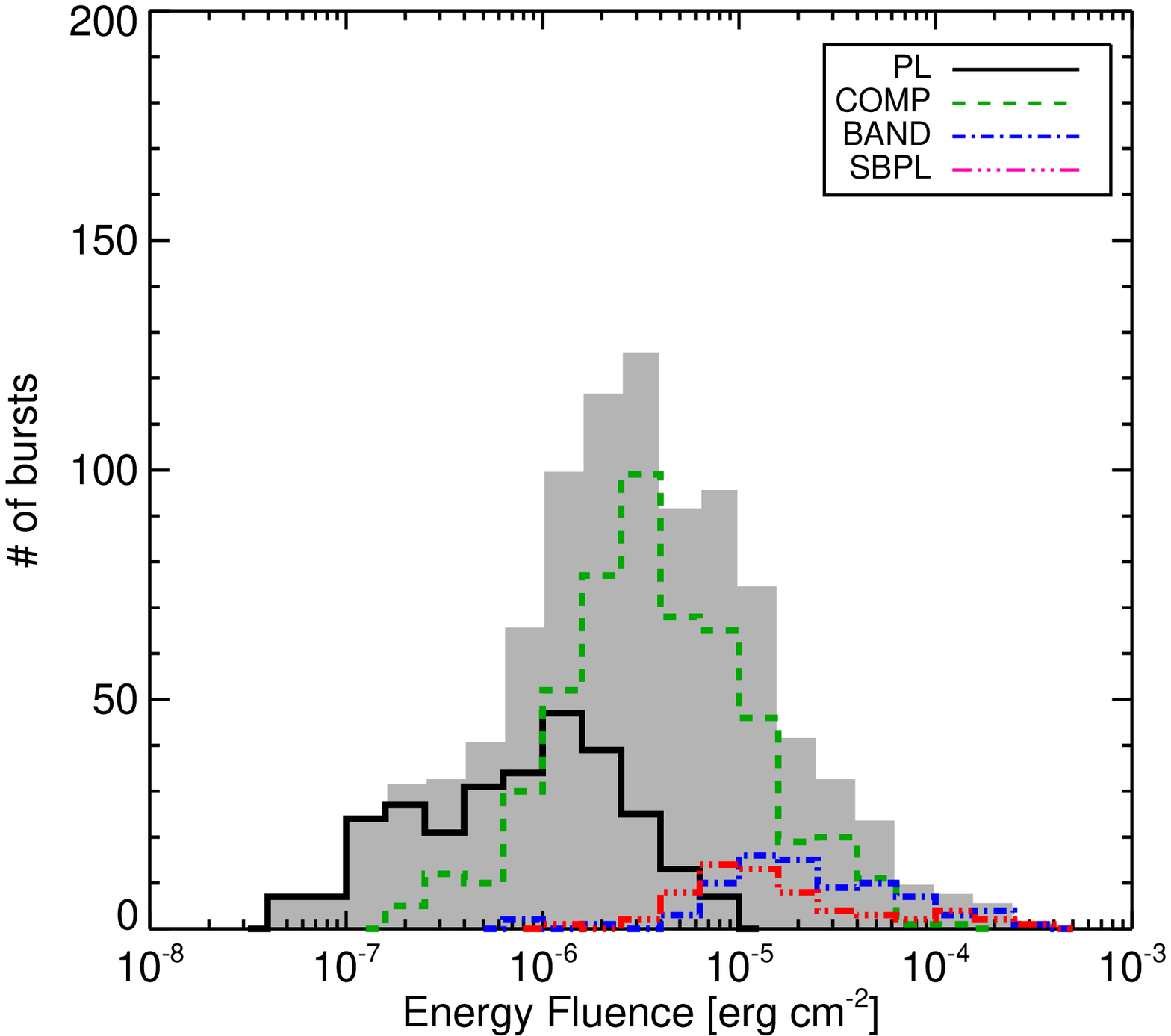}}
		\subfigure[]{\label{fig:enfluence40best}\includegraphics[scale=0.45]{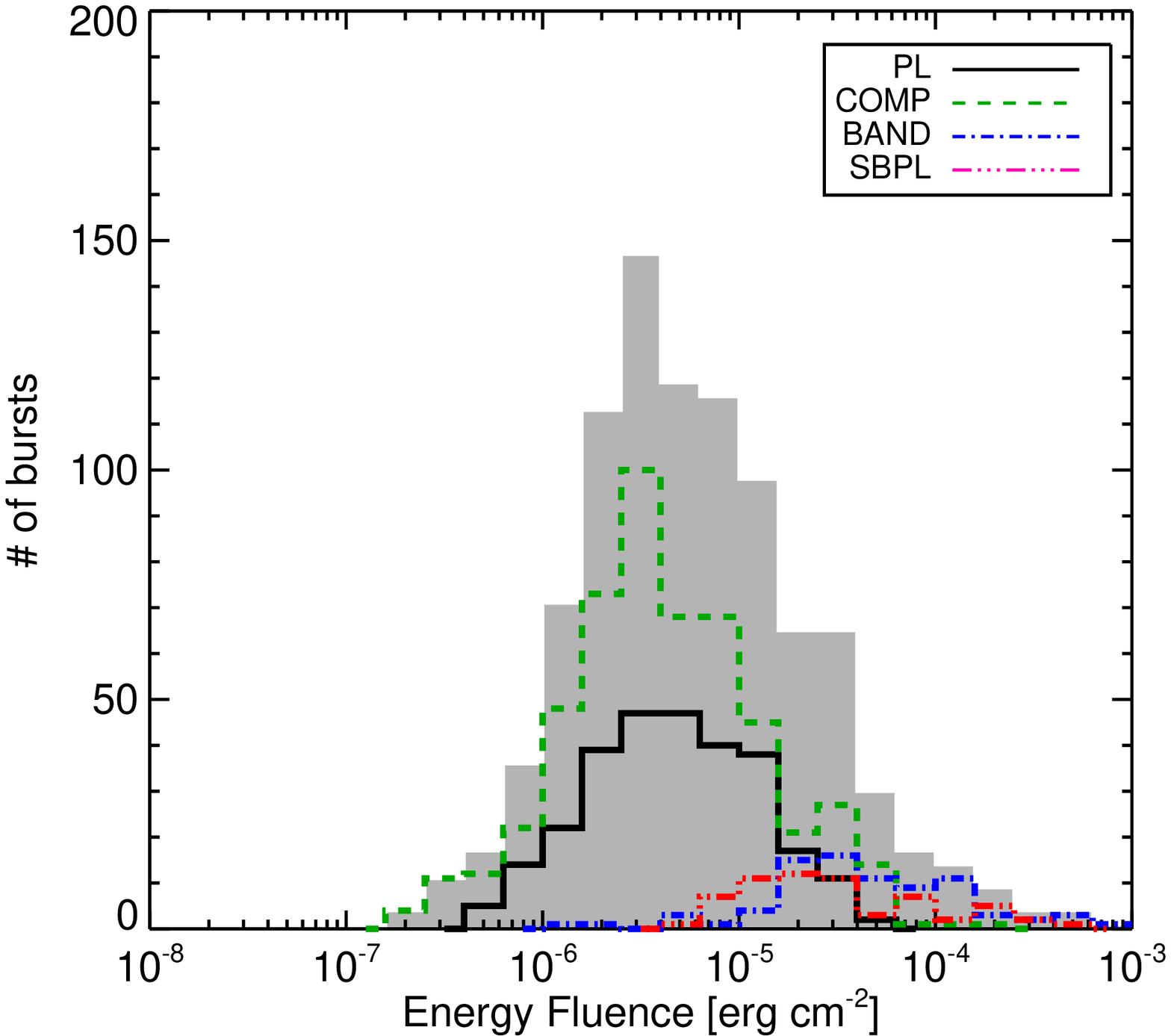}}
		\end{minipage}
	\end{center}
\caption{Distributions of energy fluence in the 10~keV--1~MeV \subref{fig:enfluence} and 10~keV--40~MeV \subref{fig:enfluence40} band from the GOOD \emph{F} spectral fits. The BEST parameter distribution for the energy fluence in both energy ranges is shown in \subref{fig:enfluencebest} and \subref{fig:enfluence40best}. The gray filled histogram shows the total distribution and the constituents are shown in colors. \label{efluence}}
\end{figure}

\begin{figure}
	\begin{center}
		\begin{minipage}[t]{1\textwidth}
		\subfigure[]{\label{fig:fluence}\includegraphics[scale=0.45]{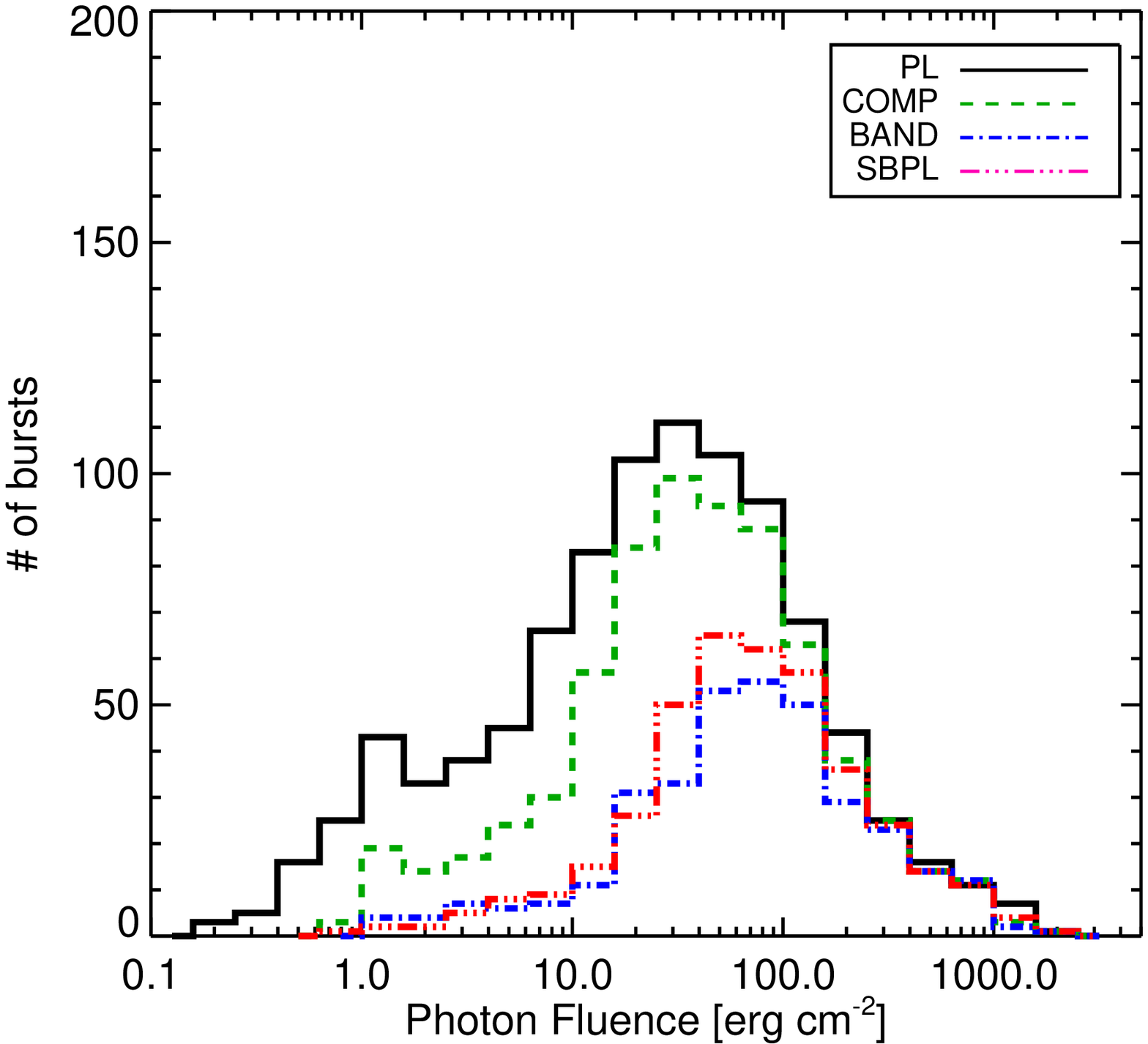}}
		\subfigure[]{\label{fig:fullfluence}\includegraphics[scale=0.45]{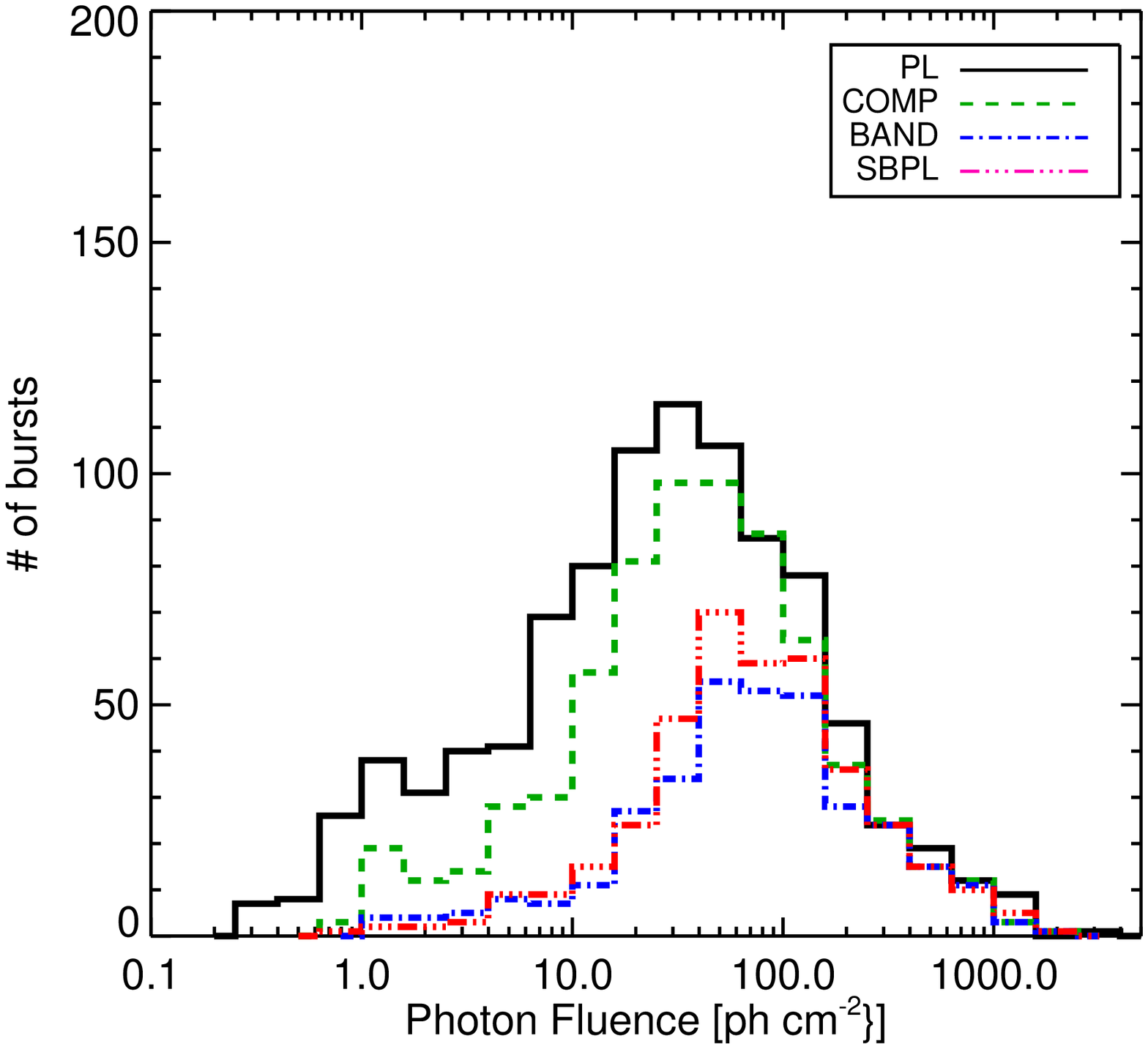}}
		\end{minipage}
		\begin{minipage}[t]{1\textwidth}
		\subfigure[]{\label{fig:fluencebest}\includegraphics[scale=0.45]{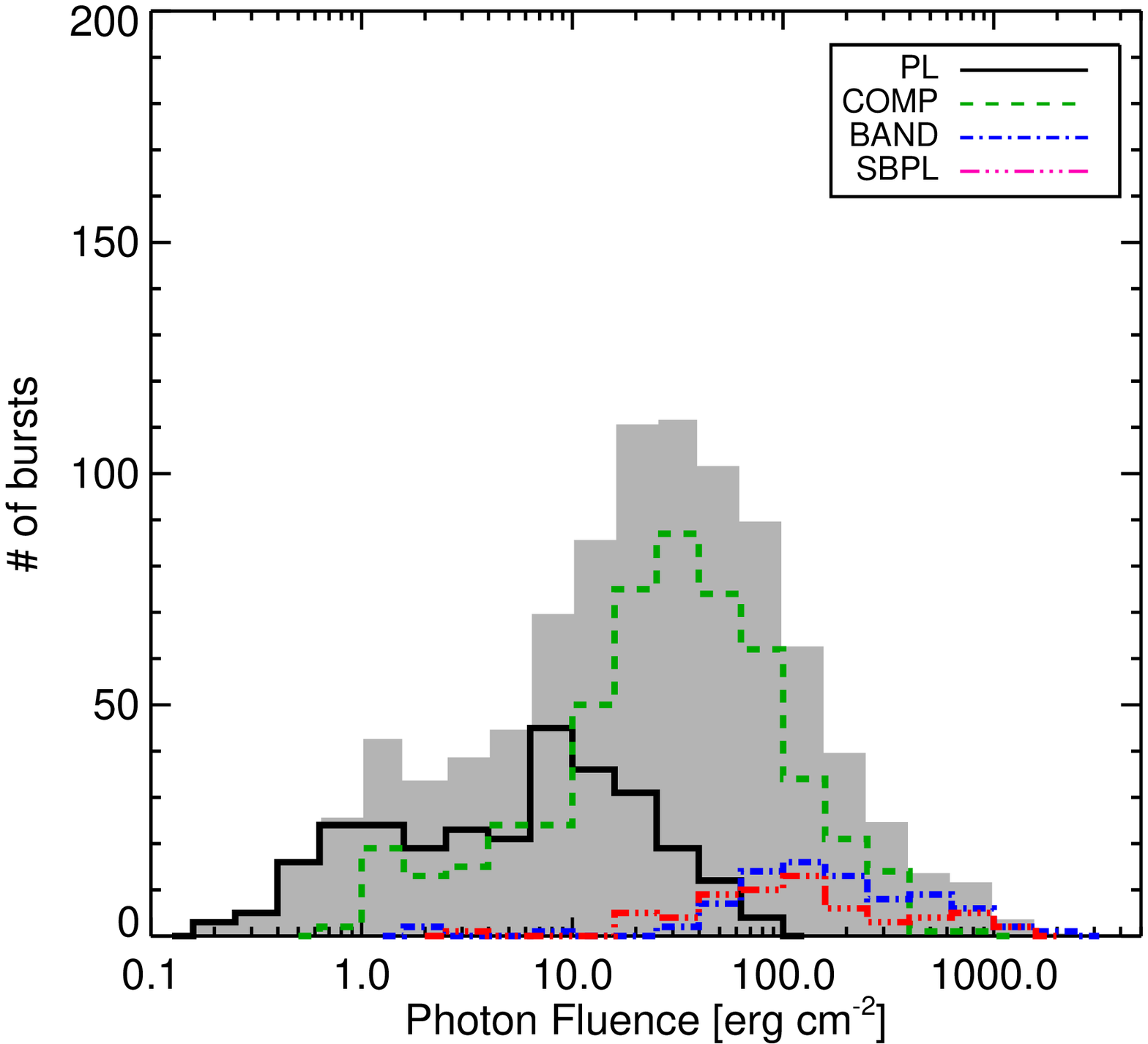}}
		\subfigure[]{\label{fig:fullfluencebest}\includegraphics[scale=0.45]{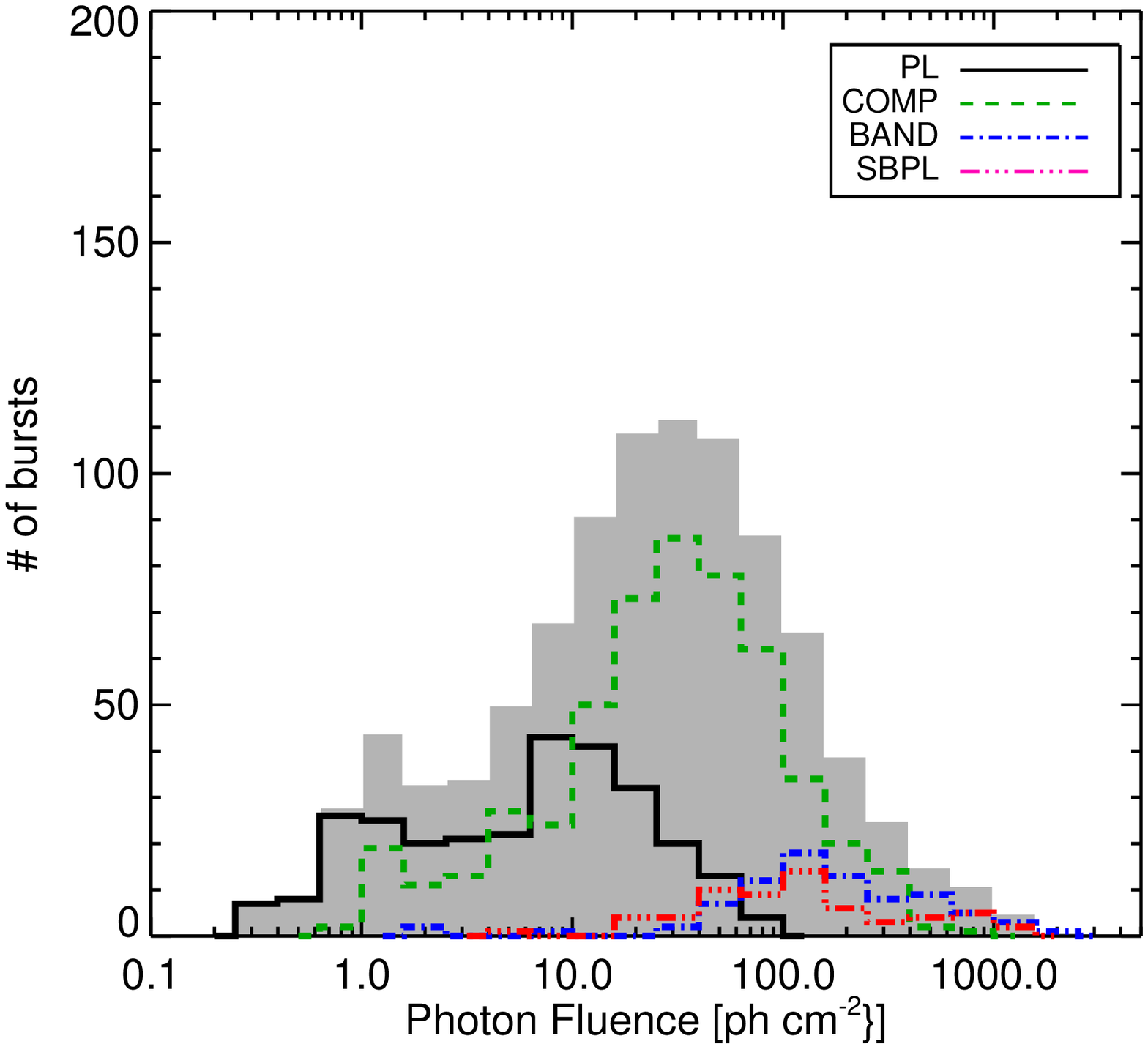}}
		\end{minipage}
	\end{center}
\caption{Distributions of photon fluence in the 10~keV--1~MeV \subref{fig:fluence} and 10~keV--40~MeV \subref{fig:fullfluence} band from the GOOD \emph{F} spectral fits. The BEST parameter distribution for the energy fluence in both energy ranges is shown in \subref{fig:fluencebest} and \subref{fig:fullfluencebest}. The gray filled histogram shows the total distribution and the constituents are shown in colors.  \label{pfluence}}
\end{figure}

The distributions for the time-averaged energy flux and photon flux are shown in Figure~\ref{enluxf} and Figure~\ref{pfluxf}, respectively.  The photon and energy fluxes of the BEST sample have a median value of around 2.4 ph cm$^{-2}$  s$^{-1}$, and $3 \times 10^{-7}$ ergs cm$^{-2}$ s$^{-1}$, respectively in the 10-1000~keV band.  When integrating over the full GBM spectral band, 10~keV - 40~MeV, the BEST energy flux distribution broadens significantly, approximating a top hat function with a small high-flux tail spanning about 2 orders of magnitude.  
Note that the low-flux cutoff is due to both the sensitivity of the instrument and the clear deviation from a three-dimensional Euclidean distribution that is observed in a typical $\log N$-$\log S$ plot \citep{kienlin13}. In any case, the flux measurements with a more sensitive instrument will likely position the peak of the flux distribution at a lower flux value \citep[see also][]{untrig}. Similarly in Figure \ref{efluence} and Figure \ref{pfluence}, the distributions for the BEST photon fluence and BEST energy fluence are depicted.  
The plots for the photon fluence appear to contain evidence of the duration bimodality of GRBs. While there is a discriminant peak at $\sim$~30 photons cm$^{-2}$ there is also a deviation from a log-normal distribution at smaller photon fluence values. Fitting the photon fluence distribution in the 10-1000 keV band with a sum of two log-normal functions, we find peak values of $31_{-23}^{+91}$ and $1.1_{-0.6}^{+1.1}\rm \ ph\ cm^{-2}$, respectively.  Similarly, the distribution of the energy fluence in the 10-1000 keV band can also be fit by the sum of two log-normal functions which result in peak values of $(3.2_{-2.5}^{+11.3})\times 10^{-6}$ and $(1.5_{-0.4}^{+0.6}) \times 10^{-7} \rm \ erg\ cm^{-2}$, respectively. It is interesting to note that while the \comp model is largely unaffected by the change in energy band due to the exponential cutoff, the \pwrlw model shifts to one order of magnitude higher energy fluence values, as it overestimates the flux at higher energies (see Figure~\ref{fig:enfluence40best}).
The brightest GRB contained in this catalog based on time-averaged photon flux is GRB 120323.507 \citep{guiriec13} with a flux of $\rm \sim 115\ ph\ s^{-1}\ cm^{-2}$ and 
the burst with the largest average energy flux is GRB 111222.619 with an energy flux of $\rm \sim1.5 \times 10^{-5}\ ergs\ s^{-1} cm^{-2}$.  
The burst with the highest energy fluence is GRB 090902.462 \citep{abdo09} with an energy fluence of $\rm \sim 2.8 \times 10^{-4}\ ergs\ cm^{-2}$ while 
the burst with the highest photon fluence is GRB 090618.353 with a photon fluence of $\rm >2300\ ph\ cm^{-2}$.

\begin{figure}
	\begin{center}
		\begin{minipage}[t]{1\textwidth}
		\subfigure[]{\label{fig:fepeakzcorr}\includegraphics[scale=0.45]{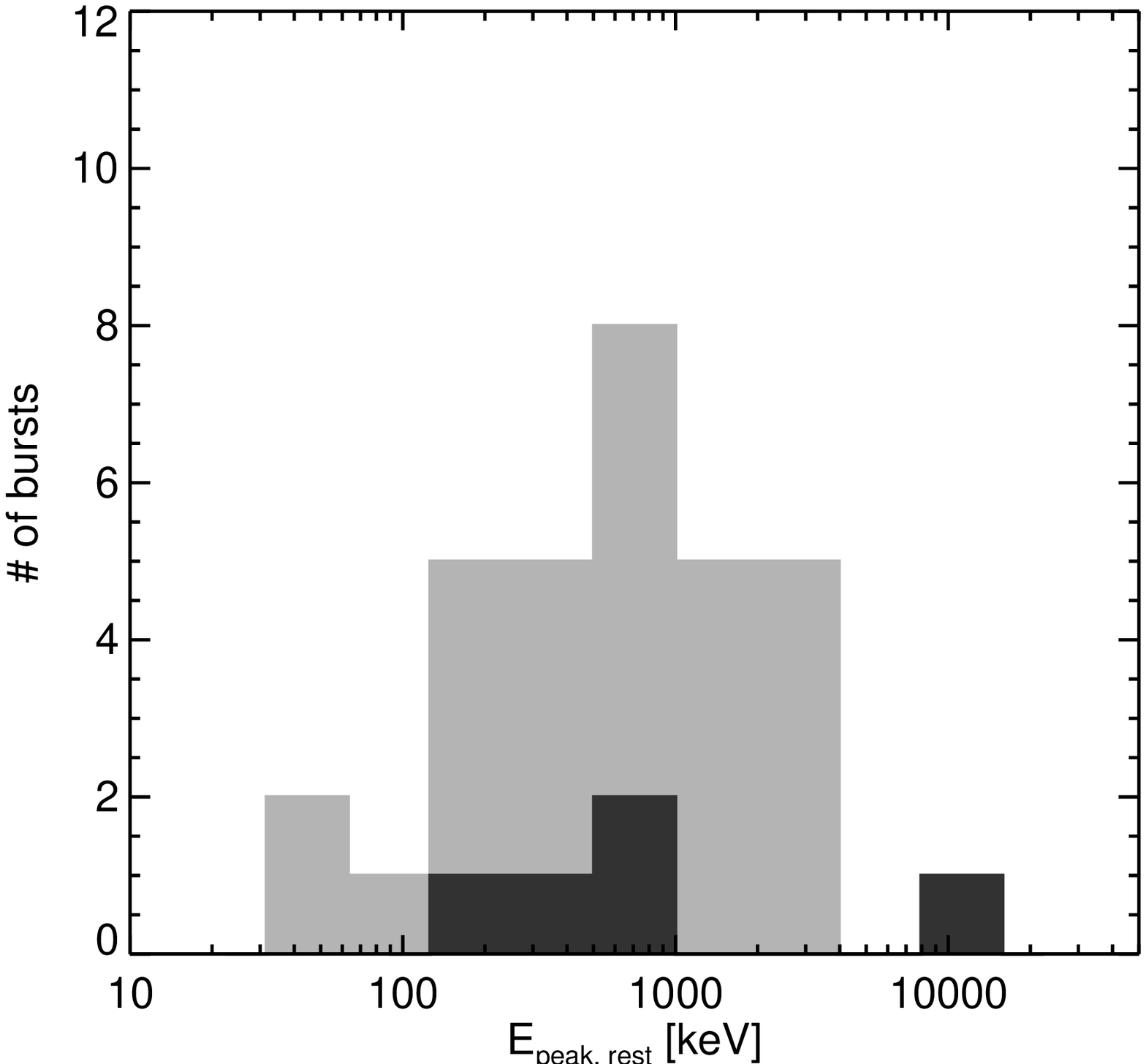}}
		\subfigure[]{\label{fig:febreakzcorr}\includegraphics[scale=0.45]{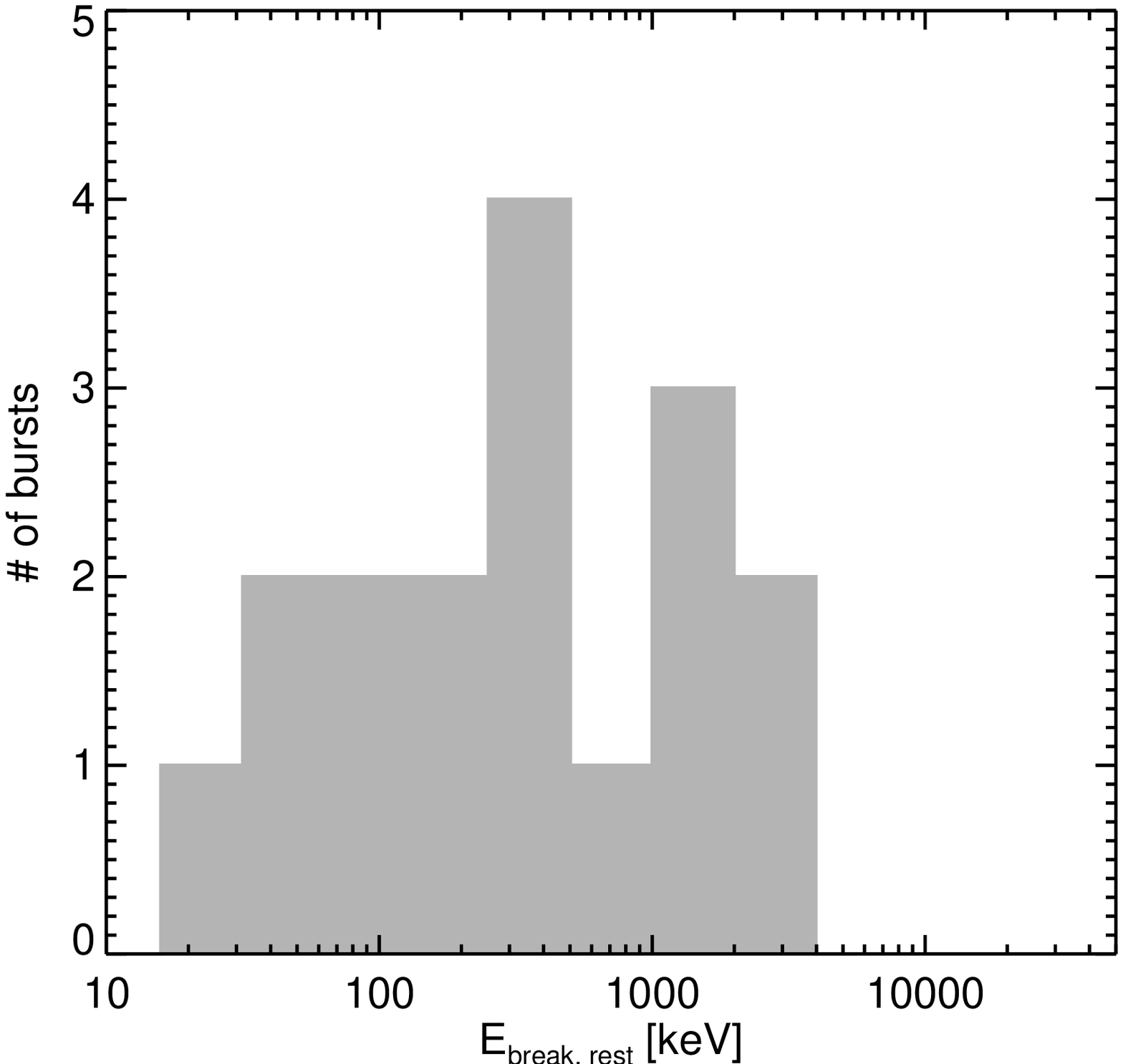}}
		\end{minipage}
		\begin{minipage}[t]{1\textwidth}
		\subfigure[]{\label{fig:fenfluxzcorr}\includegraphics[scale=0.45]{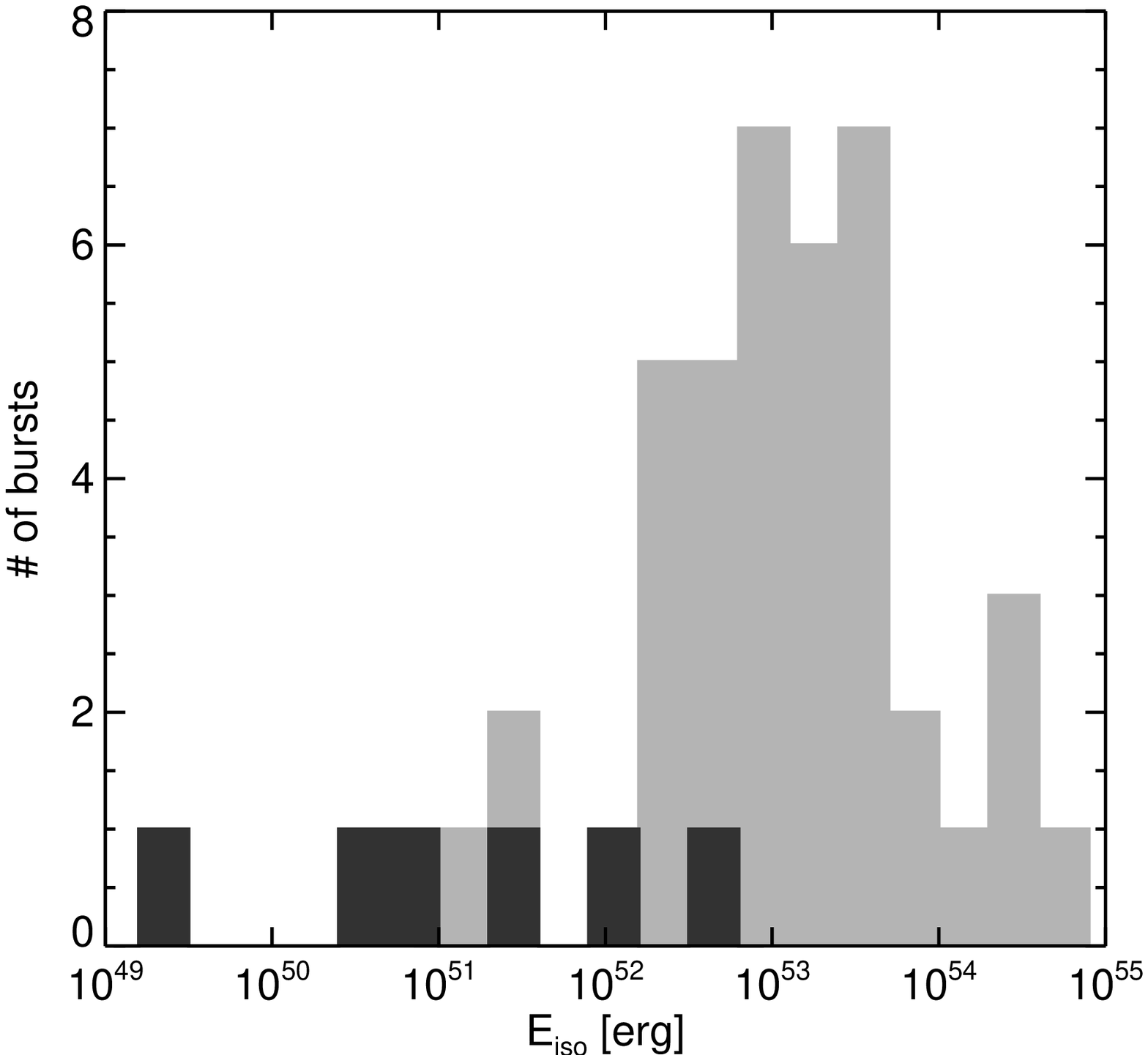}}
		\subfigure[]{\label{fig:fenfluencezcorr}\includegraphics[scale=0.45]{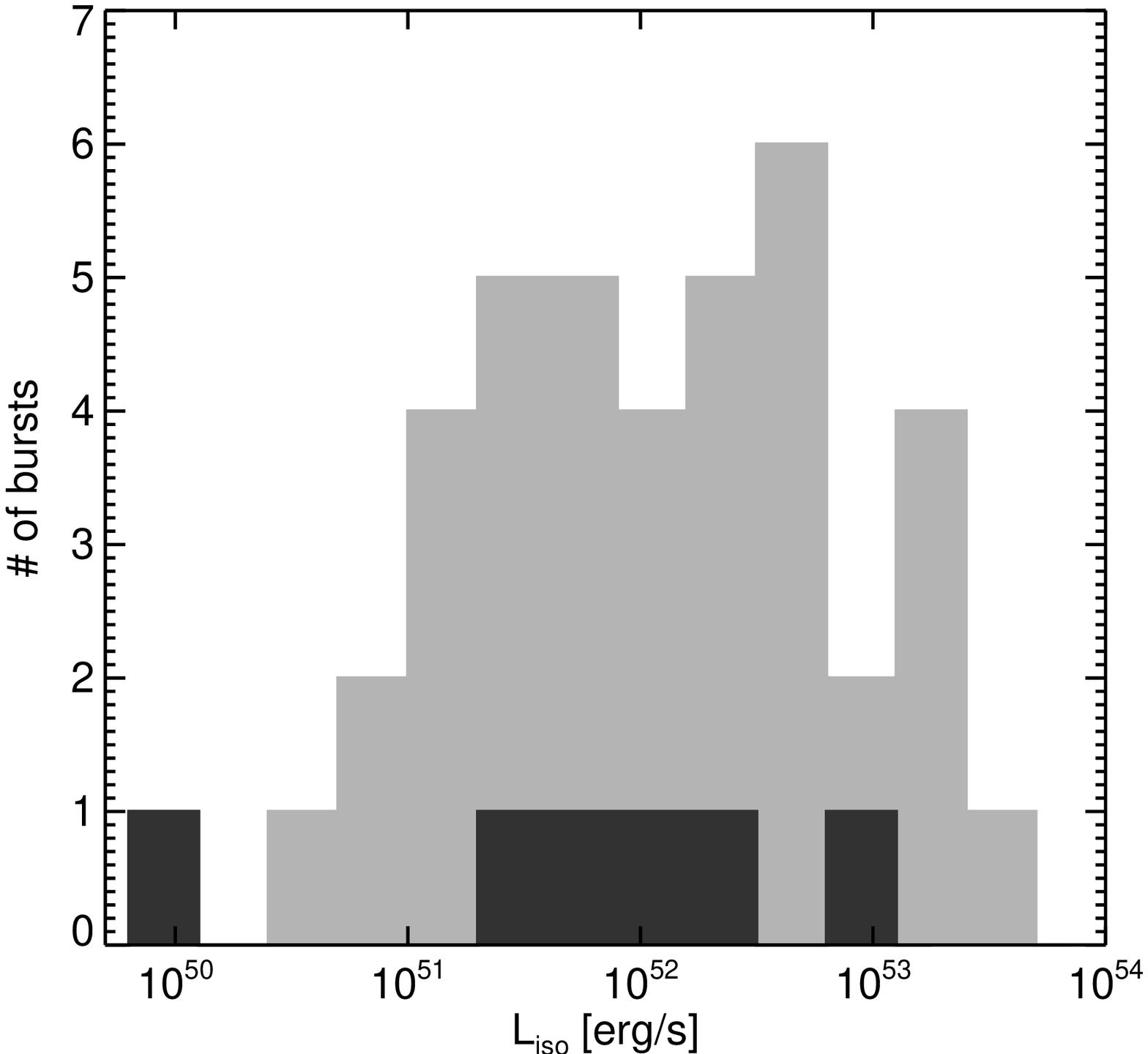}}
		\end{minipage}
	\end{center}
\caption{\subref{fig:fepeakzcorr} \eprest distribution of the \emph{F} spectra fits from the BEST sample for long and short GRBs (light gray and dark gray filled histogram, respectively). \subref{fig:febreakzcorr} Same as \subref{fig:fepeakzcorr} but for \ebrest. \subref{fig:fenfluxzcorr} and \subref{fig:fenfluencezcorr}  Distribution of \eiso and \Liso of the fluence spectral fits from the BEST sample in the rest-frame energy band 1/(1+z)~keV to 10/(1+z)~MeV.  \label{restframef}}
\end{figure}

In Figure~\ref{restframef} we present the BEST rest-frame spectral parameters for \epeak, \ebreak, \eiso and \Liso where the latter two have been determined in the rest-frame energy band between $1/(1+z)$~keV and $10/(1+z)$~MeV. Both, \eprest and \ebrest peak at roughly $600$~keV and, similarly to the distributions in the observer frame, cover just two orders of magnitude. The \eiso distribution peaks at $\sim 10^{53}$~erg and extends over more than three orders of magnitude. \Liso, on the other hand, has a median of $\sim 10^{52}$~erg/s and covers 3 orders of magnitude.

\begin{figure}
	\begin{center}
		\begin{minipage}[t]{1\textwidth}
		\subfigure[]{\label{fig:falphavsz}\includegraphics[scale=0.45]{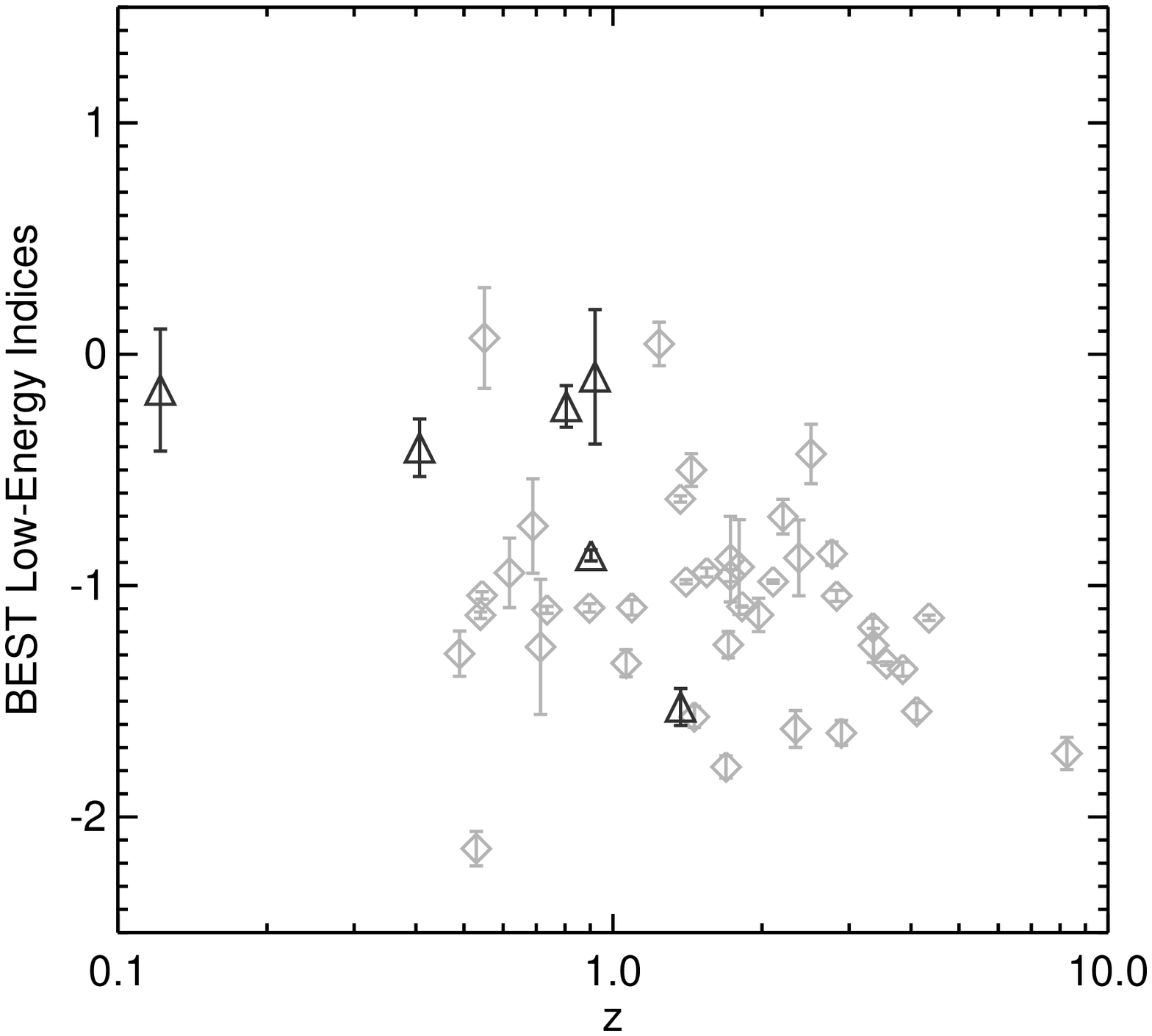}}
		\subfigure[]{\label{fig:fbetavsz}\includegraphics[scale=0.45]{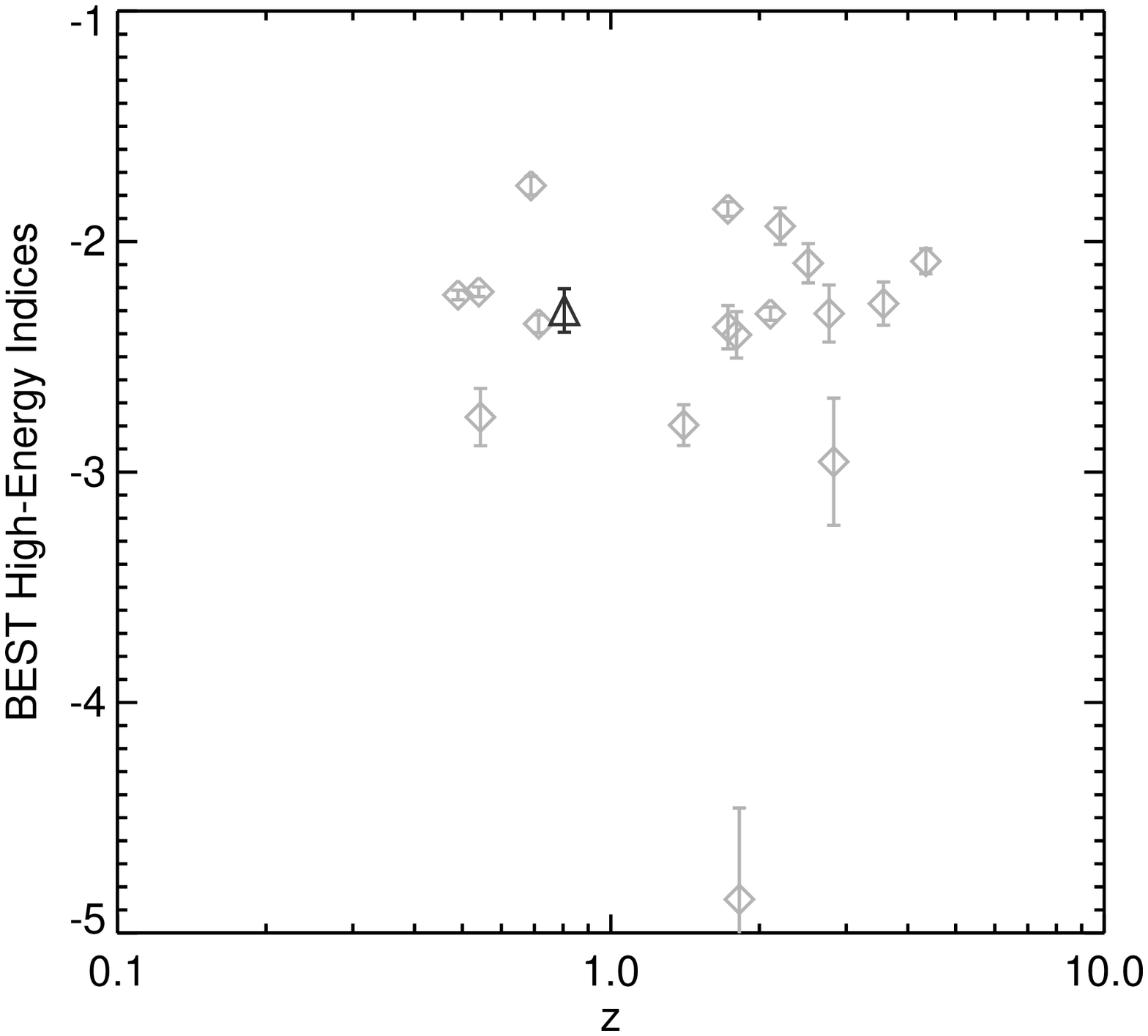}}
		\end{minipage}
		\begin{minipage}[t]{1\textwidth}
		\centering
		\subfigure[]{\label{fig:epeakvsz}\includegraphics[scale=0.45]{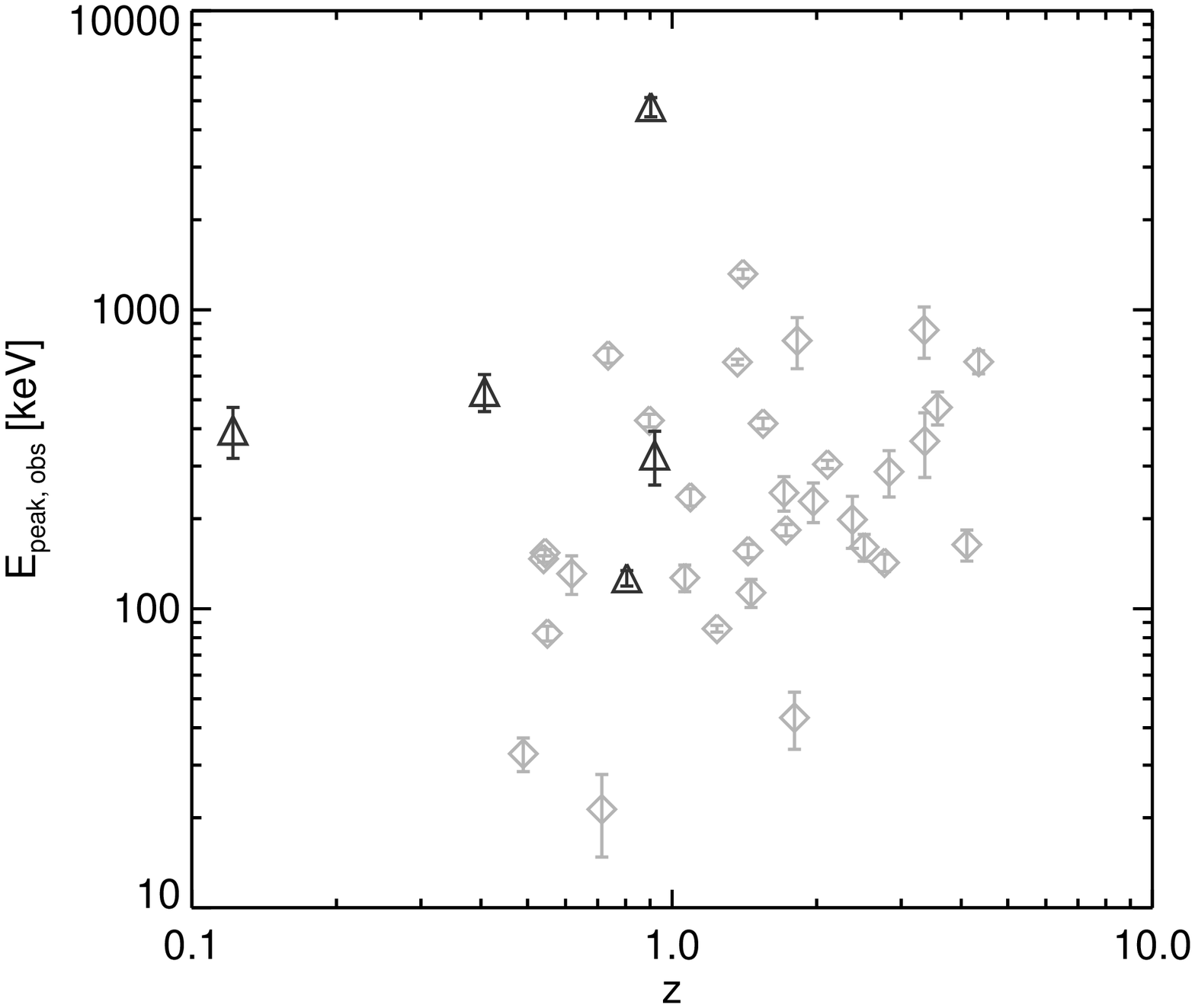}}
		\end{minipage}
	\end{center}
\caption{BEST spectral parameters as a function of the redshift of the \emph{F} spectral fits for short (black triangles) and long (light-gray diamonds) GRBs.  \label{paramsvszf}}
\end{figure}

In Figure~\ref{paramsvszf} we show the evolution of the BEST spectral parameters, $\alpha$, $\beta$ and \epeak, with redshift. While both the high-energy index and \epeak do not show a clear dependence with redshift \citep[see also][]{restframe}, the low-energy index of the long GRBs shows a trend to steeper, i.e. softer, values at higher redshifts \citep[see also][]{geng}. However, a Spearman rank correlation analysis shows that this correlation is not significant ($P=0.15$). 

\subsubsection{The break scale $\Delta$}
\begin{figure}
	\begin{center}
		\subfigure[]{\label{fig:sbplfluence2}\includegraphics[scale=0.45]{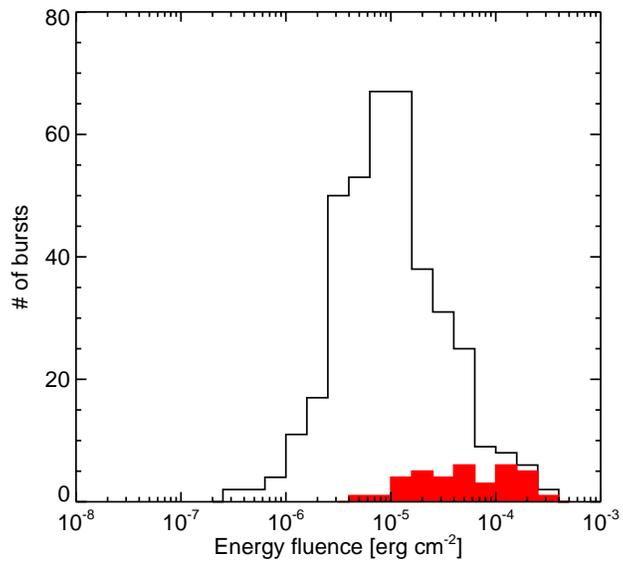}}
	\end{center}
\caption{Distribution of the energy fluence in the 10 keV--1 MeV energy range for the GOOD \sbpl \emph{F} spectral fits (black solid histogram) and the GOOD \sbpl fluence spectral fits with varying break scale $\Delta$ (red filled histogram). \label{fig:sbplfluence}}
\end{figure}

\begin{figure}
	\begin{center}
		\subfigure[]{\label{fig:breakscale}\includegraphics[scale=0.45]{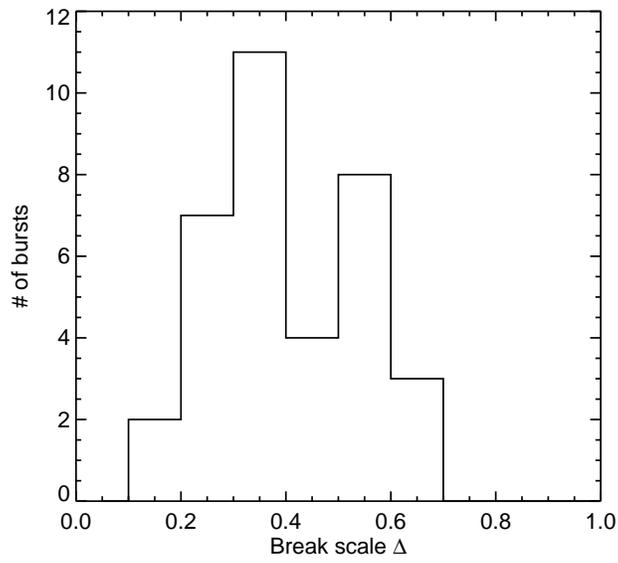}}
	\end{center}
\caption{Distribution of the break scale $\Delta$ of the \sbpl \emph{F} spectral fits. \label{fig:sbplbreakscale}}
\end{figure}

\begin{figure}
	\begin{center}
		\begin{minipage}[t]{1\textwidth}
		\subfigure[]{\label{fig:sbplalpha}\includegraphics[scale=0.45]{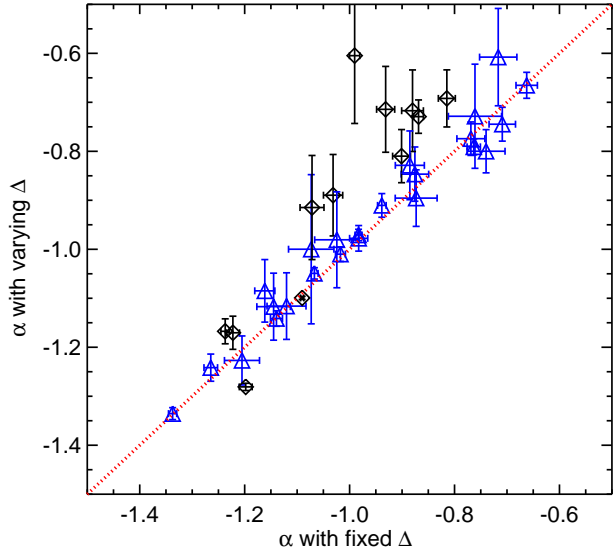}}
		\subfigure[]{\label{fig:sbplbeta}\includegraphics[scale=0.45]{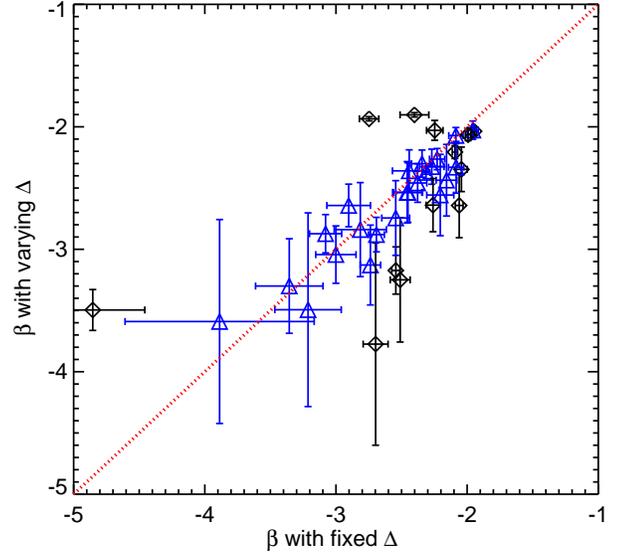}}\\
		\end{minipage}
		\begin{minipage}[t]{1\textwidth}
		\centering
		\subfigure[]{\label{fig:sbplebreak}\includegraphics[scale=0.45]{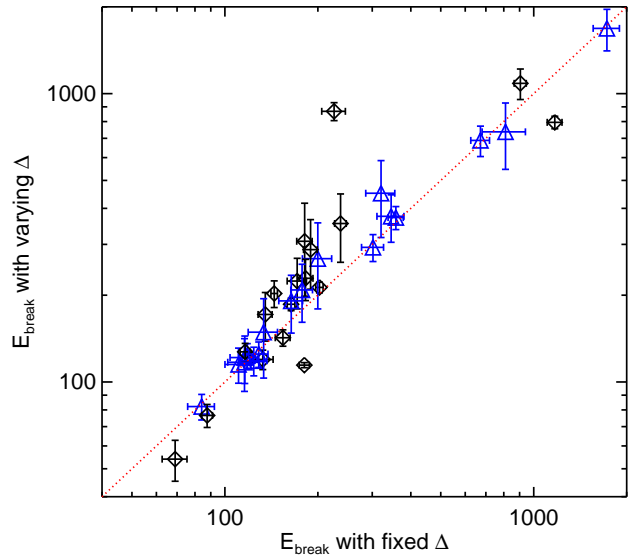}}
		\end{minipage}
	\end{center}
\caption{Comparison of \sbpl model parameters with $\Delta$ free to vary and fixed to 0.3. Values consistent within 1$\sigma$ (blue triangles) and not consistent within 1$\sigma$ (black diamonds) including a line of equality (red dashed line) are shown. \label{fig:sbplmodelparams}}
\end{figure}

In order to test the behavior of the break scale $\Delta$ of the \sbpl model, we re-fitted all GOOD \sbpl spectra obtained from the \emph{F} spectral fits. Contrary to the initial fits, we vary the break scale, thus increasing the number of free model parameters. After the fit, we applied the same quality cuts to the resulting parameters as for the GOOD \sbpl sample with the additional requirement that $\sigma_{\Delta}/\Delta \le 0.4$. Out of the initial 384 \sbpl fits only 36 fulfilled this newly added criterion, with the fit not able to constrain 5 free parameters for the bulk of the GOOD sample. As intuitively obvious, only the most fluent portion of the sample could be fitted with such a complex model (see Figure~\ref{fig:sbplfluence}), but it should be noted that the GRBs for which the break scale could be constrained are not necessarily the GRBs for which this model was selected as BEST in Section~5.2. In Figure~\ref{fig:sbplbreakscale} we show the distribution of $\Delta$. As can be seen $\Delta$ varies between 0.1 and 0.7 with an average value of $0.4\pm0.2$. 
It is instructive to investigate how the additional free fit parameter affects the results of the other model components. In Figure~\ref{fig:sbplmodelparams} we show the relations between fixed and varying \ebreak, low- and high-energy power-law indices. The low--energy power--law index seems to tend towards shallower values (particularly when the index is already shallow with the fixed $\Delta$) when letting $\Delta$ vary freely. In fact, 13 (36\%) low--energy power--law indices are not consistent within 1$\sigma$. The high-energy power--law index has a tendency to steepen and 11 (31\%) indices are not consistent within the 1$\sigma$ limit. With 8 (22\%) 1$\sigma$ outliers, the situation is similar for \ebreak. In general, \ebreak tends towards harder values when $\Delta$ is allowed to vary.

\subsection{\emph{P} spectral fits}
Similarly to the \emph{F} spectral fits, we are going to present the results of the \emph{P} spectral fits in this section.

\begin{figure}
	\begin{center}
		\subfigure[]{\label{allloidxPeak}\includegraphics[scale=0.45]{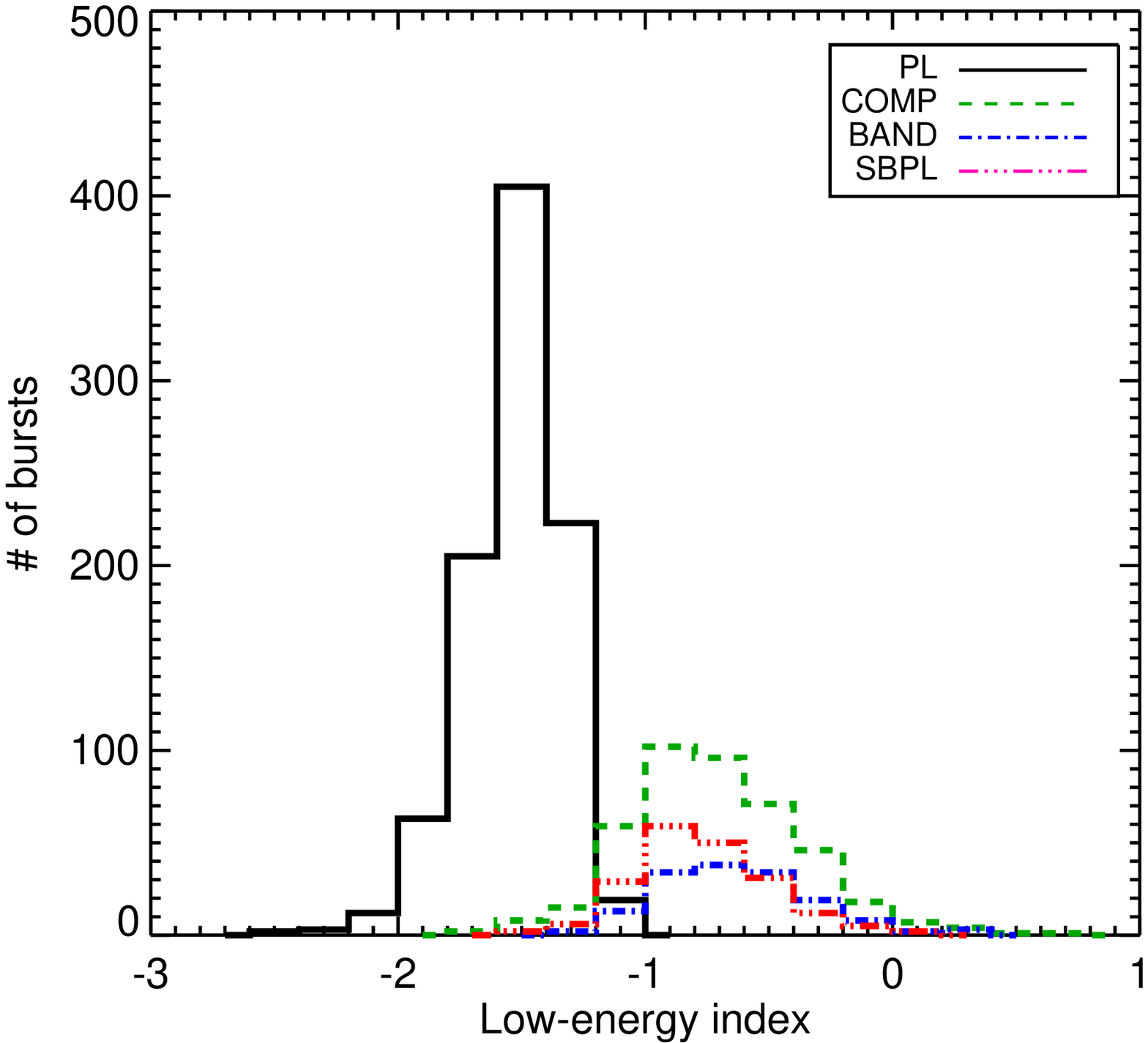}}
		\subfigure[]{\label{allloidxPeakbest}\includegraphics[scale=0.45]{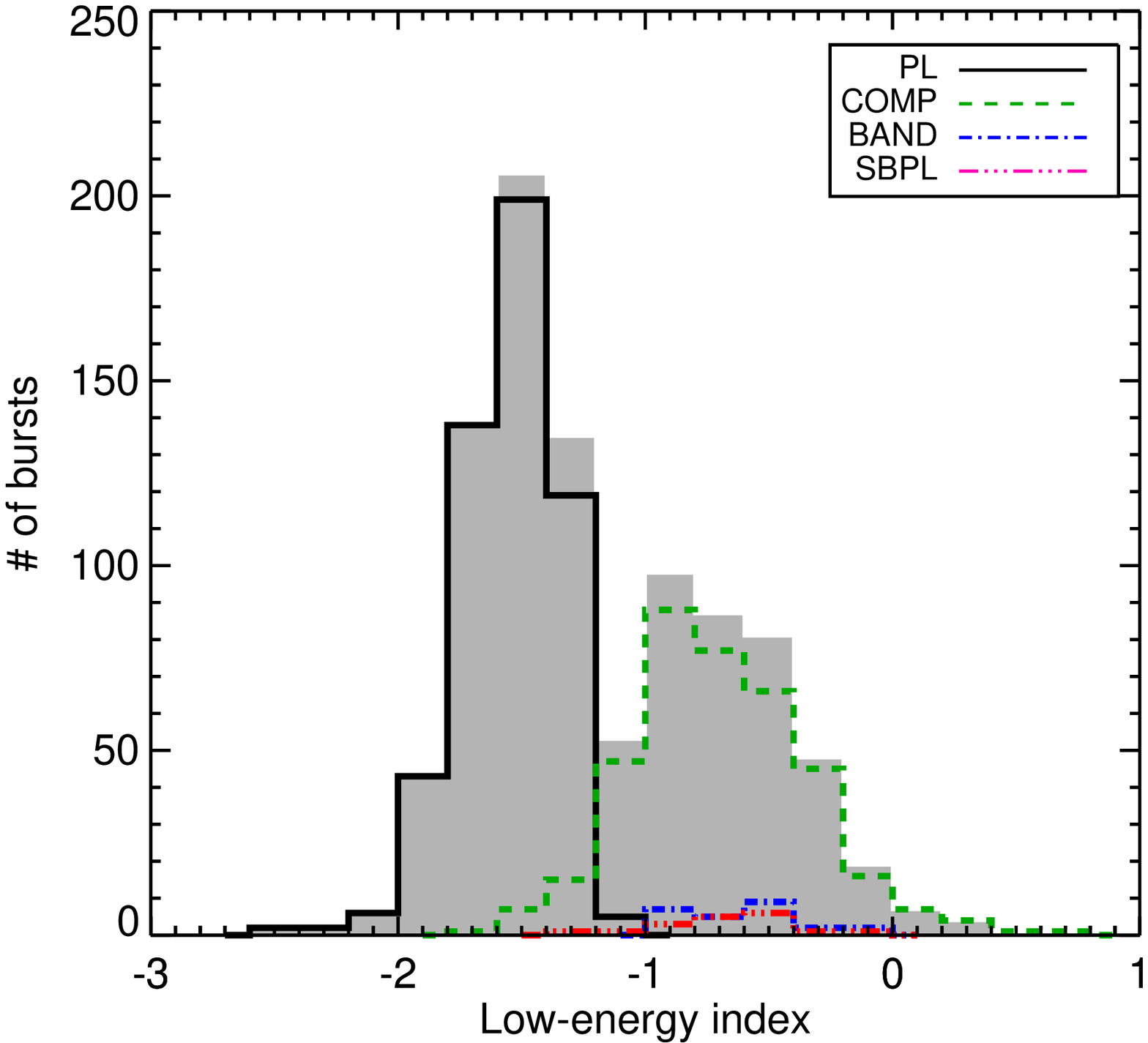}}
	\end{center}
\caption{Distribution of the low-energy indices obtained from the GOOD \emph{P} spectral fits \subref{allloidx}. The BEST parameter distribution (gray filled histogram) and its constituents for the low-energy index is shown in \subref{allloidxbest}. 
}\label{loenindexPeak}
\end{figure}

\begin{figure}
	\begin{center}
		\subfigure[]{\label{allhiidxPeak}\includegraphics[scale=0.45]{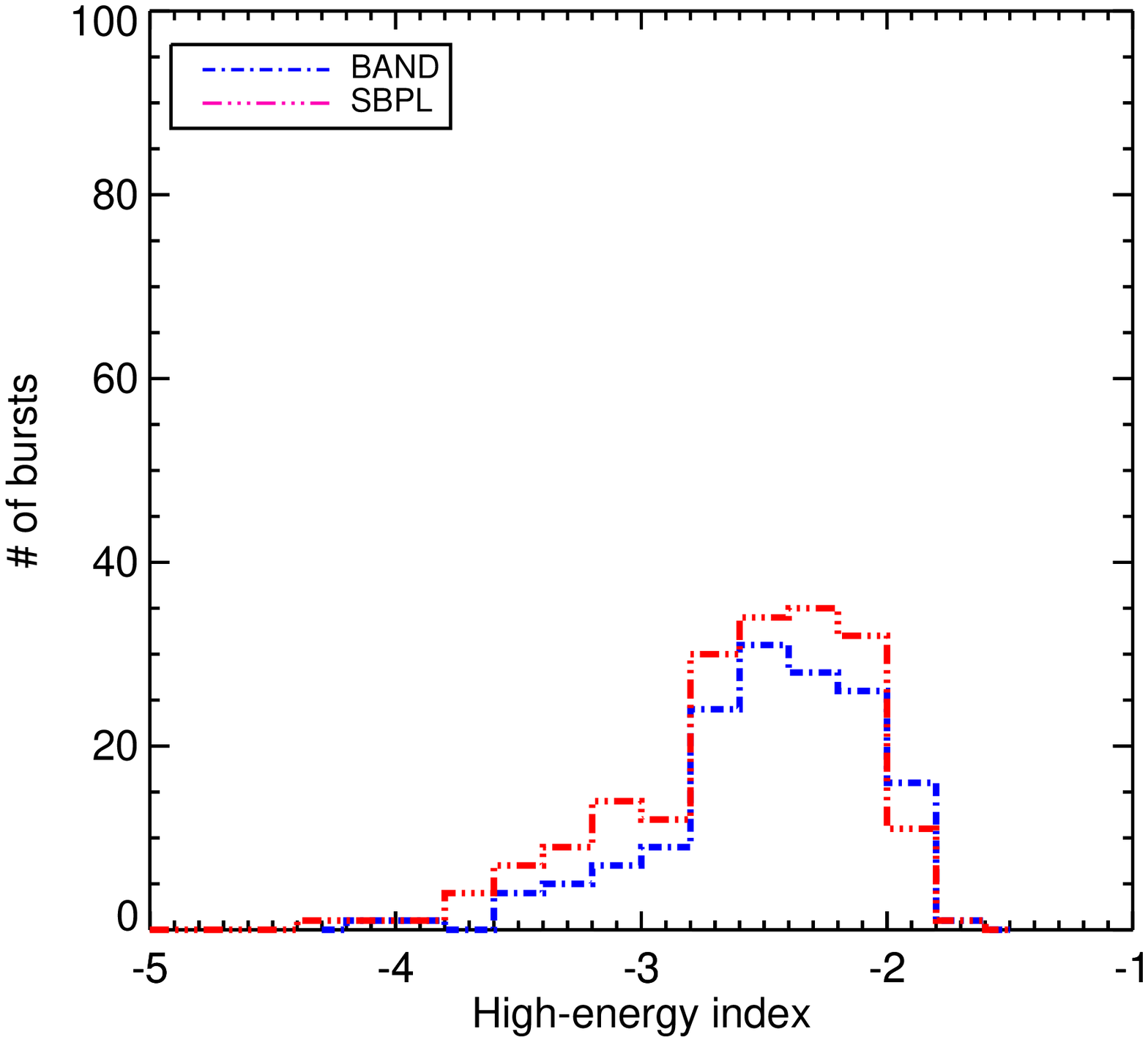}}
		\subfigure[]{\label{allhiidxPeakbest}\includegraphics[scale=0.45]{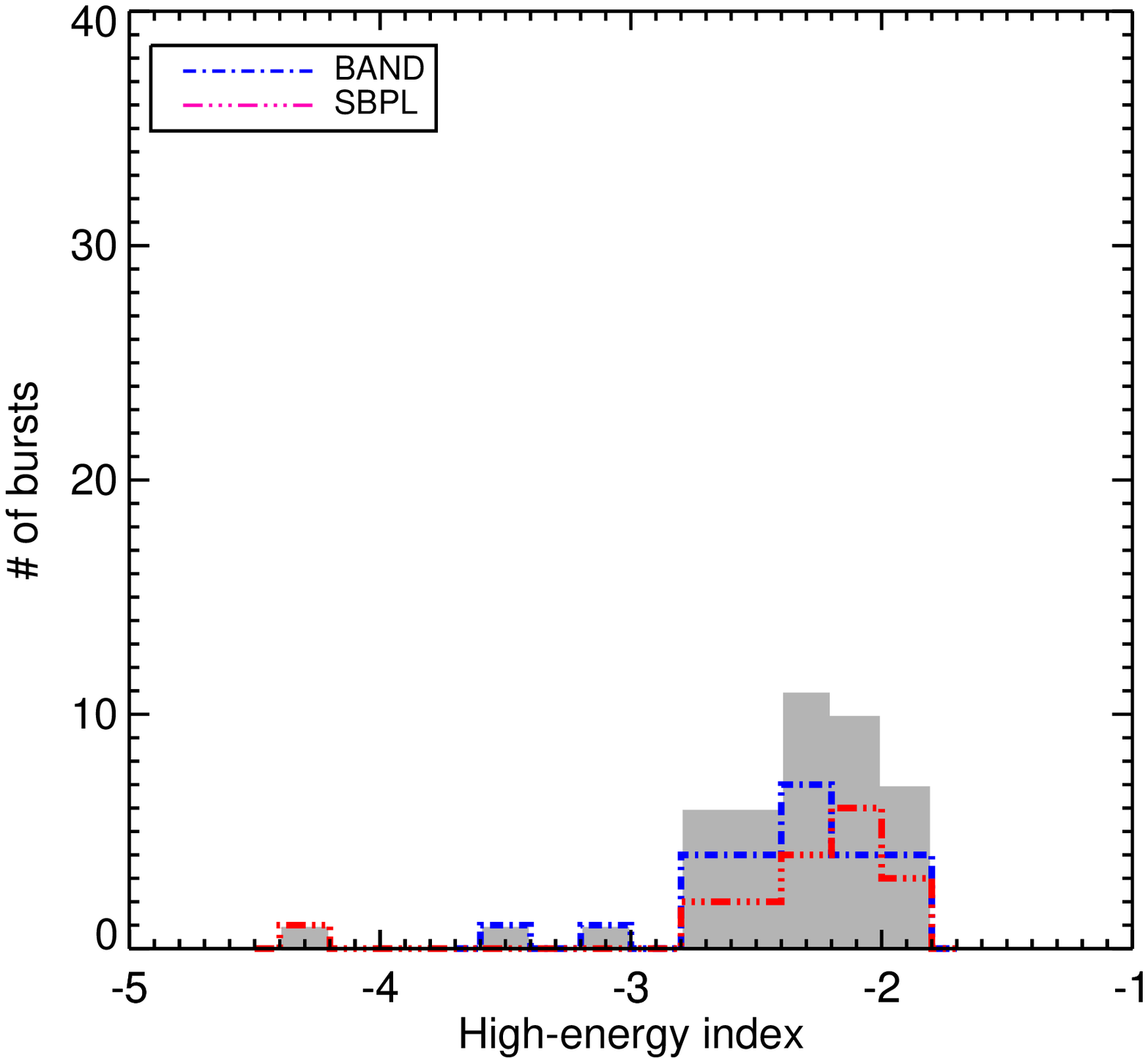}}
	\end{center}
\caption{Distribution of the high-energy indices obtained from the GOOD \emph{P} spectral fits \subref{allhiidx}. The BEST parameter distribution (gray filled histogram) and its constituents for the high-energy index is shown in \subref{allhiidxbest}. 
}\label{highenindexpeak}
\end{figure}

The \emph{P} spectral distributions have been produced by fitting the GRB spectra over the 1024 ms and 64 ms peak 
flux duration of long and short bursts respectively.  Note that the results from both long and short bursts are included in the 
following figures. The low-energy indices from the \emph{P} selections of the BEST sample, as shown in Figure \ref{allloidxPeakbest}, have a median value of about  $-1.3$ and show a bimodal distribution. This is due to the fact that more GRBs of the \emph{P} sample are best fit by the \pwrlw because, due to less photon fluence accumulation, the signal-to-noise ratio decreases. 20\% of the BEST low-energy indices show $\alpha > -2/3$ and violate the synchrotron ``line-of-death'', while an additional 32\% of the indices are $\alpha < -3/2$ and violate the synchrotron cooling limit, both of which are significantly larger percentages than those from the \emph{F} spectra. The high-energy indices in Figure \ref{highenindexpeak} peak at about $-2.2$ and  again have a long tail toward steeper indices.  The number of unconstrained high-energy indices increases when compared to the \emph{F} spectra, again likely due to the poorer statistics resulting from shorter integration times. 
As has been shown with the \emph{F} spectral fits, the \pwrlw index serves as an average between low- and high-energy indices for the \band and \sbpl functions. 

\begin{figure}
	\begin{center}
		\subfigure[]{\includegraphics[scale=0.45]{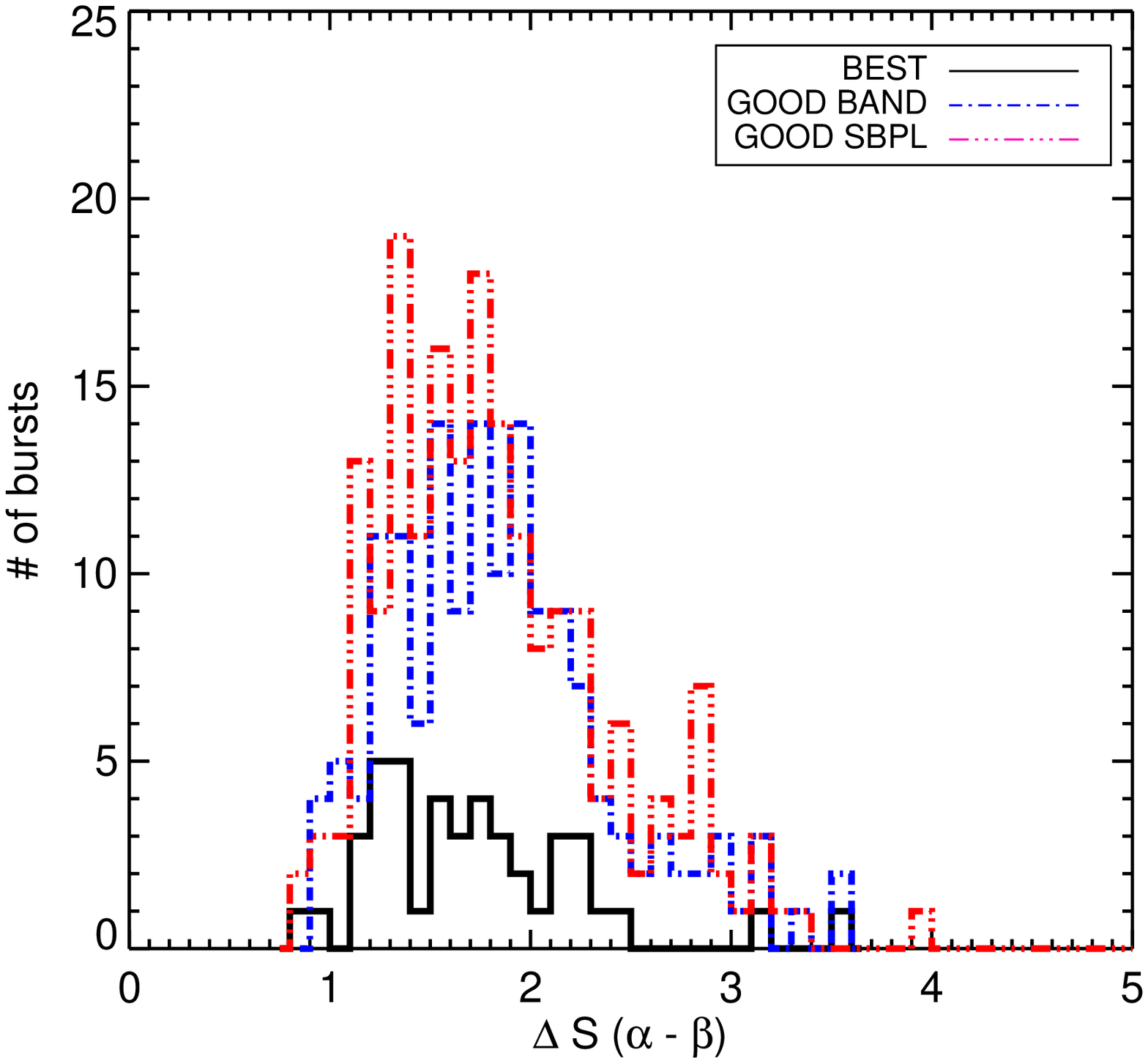}}
	\end{center}
\caption{Distributions of $\Delta S$, the difference between the low- and high-energy spectral indices ($\alpha - \beta$) for the \emph{P} spectral fits. The BEST  (black solid line), the GOOD BAND power-law indices (blue dash-dotted line) and the GOOD \sbpl power-law indices (red dash-dot-dot-dotted line) for the \emph{F} spectral fits are shown.  The first bin contains values 
less than 0, indicating that the centroid value of $\alpha$ is steeper than the centroid value of $\beta$.  \label{deltasbestpeak}}
\end{figure}

Shown in Figure \ref{deltasbestpeak} is the $\Delta S$ distribution for the \emph{P} spectra.  This distribution seems consistent with 
the BATSE results found previously \citep{Preece, Kaneko06}, but suffers from a deficit in values compared to the \emph{F} fits largely due to the inability 
of the data to sufficiently constrain the high-energy power law index. The distribution peaks at $\sim 1.3$ and has a median value of $1.64$.

\begin{figure}
	\begin{center}
		\subfigure[]{\label{allepeak}\includegraphics[scale=0.45]{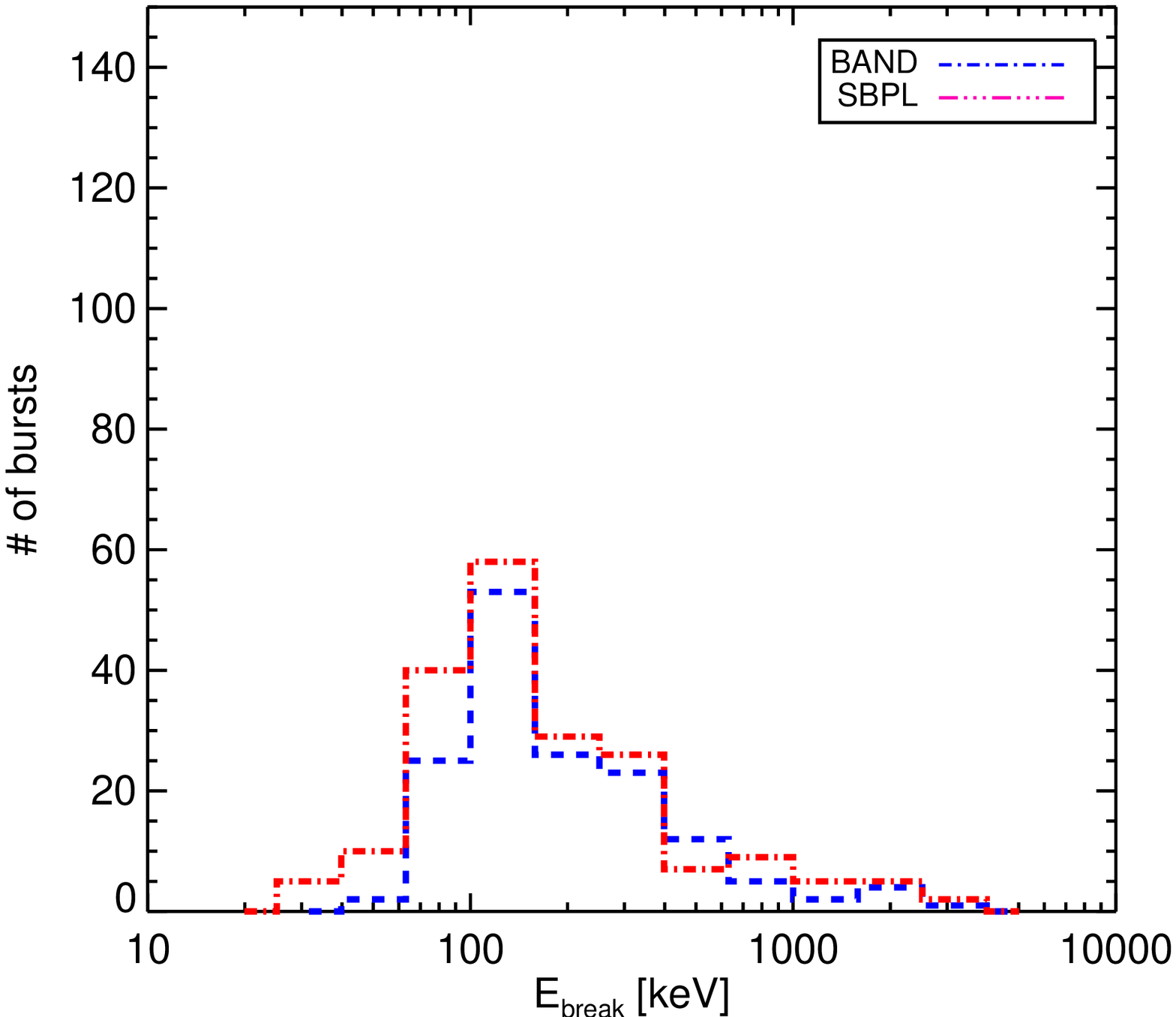}}
		\subfigure[]{\label{allepeakbest}\includegraphics[scale=0.45]{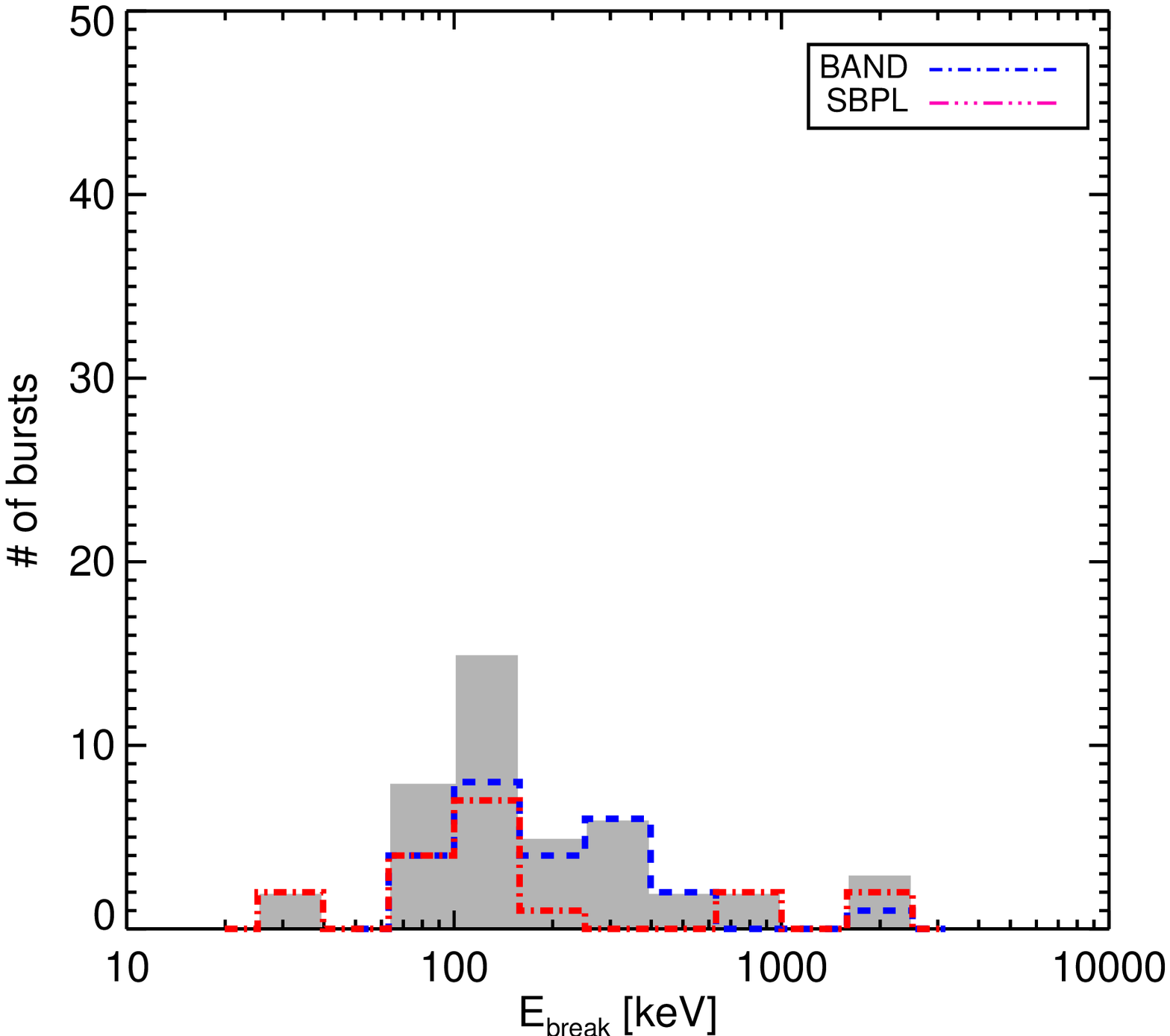}}
	\end{center}
\caption{Distribution of the \ebreak obtained from the GOOD \emph{P} spectral fits \subref{allebreak}. The BEST parameter distribution (gray filled histogram) and its constituents for \ebreak is shown in \subref{allebreakbest}. 
}\label{ebreakPeak}
\end{figure}

\begin{figure}
	\begin{center}
		\subfigure[]{\label{allepeakPeak}\includegraphics[scale=0.45]{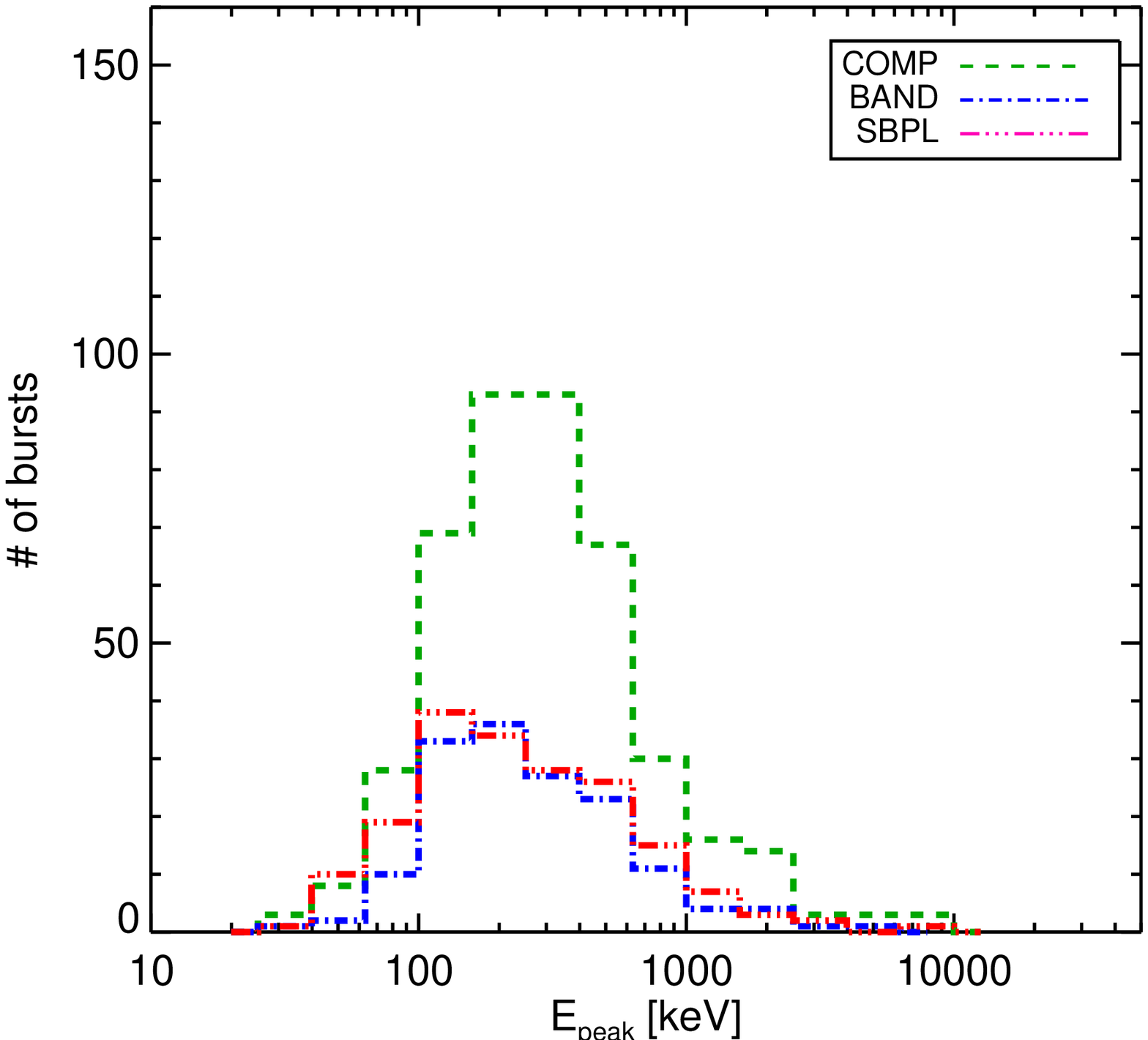}}
		\subfigure[]{\label{allepeakPeakbest}\includegraphics[scale=0.45]{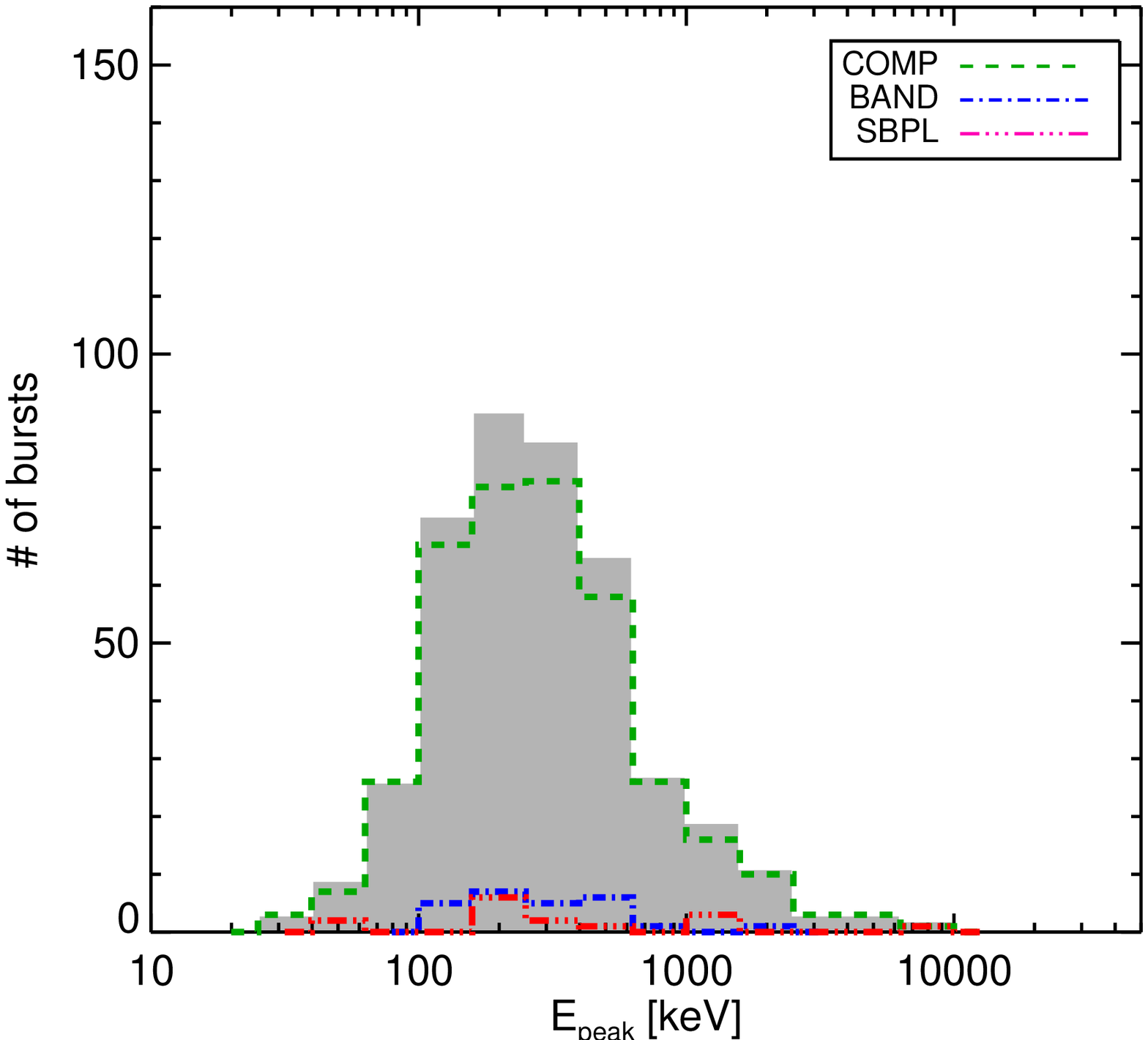}}
	\end{center}
\caption{Distribution of the \epeak obtained from the GOOD \emph{P} spectral fits \subref{allepeak}. The BEST parameter distribution (gray filled histogram) and its constituents for \epeak is shown in \subref{allepeakbest}. 
}\label{epeakPeak}
\end{figure}

In Figure \ref{ebreakPeak} and Figure \ref{epeakPeak}, we show the distributions for \ebreak and  \epeak, respectively.  As was evident from the \emph{F} 
spectra, the \ebreak from the \sbpl fits appears to peak at 100 keV.  The \epeak distribution for the BEST sample peaks around 260 keV and covers just over two orders of 
magnitude, which is consistent with previous findings \citep{Preece98, Kaneko06}.  It should be noted that the data over the 
short timescales in the \emph{P} spectra do not often favor either the \band or the \sbpl model, resulting in large parameter errors.  

\begin{figure}
	\begin{center}
		\subfigure[]{\label{allepeakvsalphaPeak}\includegraphics[scale=0.45]{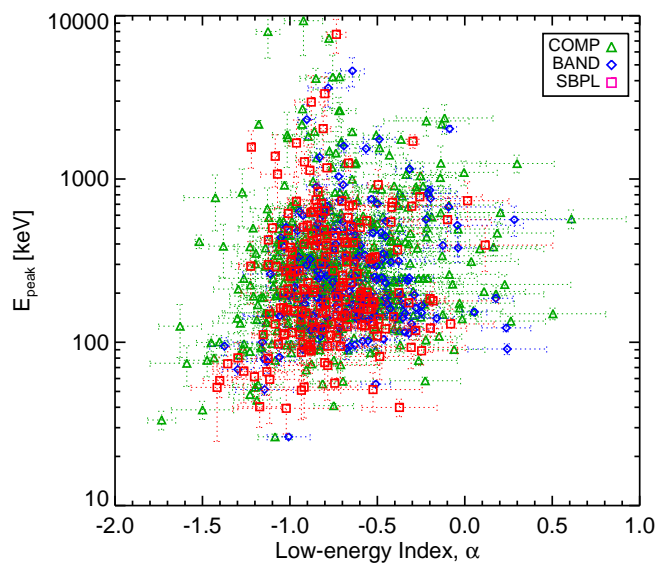}}
		\subfigure[]{\label{allepeakvsalphaPeak}\includegraphics[scale=0.45]{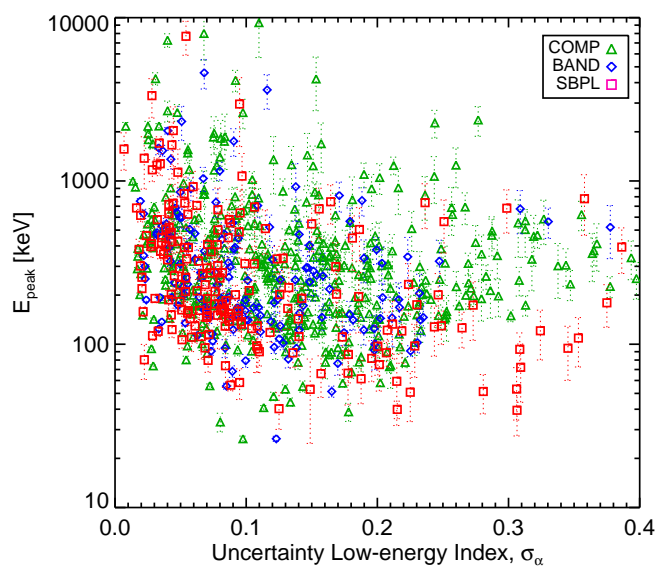}}
	\end{center}
\caption{Comparison of the low-energy index and \epeak for three models from the \emph{P} spectral fits.  This comparison 
reveals a correlation between the \epeak energy and the uncertainty on the low-energy index: generally a  lower energy 
\epeak tends to result in a less constrained low-energy index. \label{alphaepeakPeak}}
\end{figure}

Even though it is less obvious, Figure~\ref{alphaepeakPeak} shows a correlation between the \epeak and low-energy power-law index and its uncertainty which is similar to the \emph{F} spectral fits.

\begin{figure}
	\begin{center}
		\subfigure[]{\label{alldeltaepPeak}\includegraphics[scale=0.45]{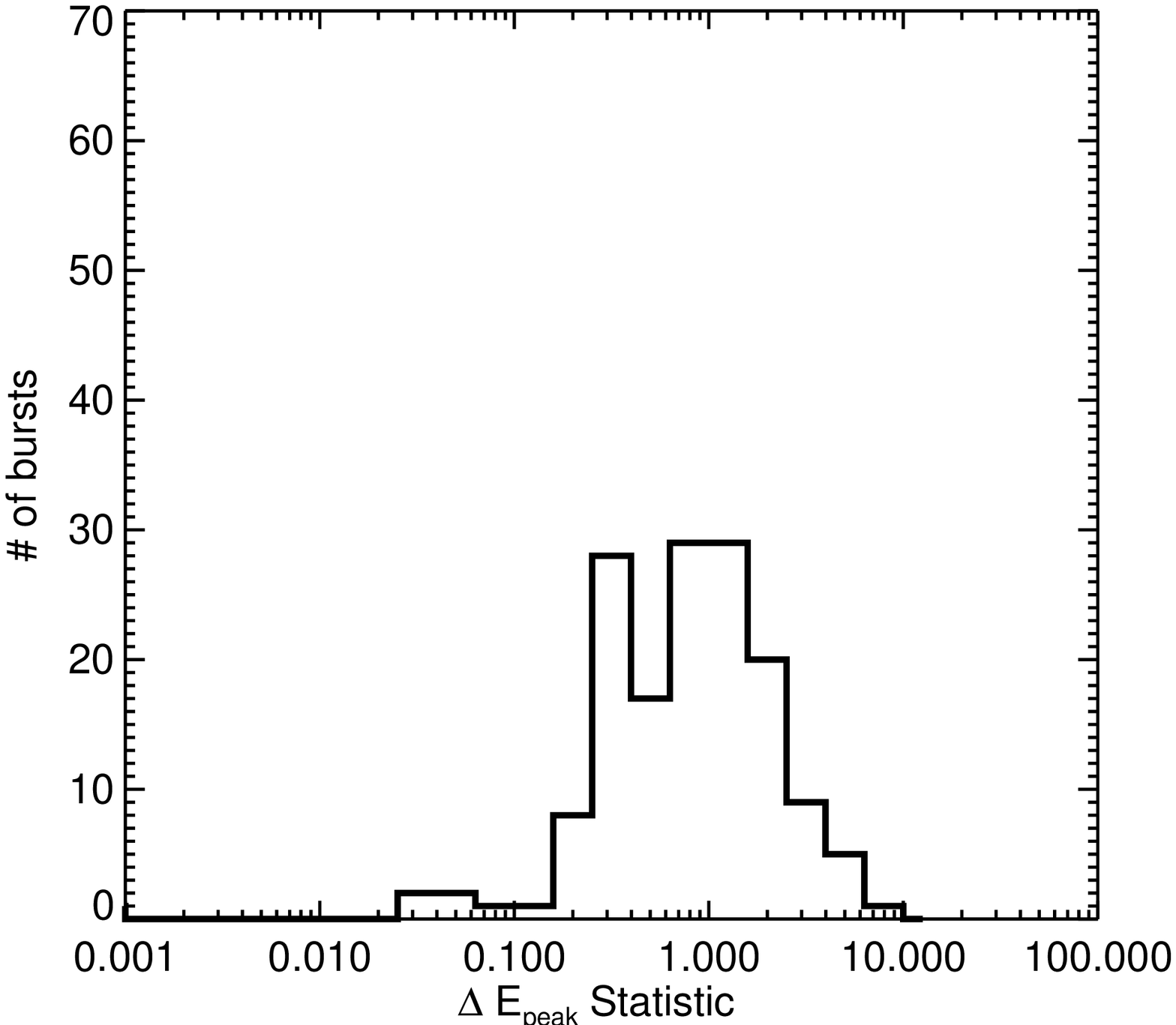}}
	\end{center}
\caption{Distribution of the $\Delta E_{\rm peak}$ statistic for the \comp and \band models from \emph{P} spectral fits.  A value less 
than 1 indicates the \epeak values are within errors, while a value larger than 1 indicates the \epeak values are not 
within errors. \label{deltaepeakPeak}}
\end{figure}

In addition, we 
calculate and show the $\Delta E_{\rm peak}$ statistic in Figure \ref{deltaepeakPeak} and, as was the case with the \emph{F} spectra, 
$\sim$46\% of the \band and \comp \epeak values are found to be outside the combined errors. If the allowed error range is extended to 3$\sigma$ then only 8\% of the \epeak values are not consistent.

\begin{figure}
	\begin{center}
		\begin{minipage}[t]{1\textwidth}
		\subfigure[]{\label{allefluxPeak}\includegraphics[scale=0.45]{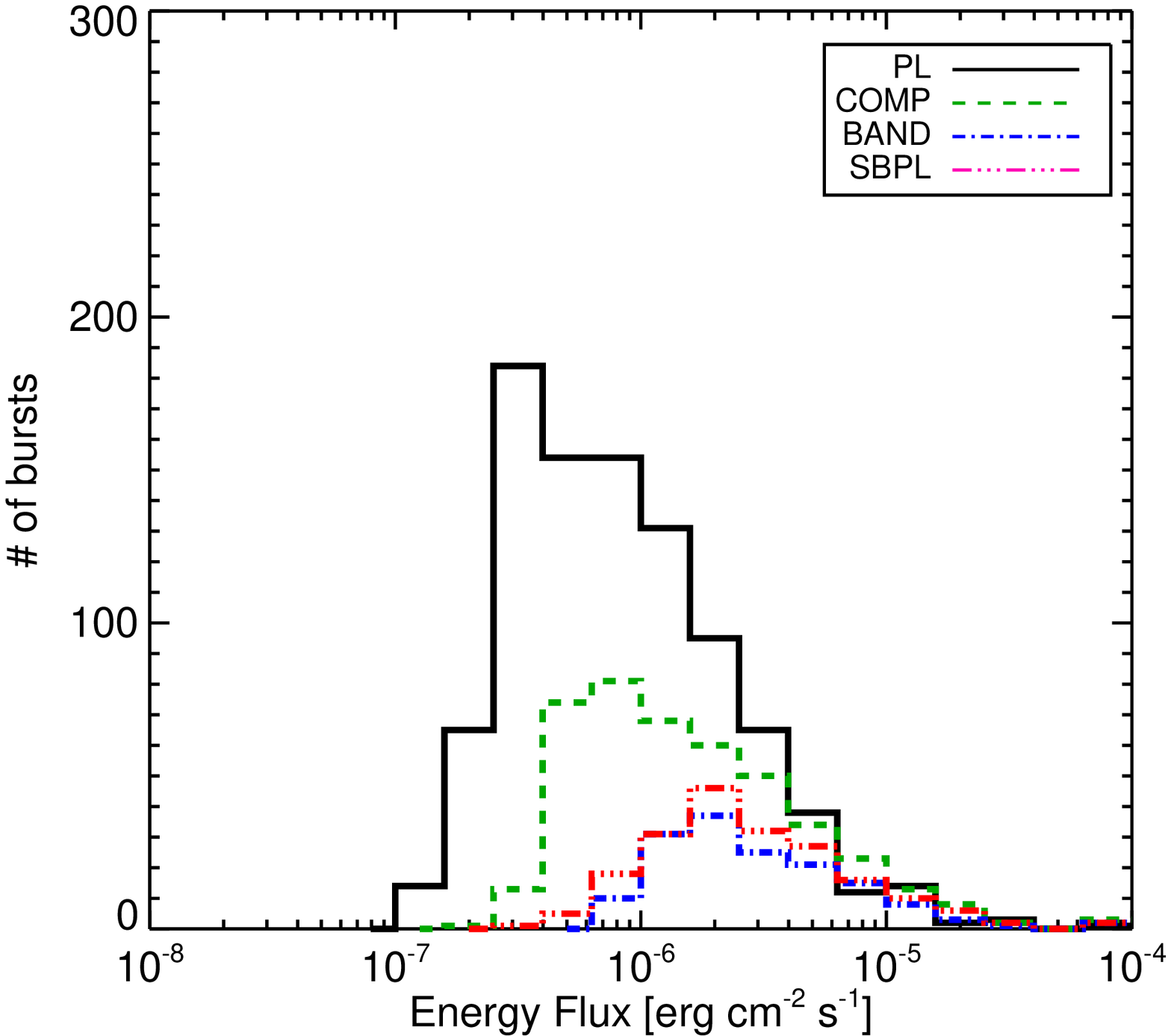}}
		\subfigure[]{\label{allefluxPeak40MeV}\includegraphics[scale=0.45]{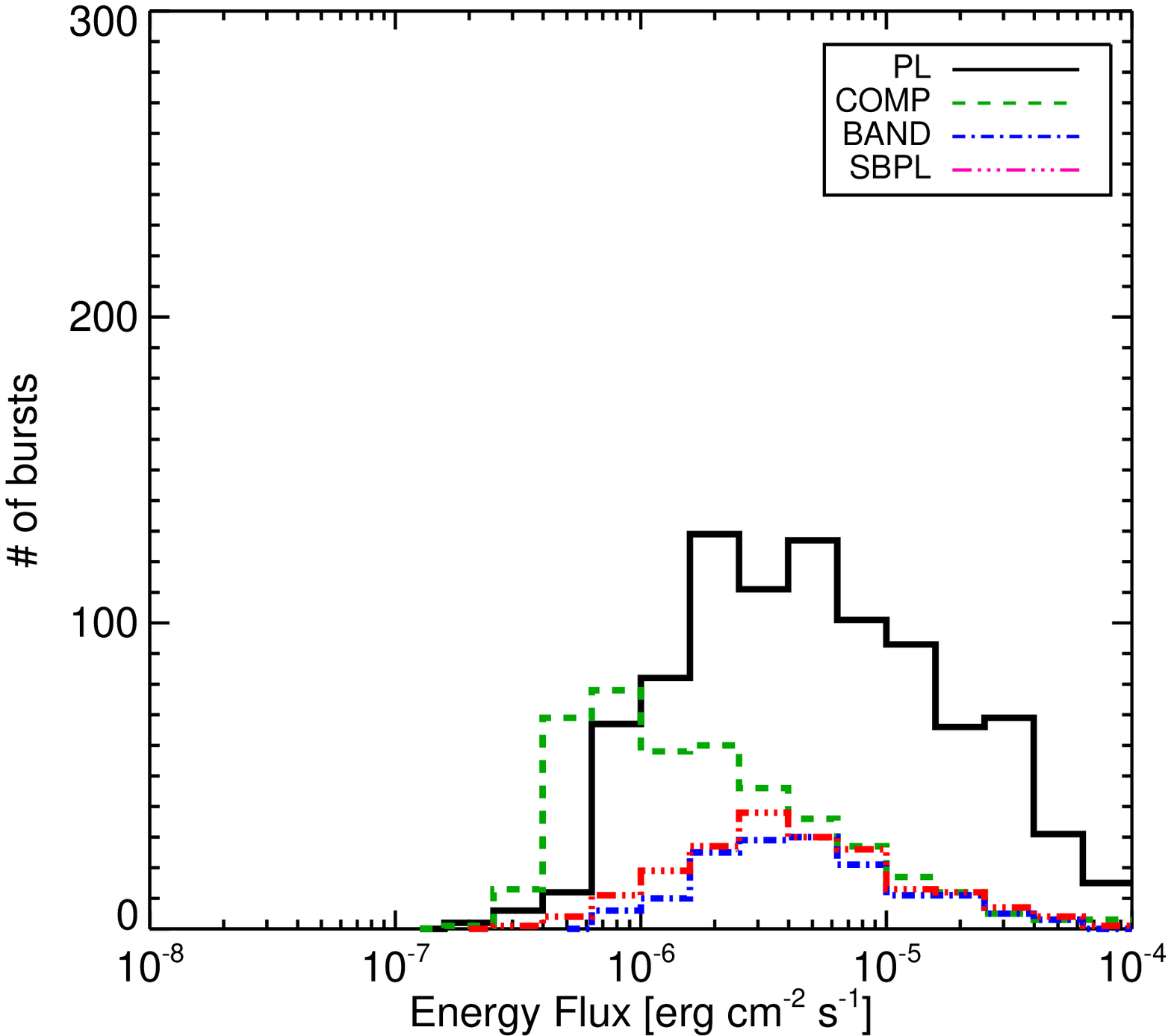}}
		\end{minipage}
		\begin{minipage}[t]{1\textwidth}
		\subfigure[]{\label{allefluxPeakbest}\includegraphics[scale=0.45]{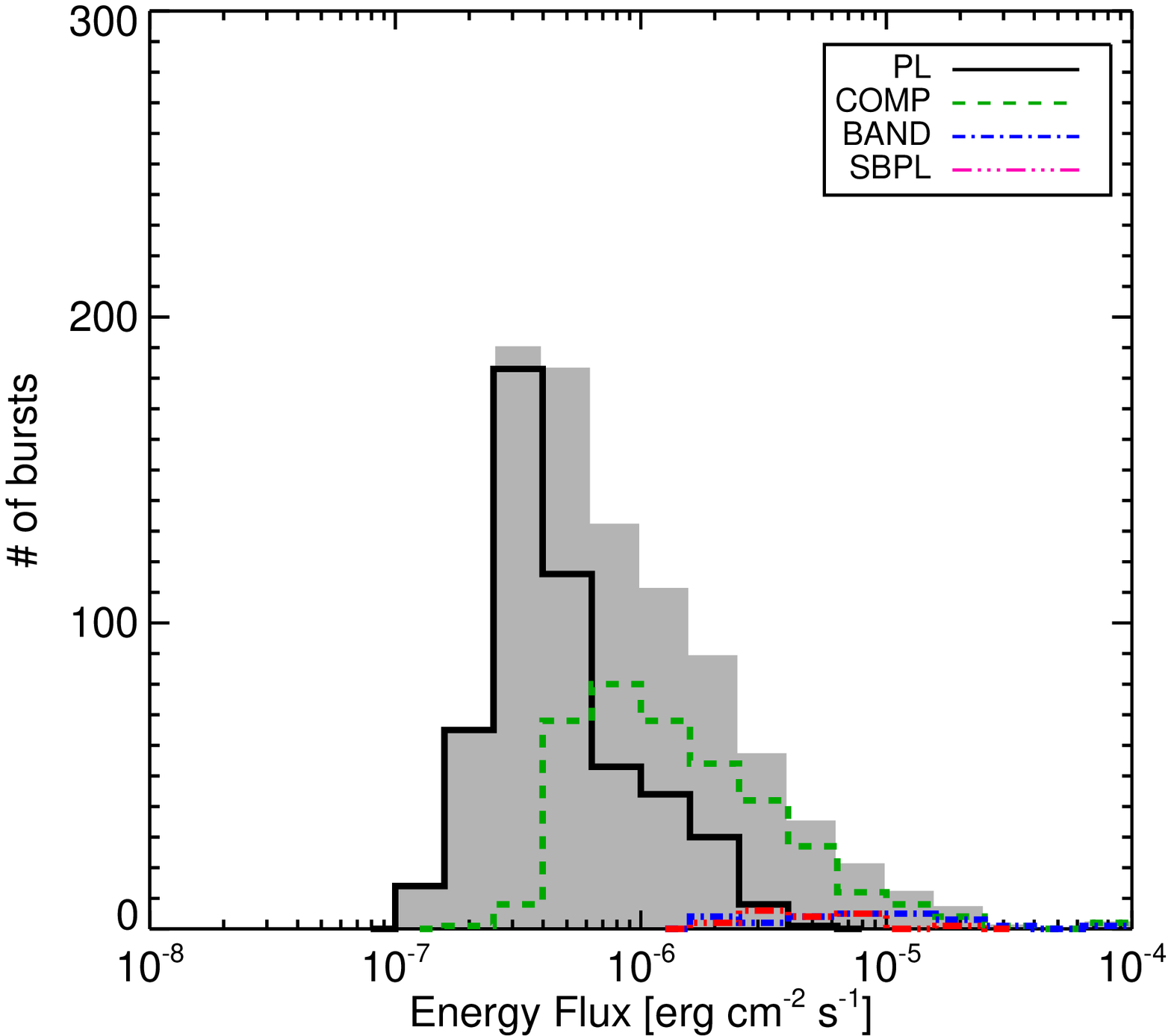}}
		\subfigure[]{\label{allefluxPeak40MeVbest}\includegraphics[scale=0.45]{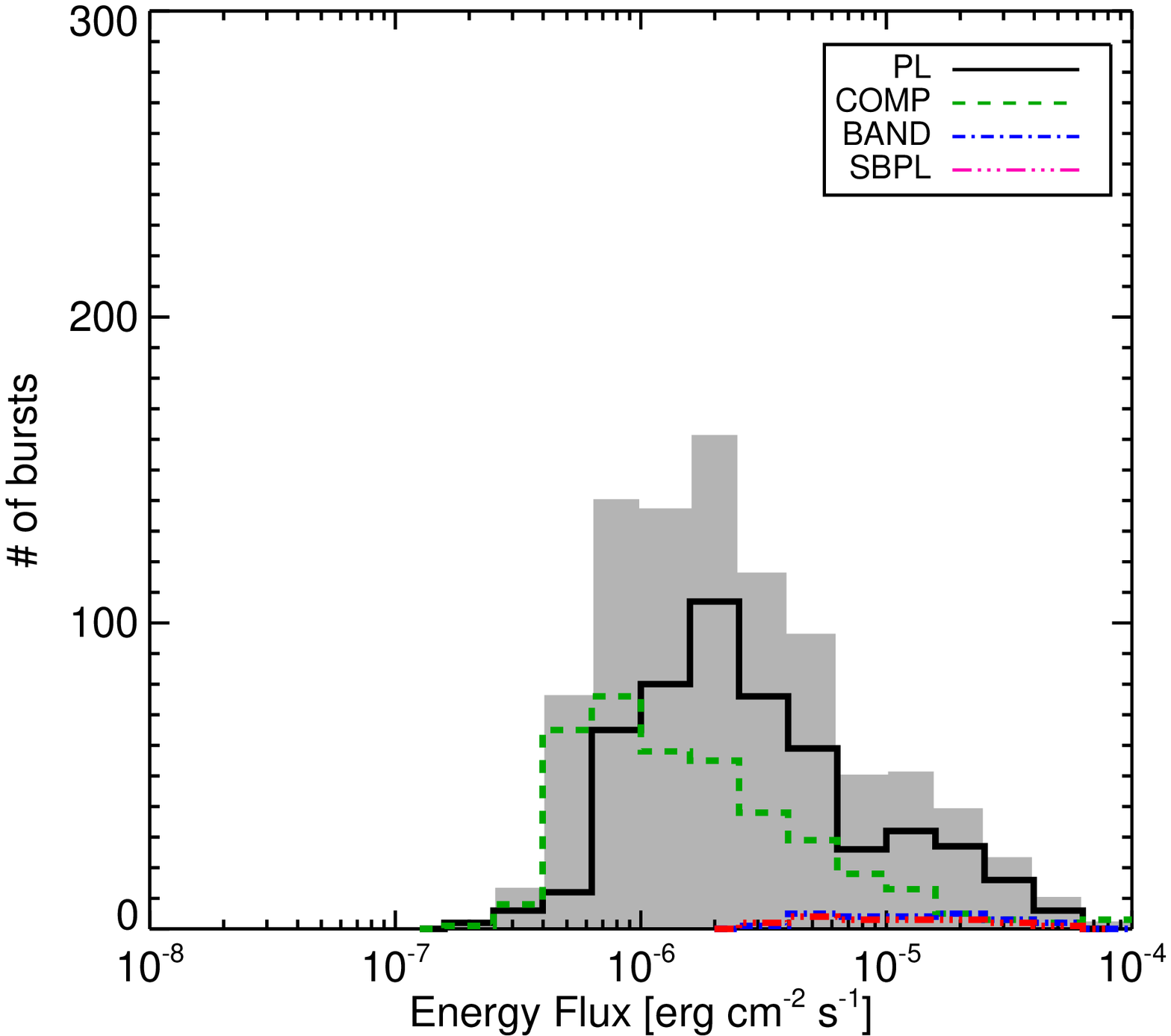}}
		\end{minipage}
	\end{center}
\caption{Distributions of energy flux in the 10~keV--1~MeV \subref{allefluxf} and 10~keV--40~MeV \subref{allefluxf40MeV} band from the GOOD \emph{P} spectral fits. Note that the plotted distributions contain the flux on two different timescales: 1024 ms (long GRBs) and 64 ms (short GRBs). The BEST parameter distribution for the energy flux in both energy ranges is shown in \subref{allefluxfbest} and \subref{allefluxf40MeVbest}. The gray filled histogram shows the total distribution and the constituents are shown in colors. \label{enfluxPeak}}
\end{figure}

\begin{figure}
	\begin{center}
		\begin{minipage}[t]{1\textwidth}
		\subfigure[]{\label{allpfluxPeak}\includegraphics[scale=0.45]{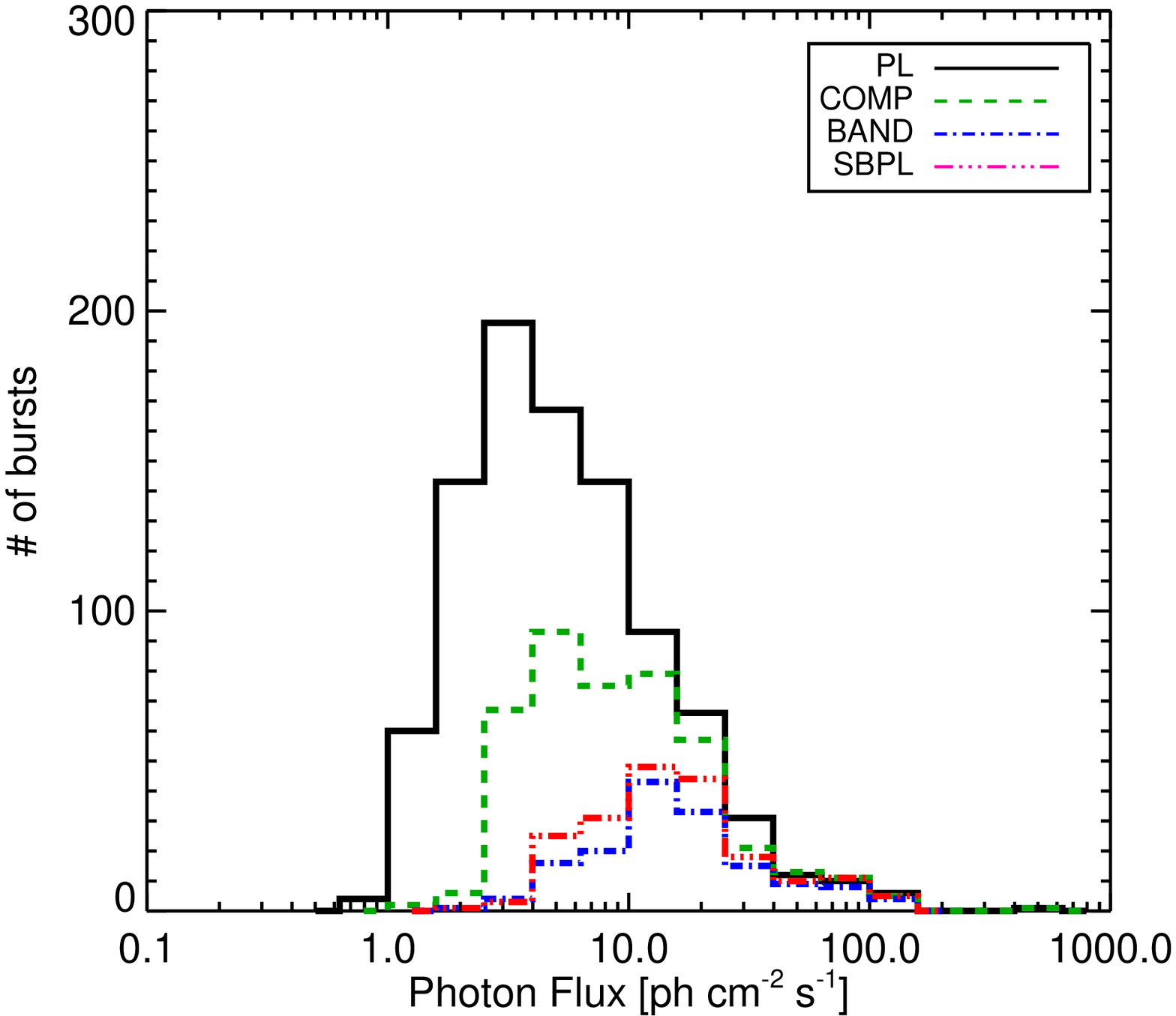}}
		\subfigure[]{\label{allpfluxPeak40MeV}\includegraphics[scale=0.45]{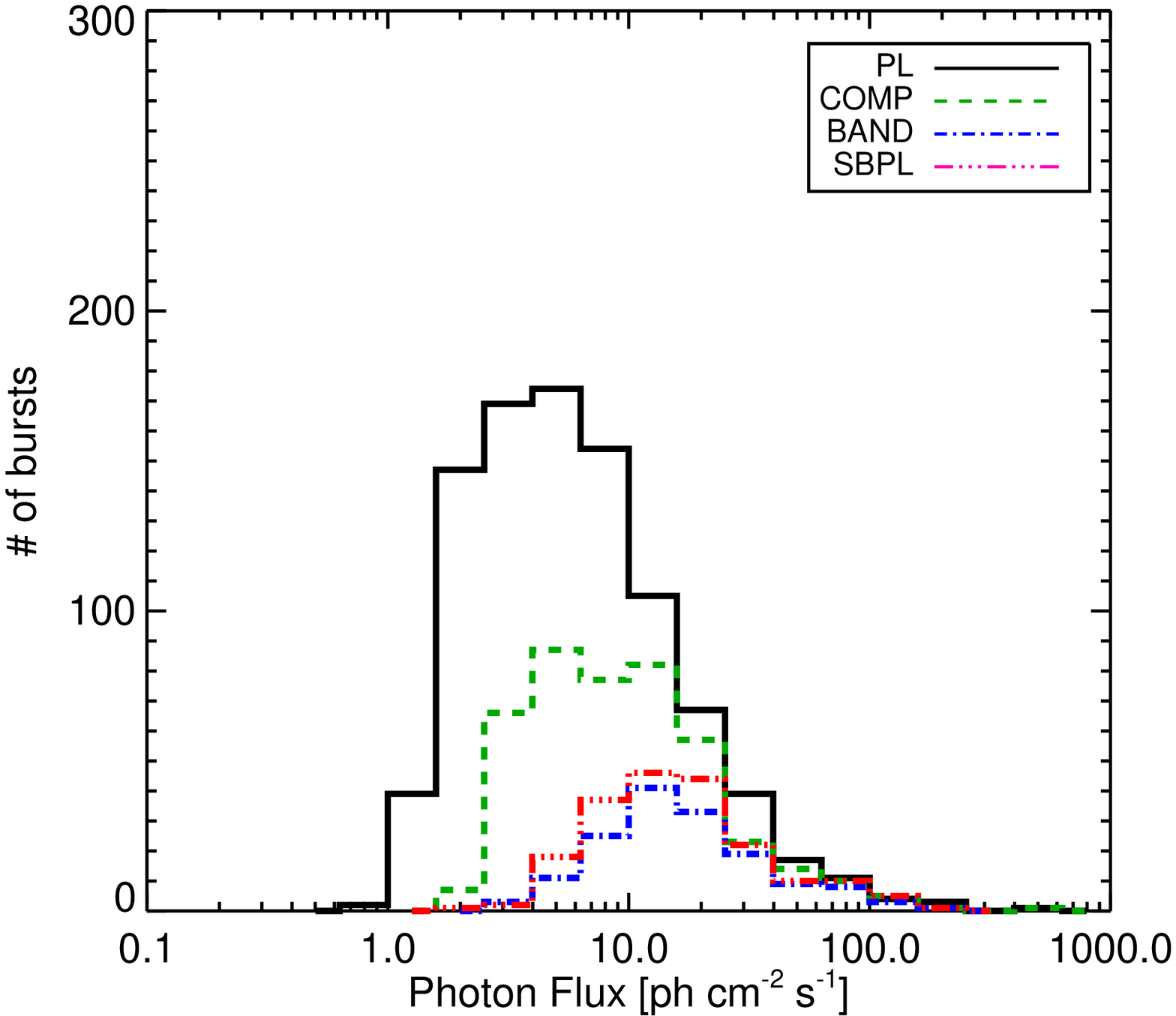}}
		\end{minipage}
		\begin{minipage}[t]{1\textwidth}
		\subfigure[]{\label{allpfluxPeakbest}\includegraphics[scale=0.45]{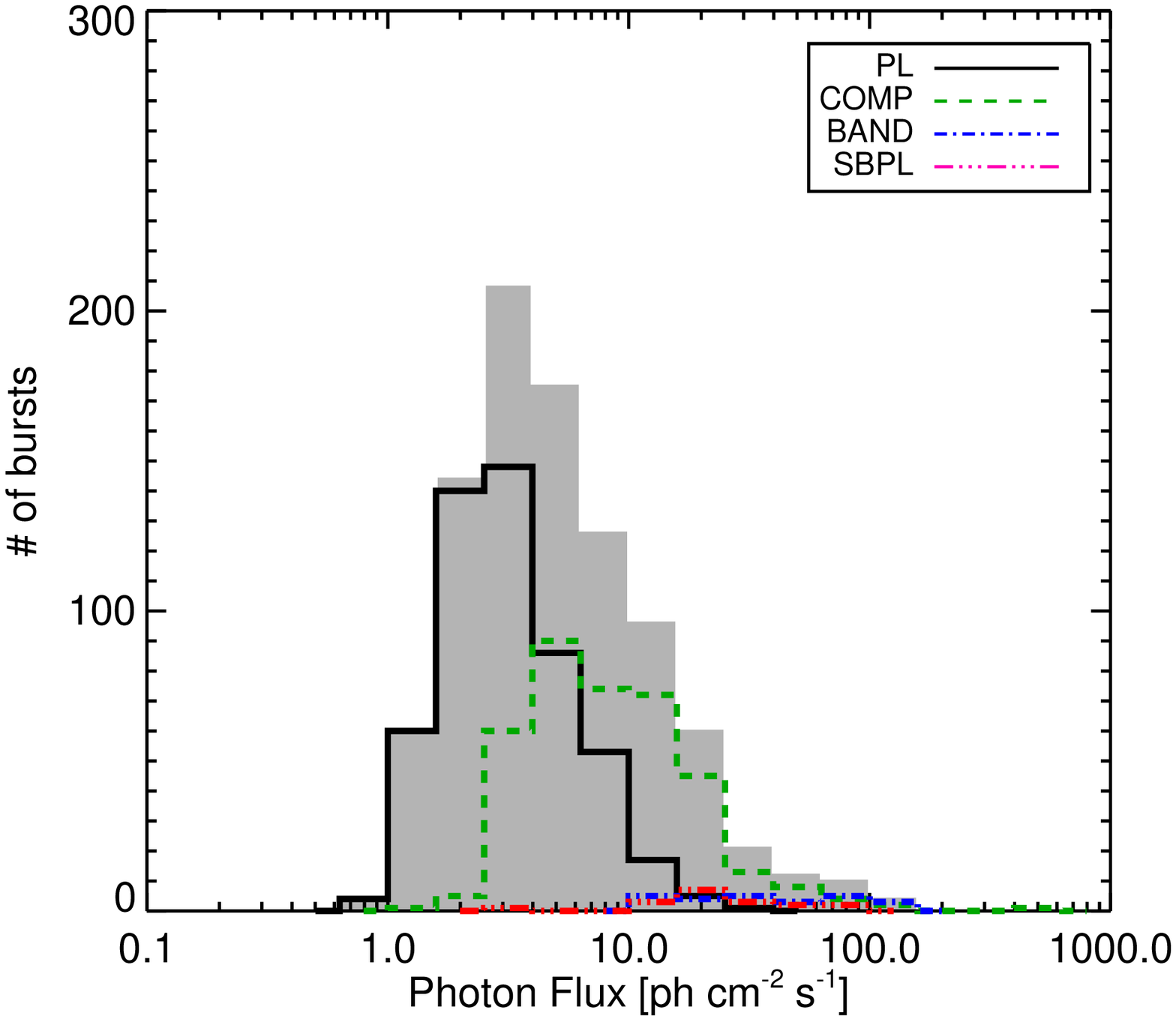}}
		\subfigure[]{\label{allpfluxPeak40MeVbest}\includegraphics[scale=0.45]{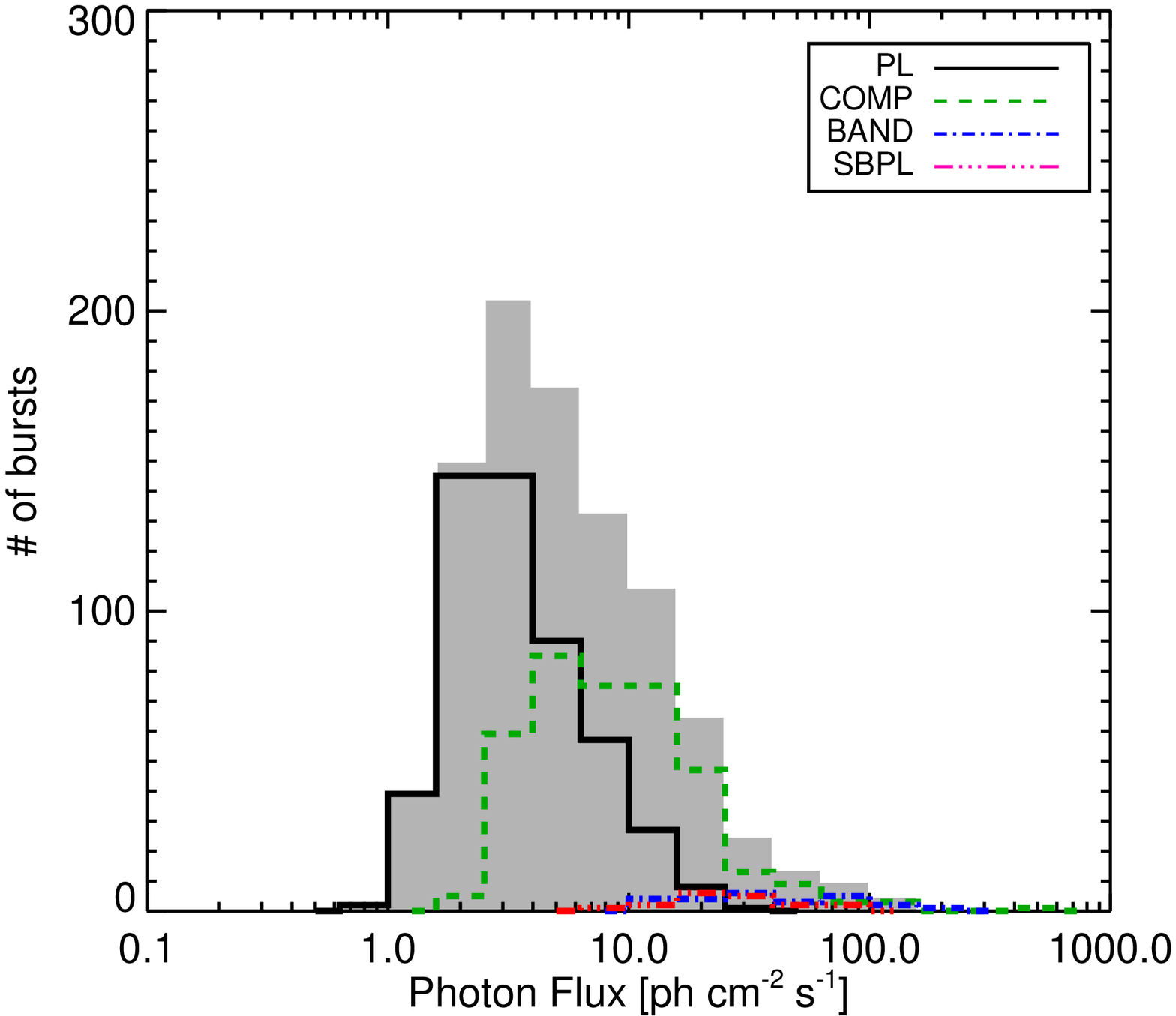}}
		\end{minipage}
	\end{center}
\caption{Distributions of photon flux in the 10~keV--1~MeV \subref{allpfluxf} and 10~keV--40~MeV \subref{allpfluxf40MeV} band from the GOOD \emph{P} spectral fits. Note that the plotted distributions contain the flux on two different timescales: 1024 ms (long GRBs) and 64 ms (short GRBs). The BEST parameter distribution for the photon flux in both energy ranges is shown in \subref{allpfluxfbest} and \subref{allpfluxf40MeVbest}. The gray filled histogram shows the total distribution and the constituents are shown in colors. \label{pfluxPeak}}
\end{figure}

The distributions for the peak energy flux and photon flux are shown in Figures~\ref{enfluxPeak} and Figure~\ref{pfluxPeak}, respectively.  The photon flux of the BEST sample peaks around 4.5 photons cm$^{-2}$  s$^{-1}$, and the energy flux of the BEST sample peaks at $6.4 \times 10^{-7}$ ergs cm$^{-2}$ s$^{-1}$ in the 10-1000 keV band.  When integrating over the full GBM spectral band, 10 keV-40 MeV, the dispersion in the energy flux increases and approximates a top hat function with a small high-flux tail spanning about 2 orders of magnitude. 
The brightest short GRB in terms of peak photon flux is GRB 120323.507 at $\rm > 580\ ph\ s^{-1}\ cm^{-2}$ while the most energetic short GRB is GRB 090227.772 with an energy flux of $\rm \sim8.3 \times 10^{-5}\ ergs\ s^{-1} cm^{-2}$. 

\begin{figure}
	\begin{center}
		\begin{minipage}[t]{1\textwidth}
		\subfigure[]{\label{fig:pepeakzcorr}\includegraphics[scale=0.45]{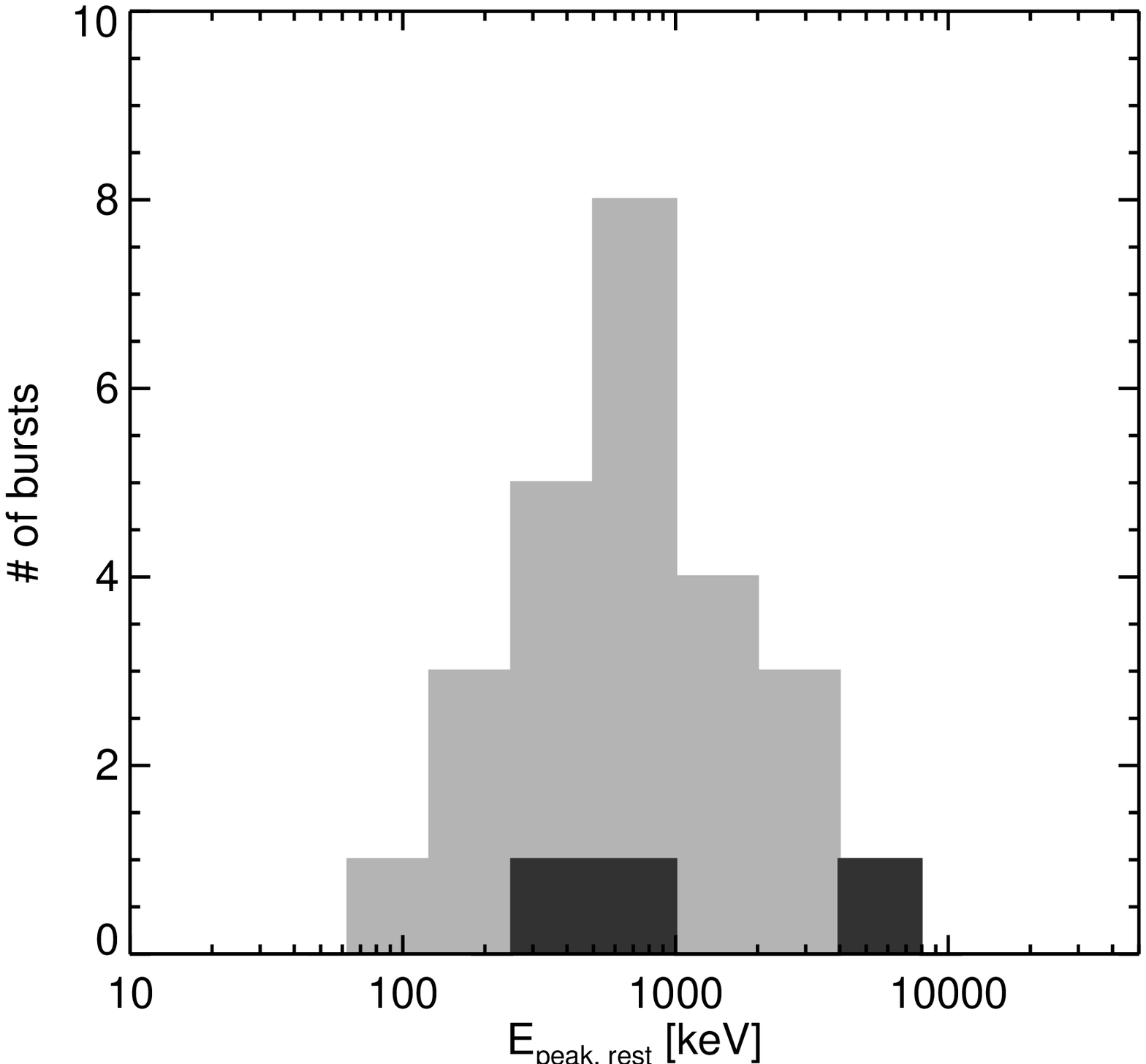}}
		\subfigure[]{\label{fig:pebreakzcorr}\includegraphics[scale=0.45]{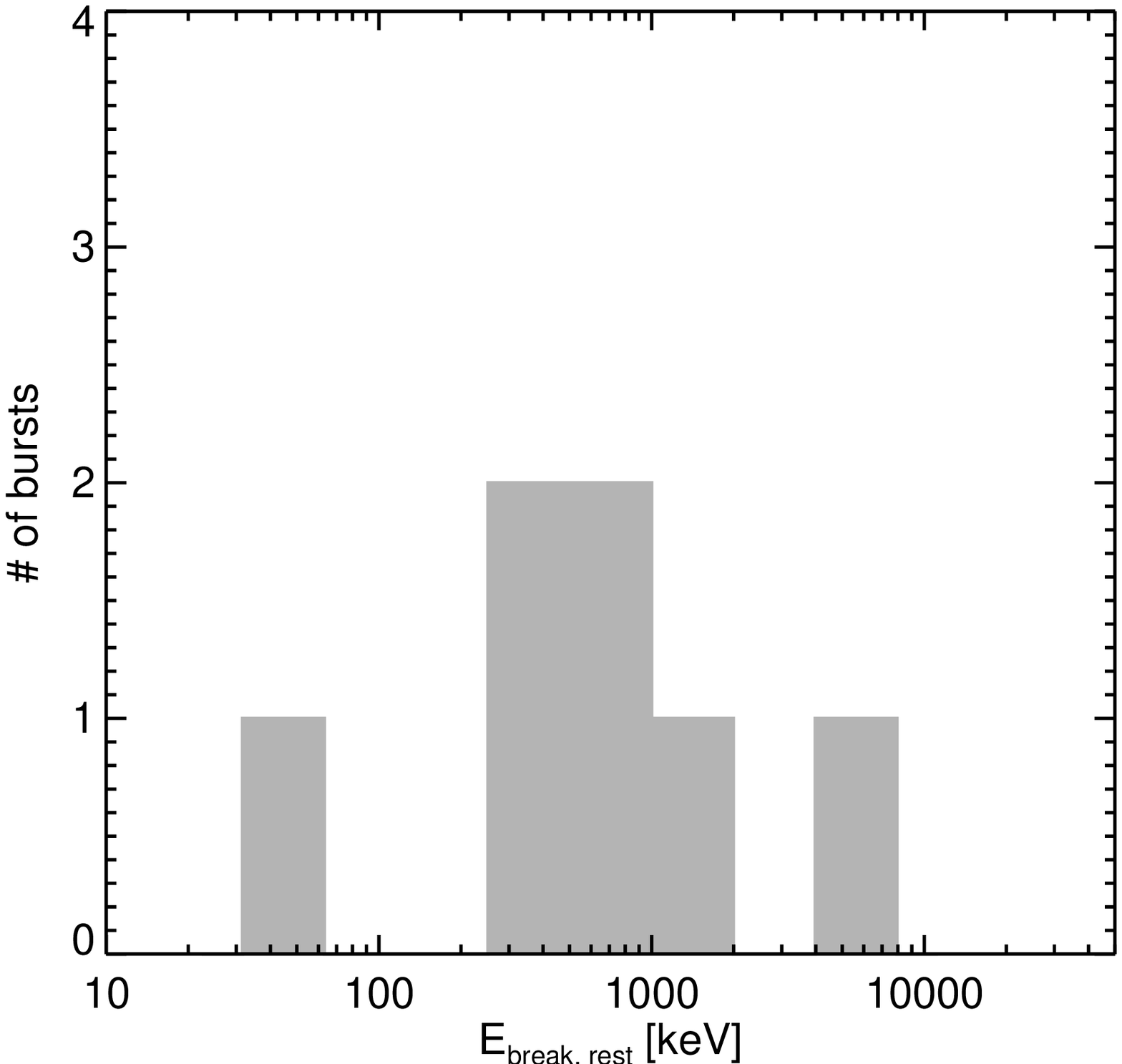}}
		\end{minipage}
		\begin{minipage}[t]{1\textwidth}
		\subfigure[]{\label{fig:penfluxzcorr}\includegraphics[scale=0.45]{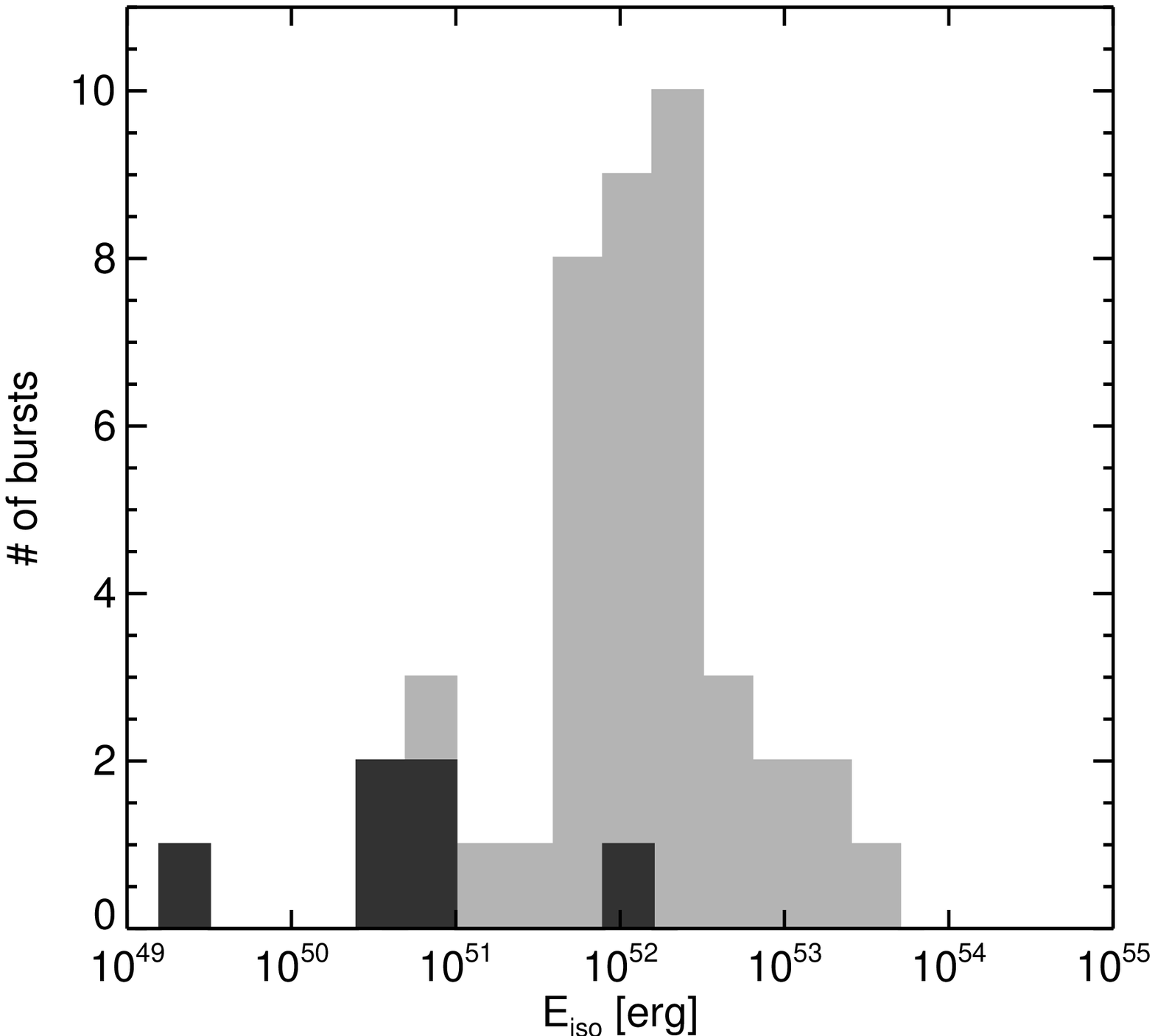}}
		\subfigure[]{\label{fig:penfluencezcorr}\includegraphics[scale=0.45]{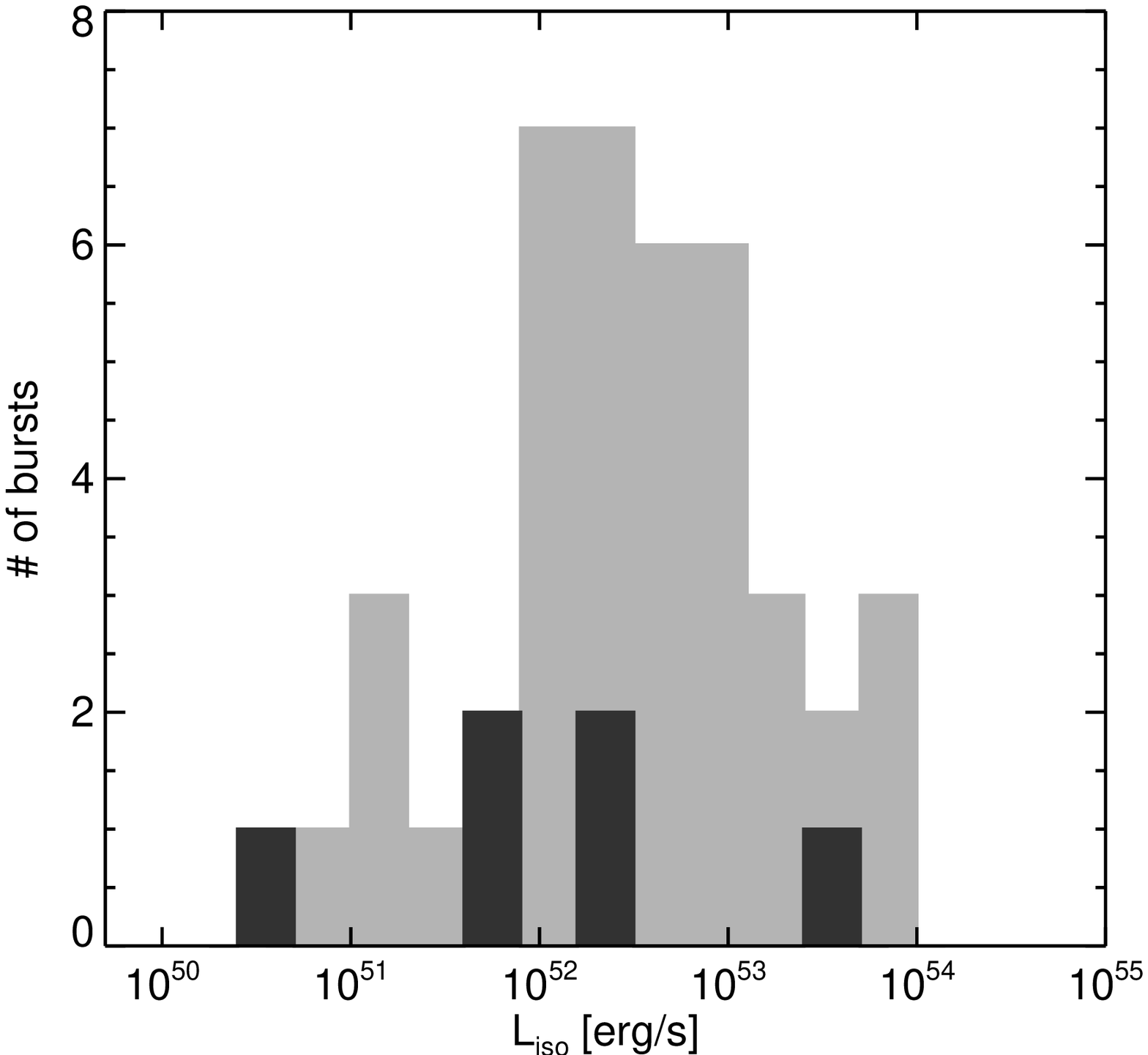}}
		\end{minipage}
	\end{center}
\caption{\subref{fig:pepeakzcorr} \eprest distribution of the \emph{P} spectra fits from the BEST sample for long and short GRBs (light gray and dark gray filled histogram, respectively). \subref{fig:pebreakzcorr} Same as \subref{fig:fepeakzcorr} but for \ebrest. \subref{fig:penfluxzcorr} and \subref{fig:penfluencezcorr}  Distribution of \eiso and \Liso of the fluence spectral fits from the BEST sample in the rest-frame energy band 1/(1+z)~keV to 10/(1+z)~MeV.  \label{restframep}}
\end{figure}

In Figure~\ref{restframep} we present the rest-frame spectral parameters for \epeak, \ebreak, \eiso and \Liso of the \emph{P} spectral fits where the latter two have been determined in the rest-frame energy band between $1/(1+z)$~keV and $10/(1+z)$~MeV. Both, \eprest and \ebrest peak at roughly $600$~keV and, similarly to the distributions in the observer frame, cover just two orders of magnitude. The \eiso distribution is narrower than the one obtained from the \emph{F} spectral fits and has a median of $\sim 10^{52}$~erg. \Liso, on the other hand, has a median of $\sim 3\times10^{52}$~erg/s and covers $\sim 3$ orders of magnitude.

\begin{figure}
	\begin{center}
		\begin{minipage}[t]{1\textwidth}
		\subfigure[]{\label{fig:falphavsz}\includegraphics[scale=0.45]{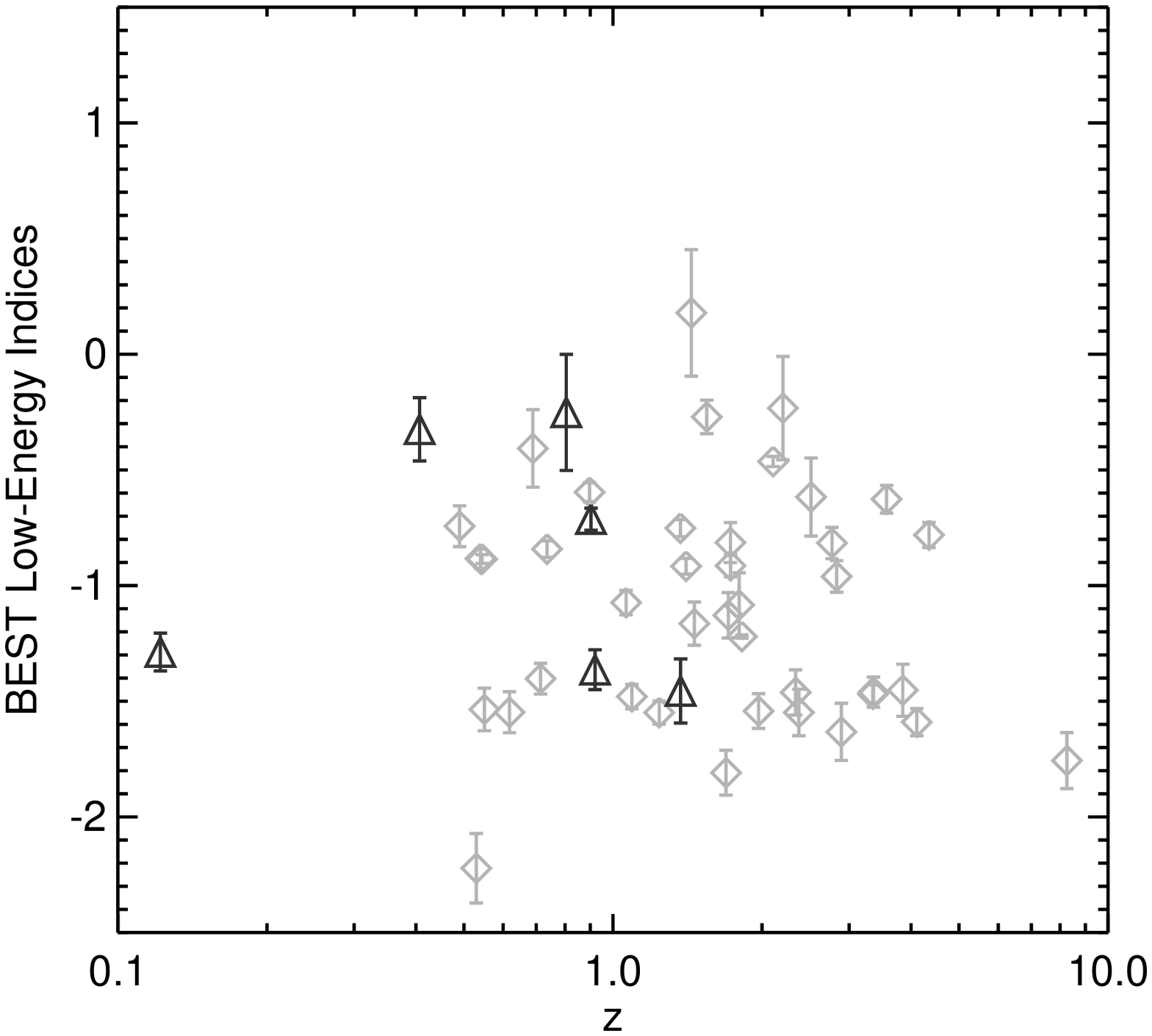}}
		\subfigure[]{\label{fig:fbetavsz}\includegraphics[scale=0.45]{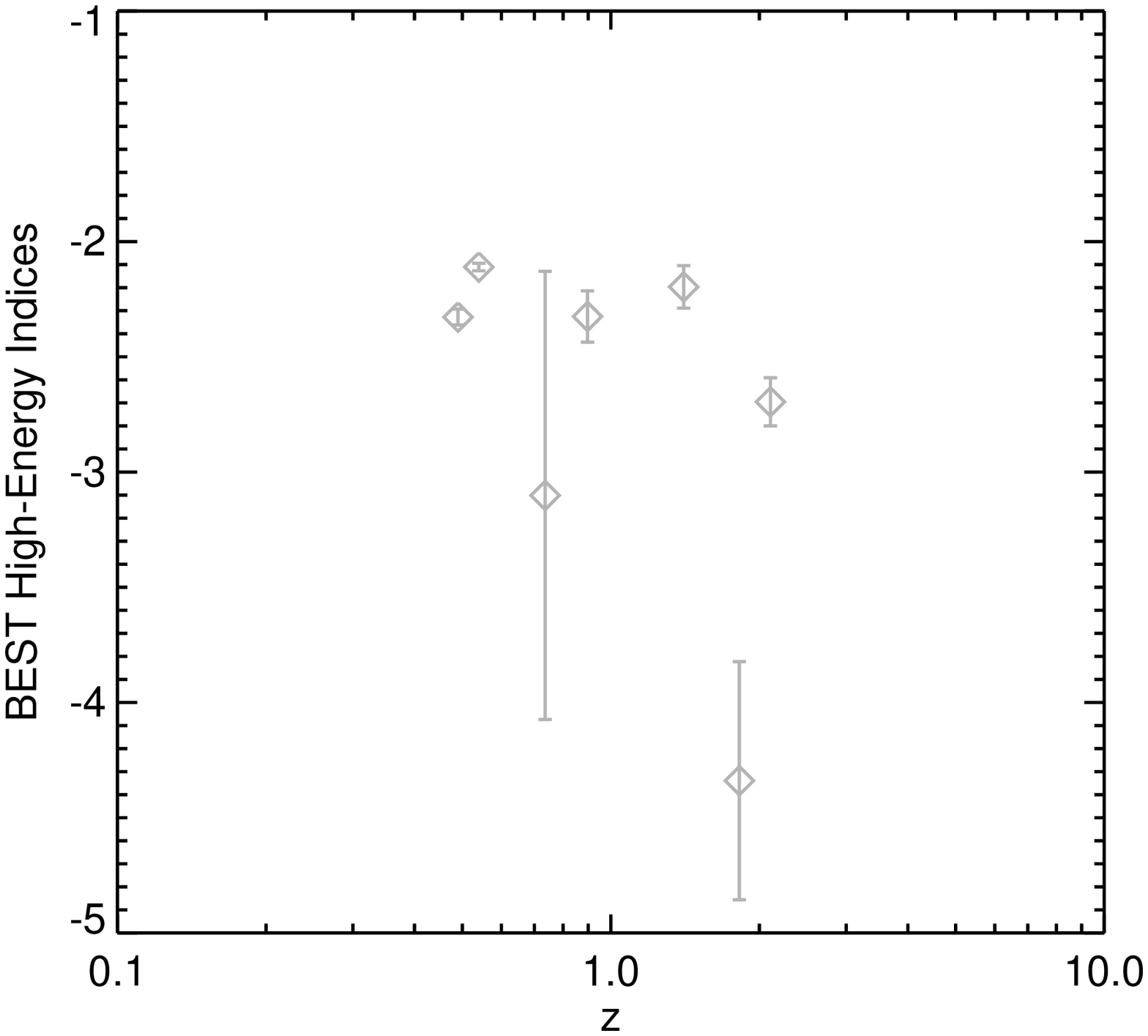}}
		\end{minipage}
		\begin{minipage}[t]{1\textwidth}
		\centering
		\subfigure[]{\label{fig:epeakvsz}\includegraphics[scale=0.45]{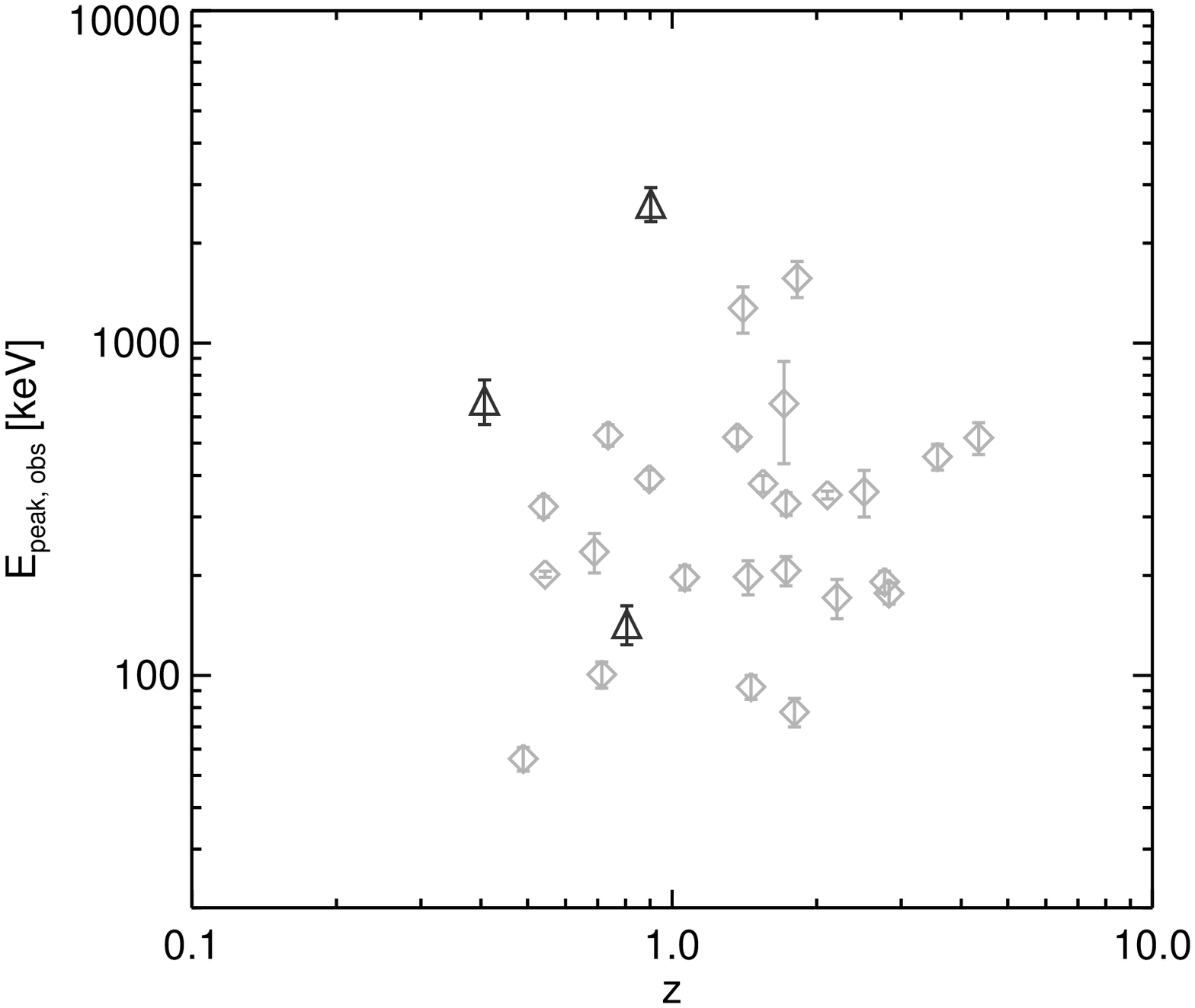}}
		\end{minipage}
	\end{center}
\caption{BEST spectral parameters as a function of the redshift of the \emph{P} spectral fits for short (black triangles) and long (light-gray diamonds) GRBs.  \label{paramsvszp}}
\end{figure}

In Figure~\ref{paramsvszp} we show the evolution of the BEST spectral parameters, $\alpha$, $\beta$ and \epeak, with redshift. None of the parameters shows a dependence with redshift.

\subsection{Comparing the \emph{F} with the \emph{P} spectral fits}

\begin{figure}
	\begin{center}
		\begin{minipage}[t]{1\textwidth}
		\subfigure[]{\label{pffalpha}\includegraphics[scale=0.45]{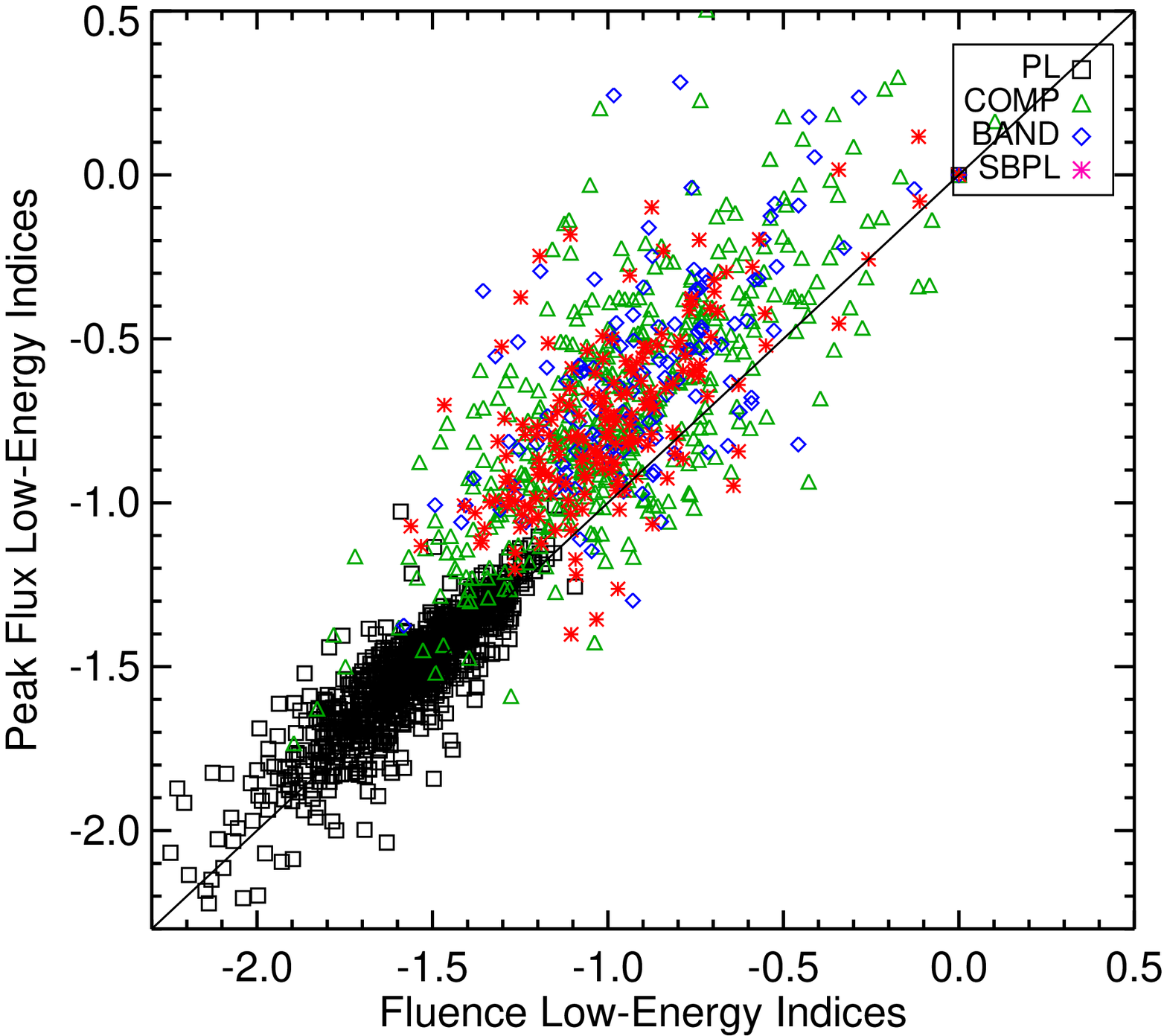}}
		\subfigure[]{\label{pffbeta}\includegraphics[scale=0.45]{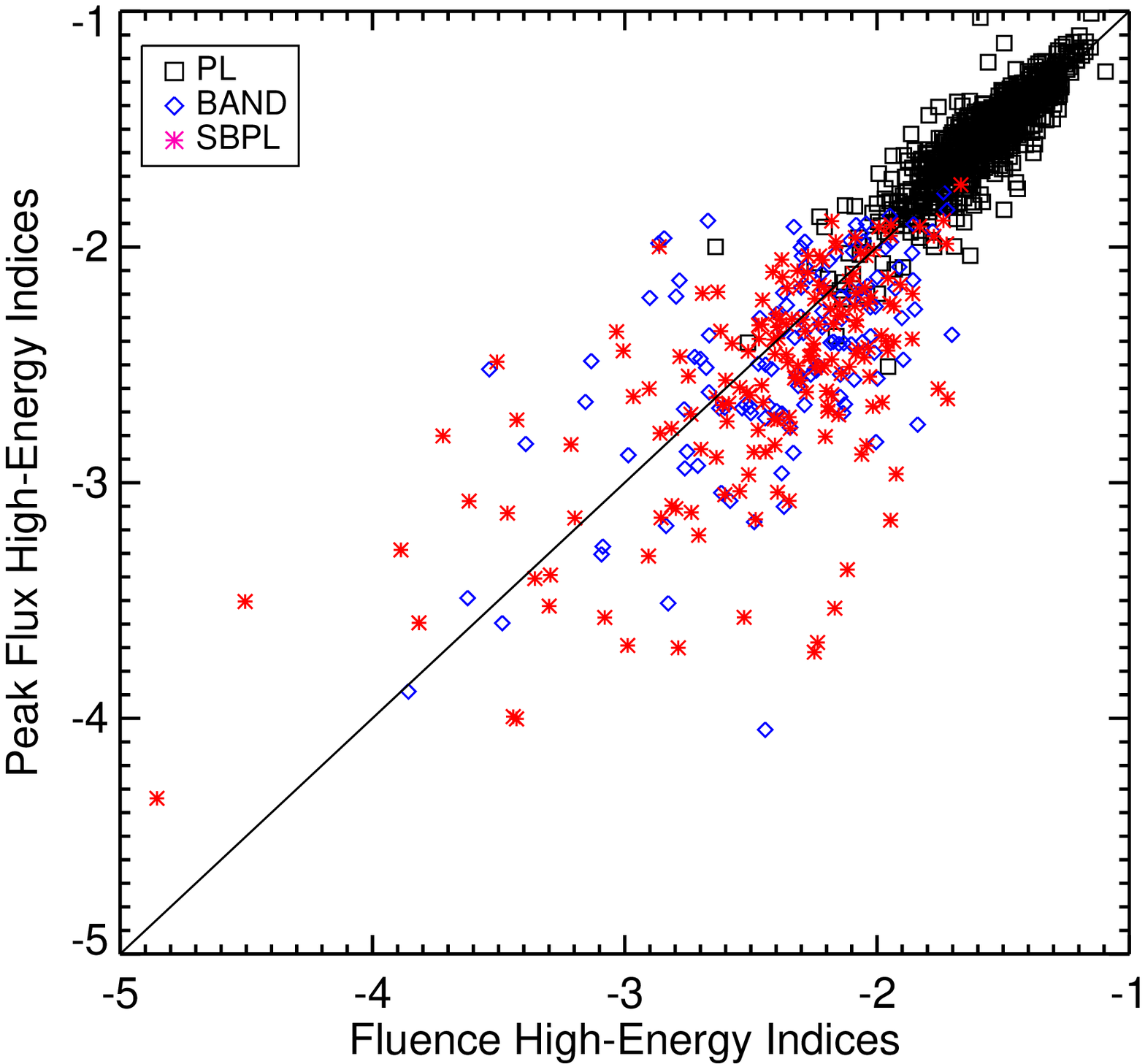}}\\
		\end{minipage}
		\begin{minipage}[t]{1\textwidth}
		\centering
		\subfigure[]{\label{pffepeak}\includegraphics[scale=0.45]{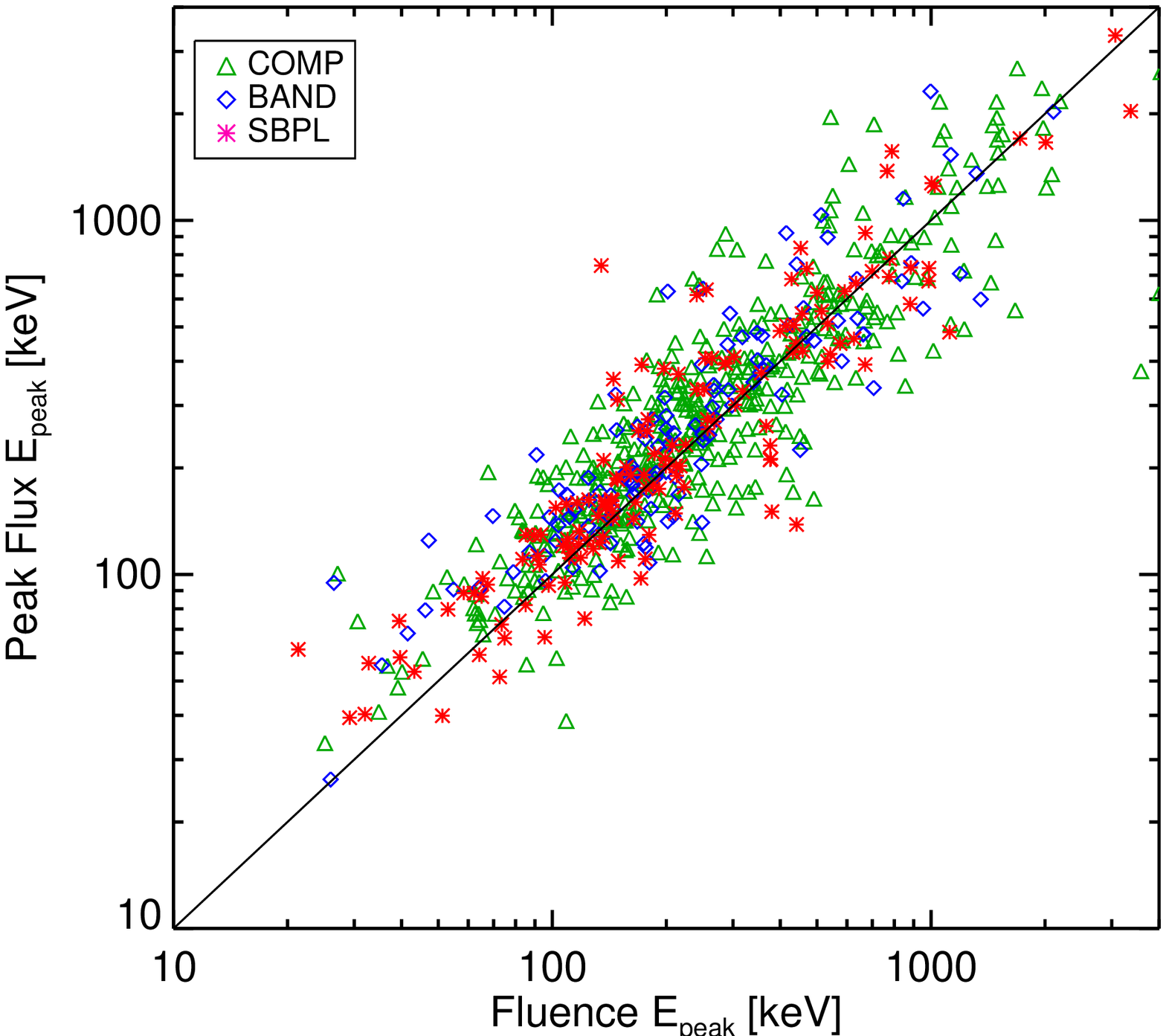}}
		\end{minipage}
	\end{center}
\caption{\emph{P} spectral parameters as a function of the fluence spectral parameters for the GOOD sample. For all three parameters there is 
evidence for a strong correlation between the parameters found for the \emph{F} spectra and those for the \emph{P} spectra.  Note that the \pwrlw index is shown in both \subref{pffalpha} and \subref{pffbeta} for comparison. \label
{pff}}
\end{figure}

When studying the two types of spectra in this catalog, it is instructive to study the similarities and differences between the 
resulting parameters.  Plotted in Figure~\ref{pff} are the low-energy indices, high-energy indices, and \epeak energies of the 
\emph{P} spectra as a function of the corresponding parameters from the \emph{F} spectra.  Most of the \emph{P} spectral 
parameters correlate with the \emph{F} spectral parameters on the order of unity.  There are particular regions in each plot where 
outliers exist, and these areas indicate that either the GRB spectrum is poorly sampled or that significant spectral 
evolution exists in the \emph{F} measurement of the spectrum that skews the time-integrated spectral values.  Examples of the former case are when the low-energy index is $\gtrsim-1.2$ or the high-energy index is steeper than average ($\lesssim-2$).  

It is common belief that \epeak is significantly larger at the peak of the GRB, compared to the average \epeak. Figure~\ref{pffepeak}, and  similar results found in \cite{goldstein12} and \cite{Nava}, seems to contradict this belief with marginal difference between \epeak measured at the peak of the photon flux and \epeak measured over the full duration of the burst. 

However, time-resolved spectral analyses in the past have shown \citep[e.g.][for BATSE and \gbm GRBs, respectively]{Kaneko06, lu12} that two different \epeak-evolution patterns can (co-)exist in a single GRB:
\begin{enumerate}
\item \epeak can evolve from hard-to-soft and/or 
\item \epeak shows a tracking behavior with respect to the photon flux. 
\end{enumerate}

If a GRB evolves from hard-to-soft, the peak of the photon flux does not necessarily correspond to the highest \epeak value \citep[see also][]{crider99}. Only when \epeak tracks the intensity its highest value is indeed expected at the peak photon flux.

In addition, the time-averaged \epeak is also dependent on the ratio of the peak photon flux versus the average photon flux of the GRB. A larger ratio, i.e. a higher photon flux at the peak of the burst, skews the average \epeak toward the \epeak found at peak flux. Therefore, a marginal difference between \epeak of the \emph{F} spectral fits and \emph{P} spectral is expected.

In a forthcoming catalog \citep{yu14}, the results of a time-resolved study of \gbm GRBs will be presented.

\begin{figure}
	\begin{center}
		\begin{minipage}[t]{1\textwidth}
		\subfigure[]{\label{acctimealpha}\includegraphics[scale=0.45]{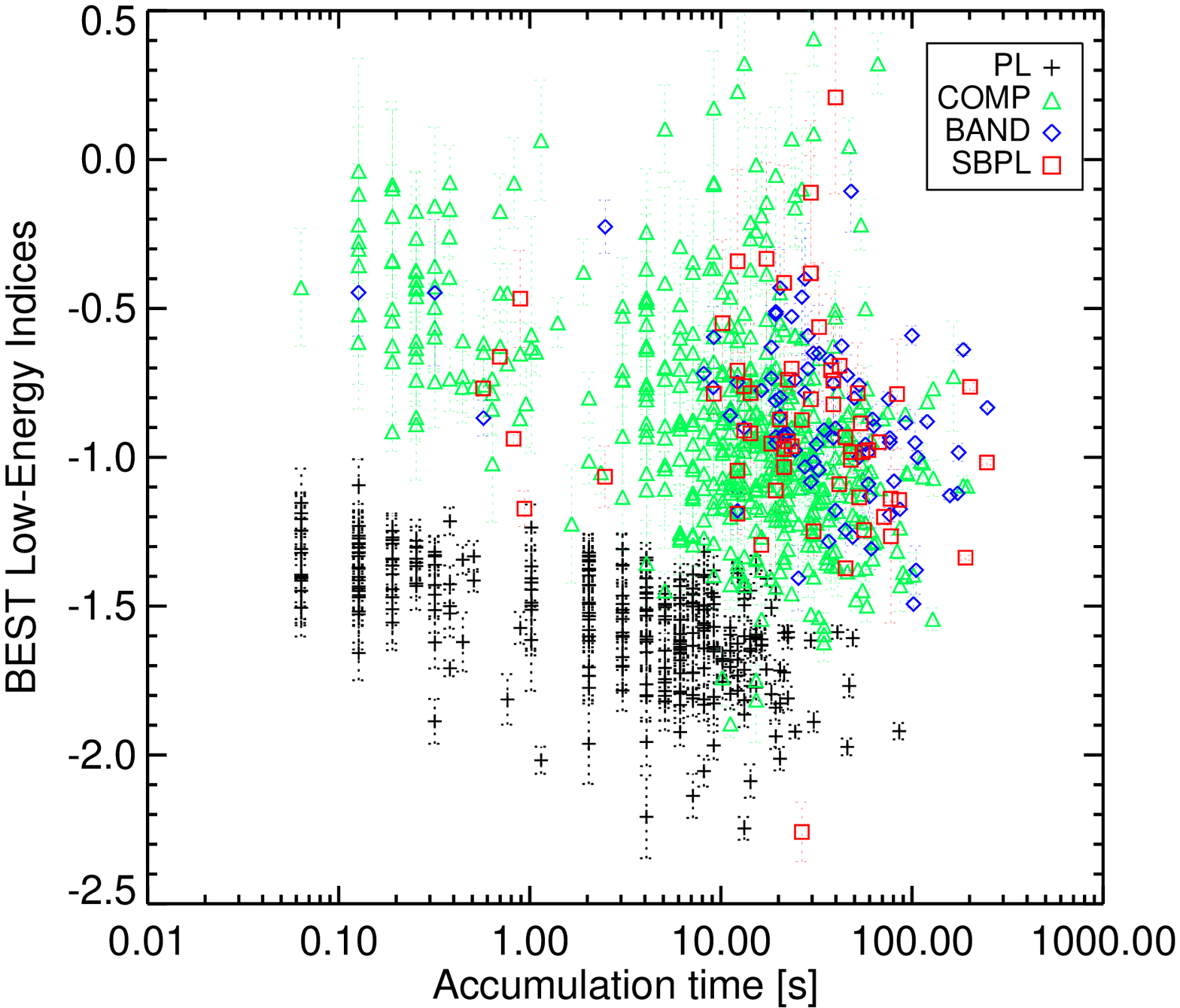}}
		\subfigure[]{\label{acctimebeta}\includegraphics[scale=0.45]{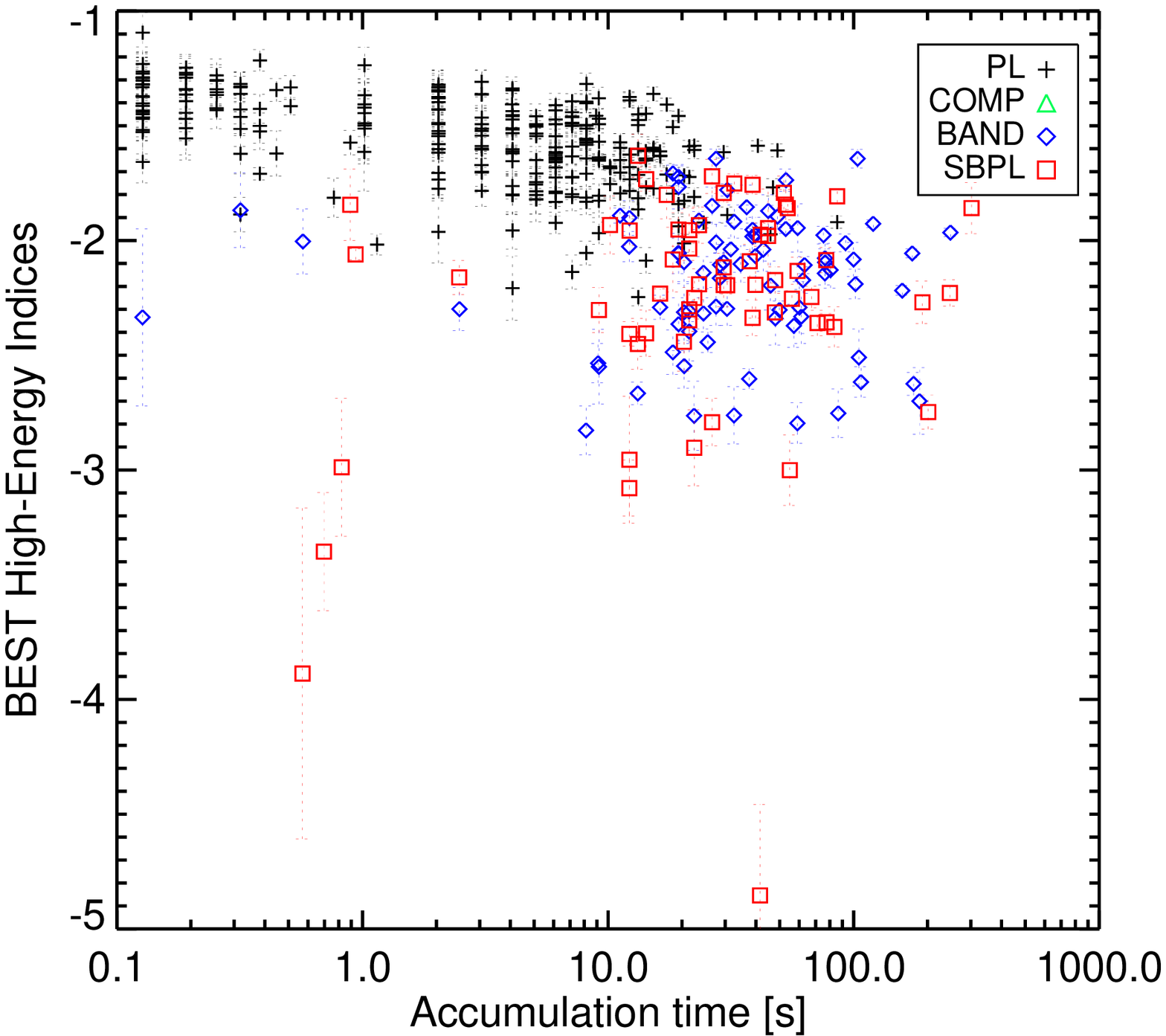}}\\
		\end{minipage}
		\begin{minipage}[t]{1\textwidth}
		\centering
		\subfigure[]{\label{acctimeepeak}\includegraphics[scale=0.45]{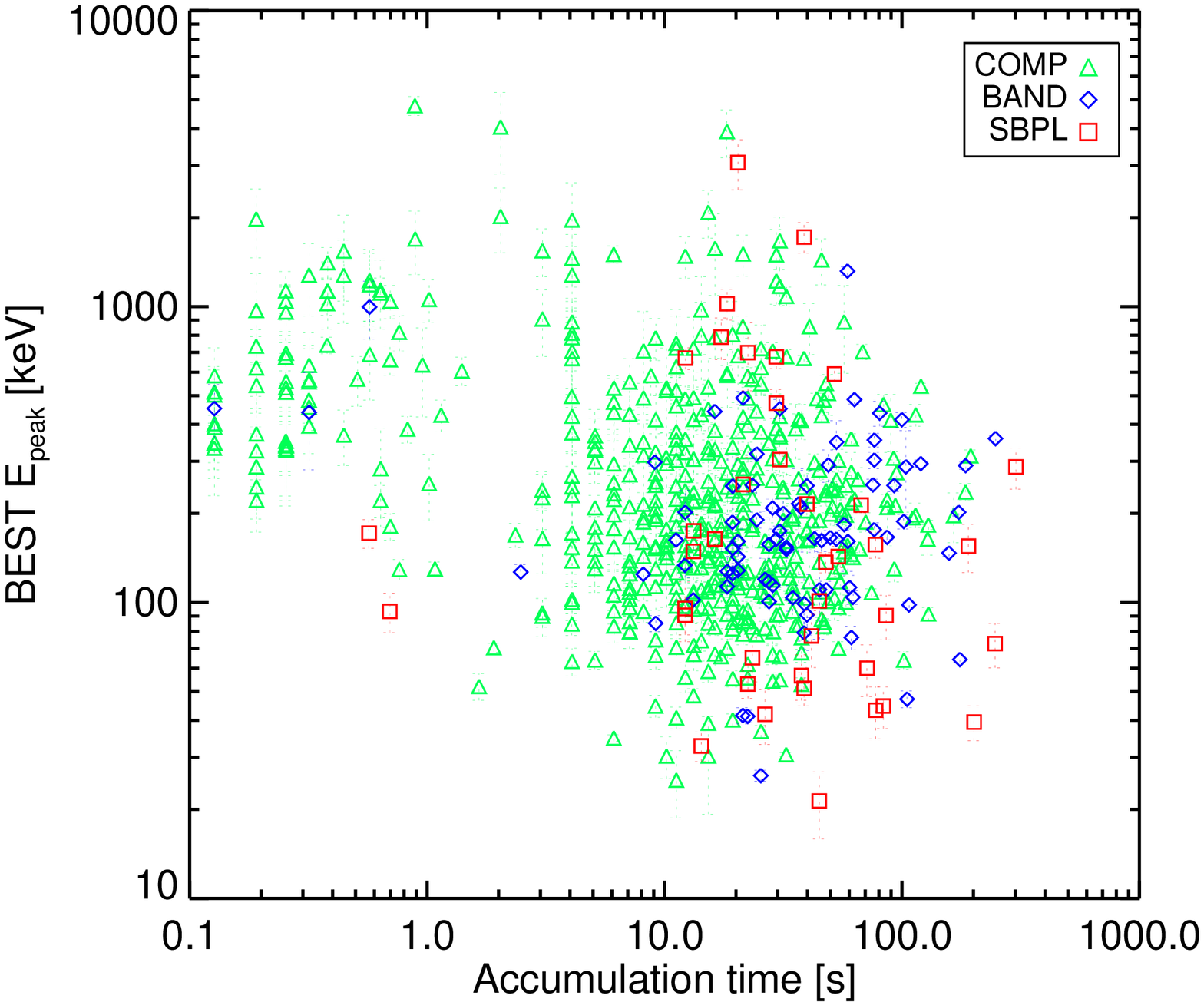}}
		\end{minipage}
	\end{center}
\caption{BEST spectral parameters as a function of the accumulation time for \emph{F} spectral fits.  Note that the \pwrlw index is shown in both 
\subref{acctimealpha} and \subref{acctimebeta} for comparison. \label{acctimeparms}}
\end{figure}

\begin{figure}
	\begin{center}
		\begin{minipage}[t]{1\textwidth}
		\subfigure[]{\label{fluencealpha}\includegraphics[scale=0.45]{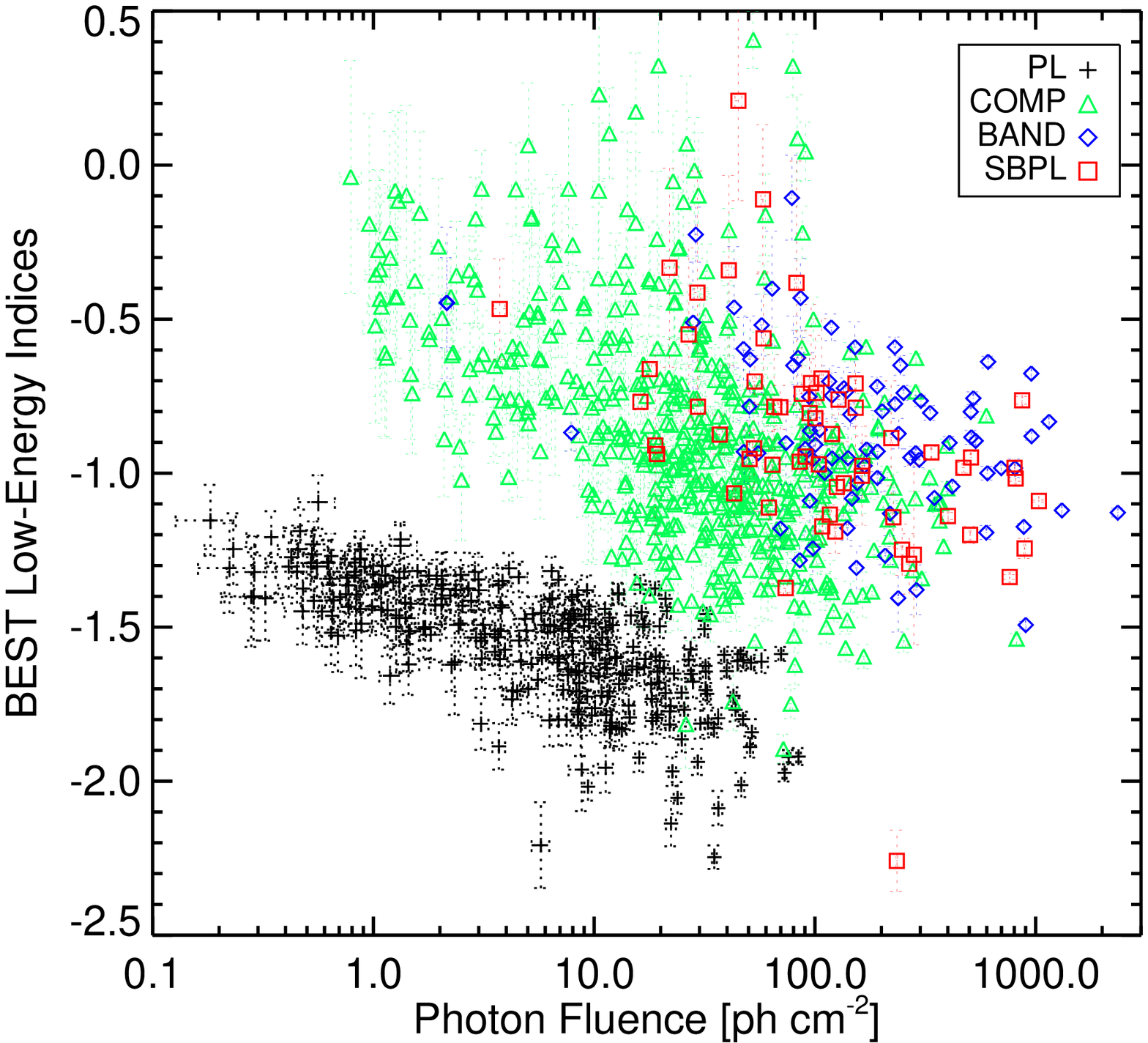}}
		\subfigure[]{\label{fluencebeta}\includegraphics[scale=0.45]{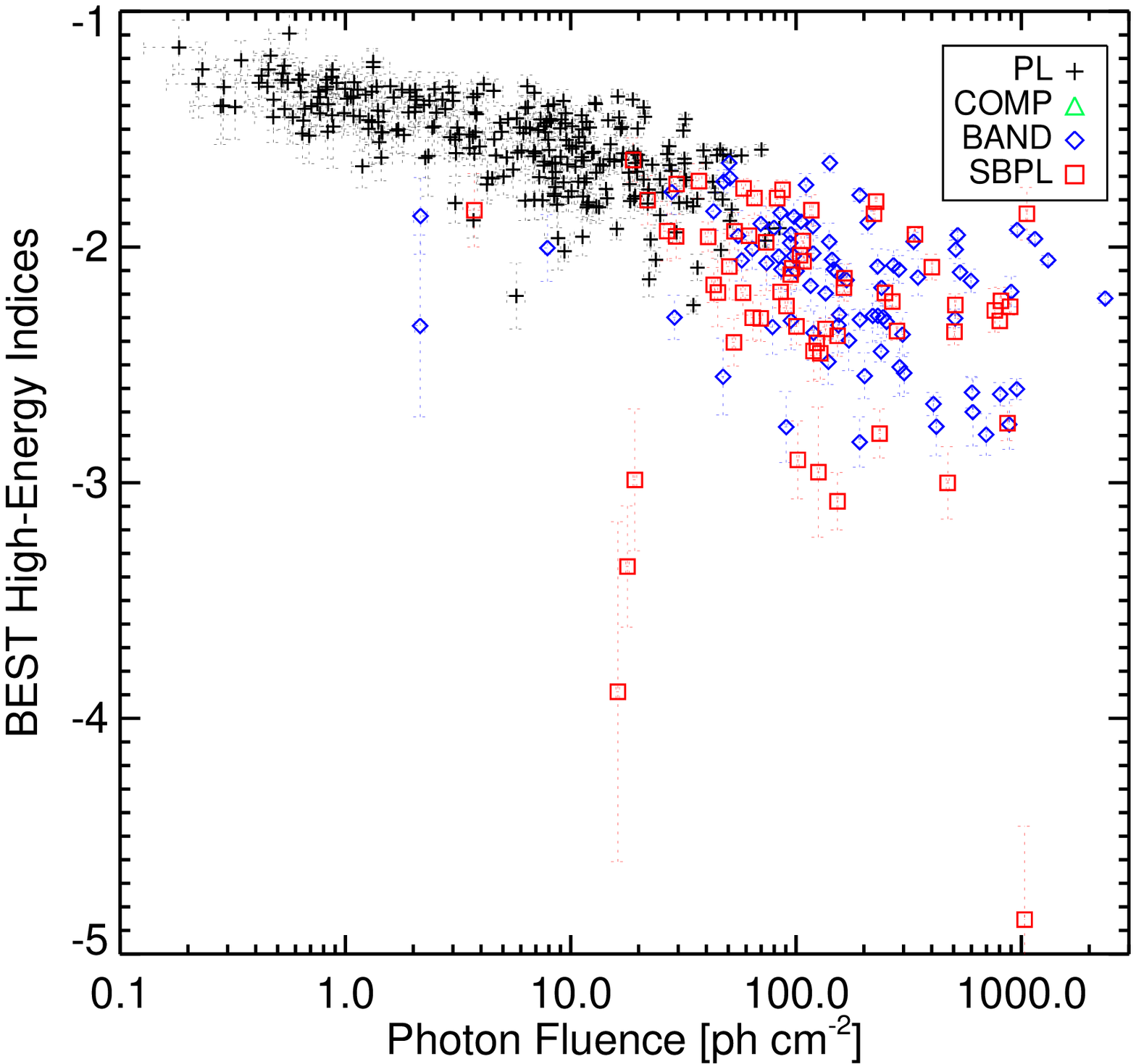}}\\
		\end{minipage}
		\begin{minipage}[t]{1\textwidth}
		\centering
		\subfigure[]{\label{fluenceepeak}\includegraphics[scale=0.45]{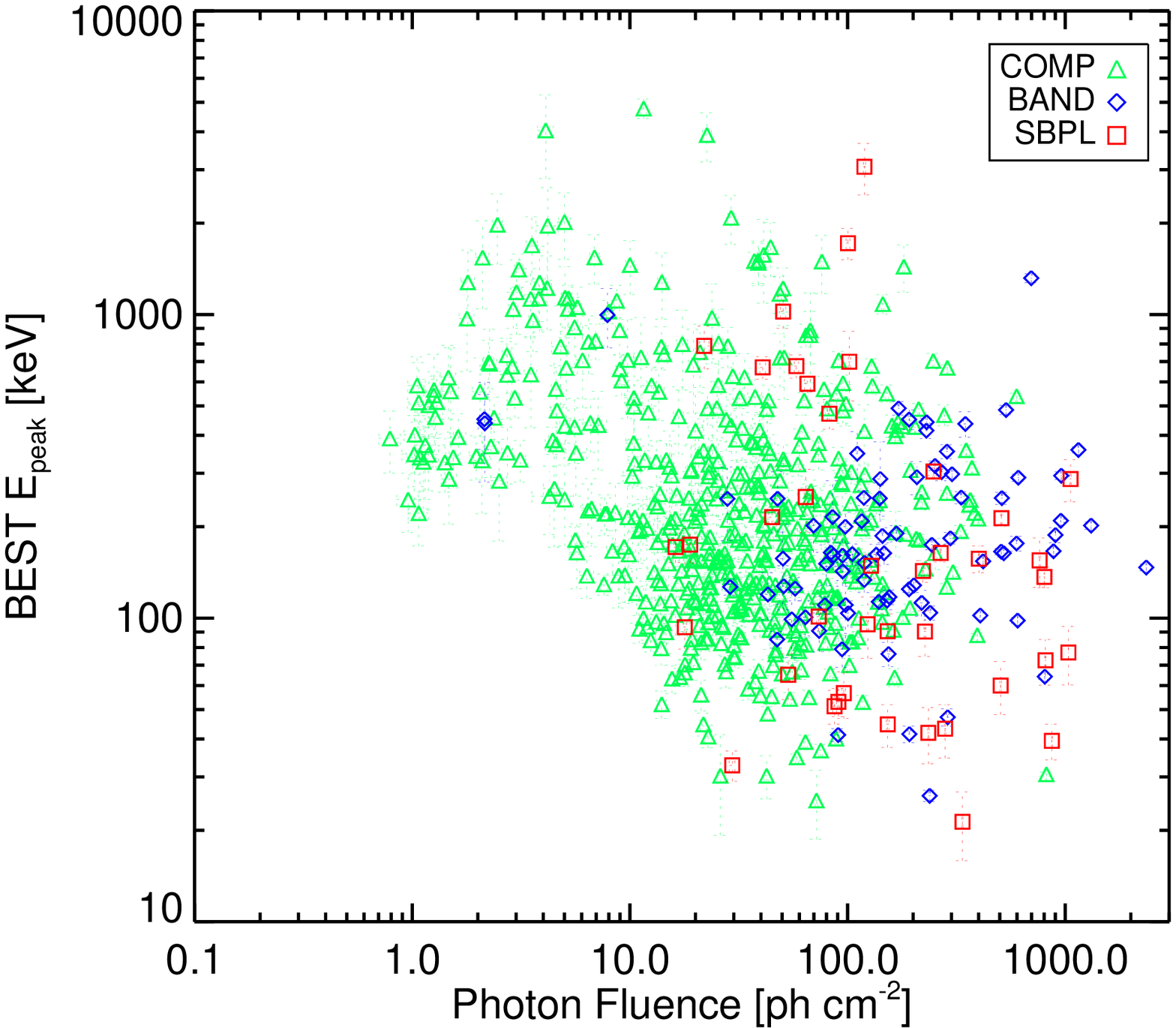}}
		\end{minipage}
	\end{center}
\caption{BEST spectral parameters as a function of the model photon fluence of the  \emph{F} spectral fits.  Note that the \pwrlw index is shown in both 
\subref{fluencealpha} and \subref{fluencebeta} for comparison. \label{fluenceparms}}
\end{figure}

\begin{figure}
	\begin{center}
		\begin{minipage}[t]{1\textwidth}
		\subfigure[]{\label{fluxalpha}\includegraphics[scale=0.45]{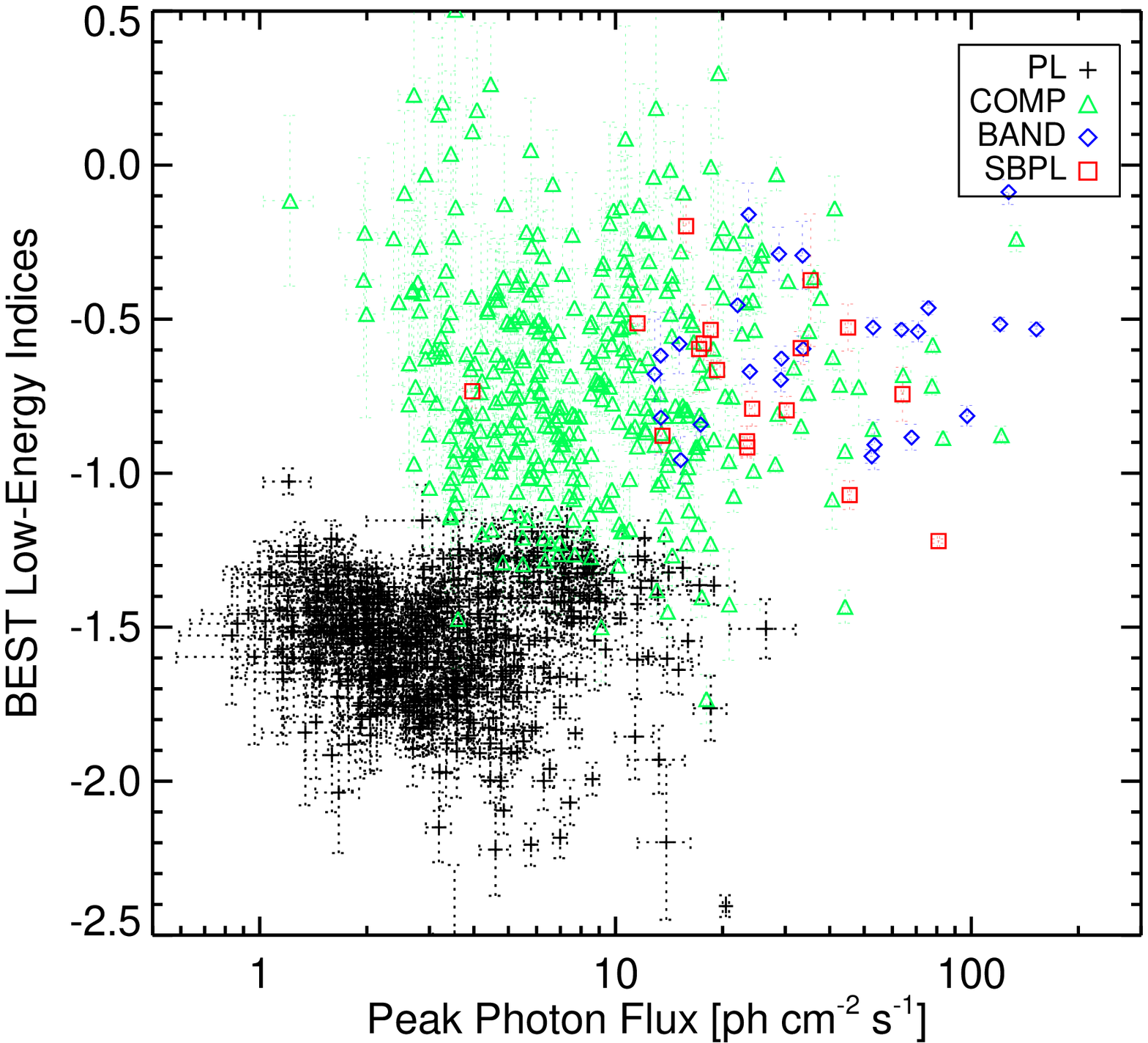}}
		\subfigure[]{\label{fluxbeta}\includegraphics[scale=0.45]{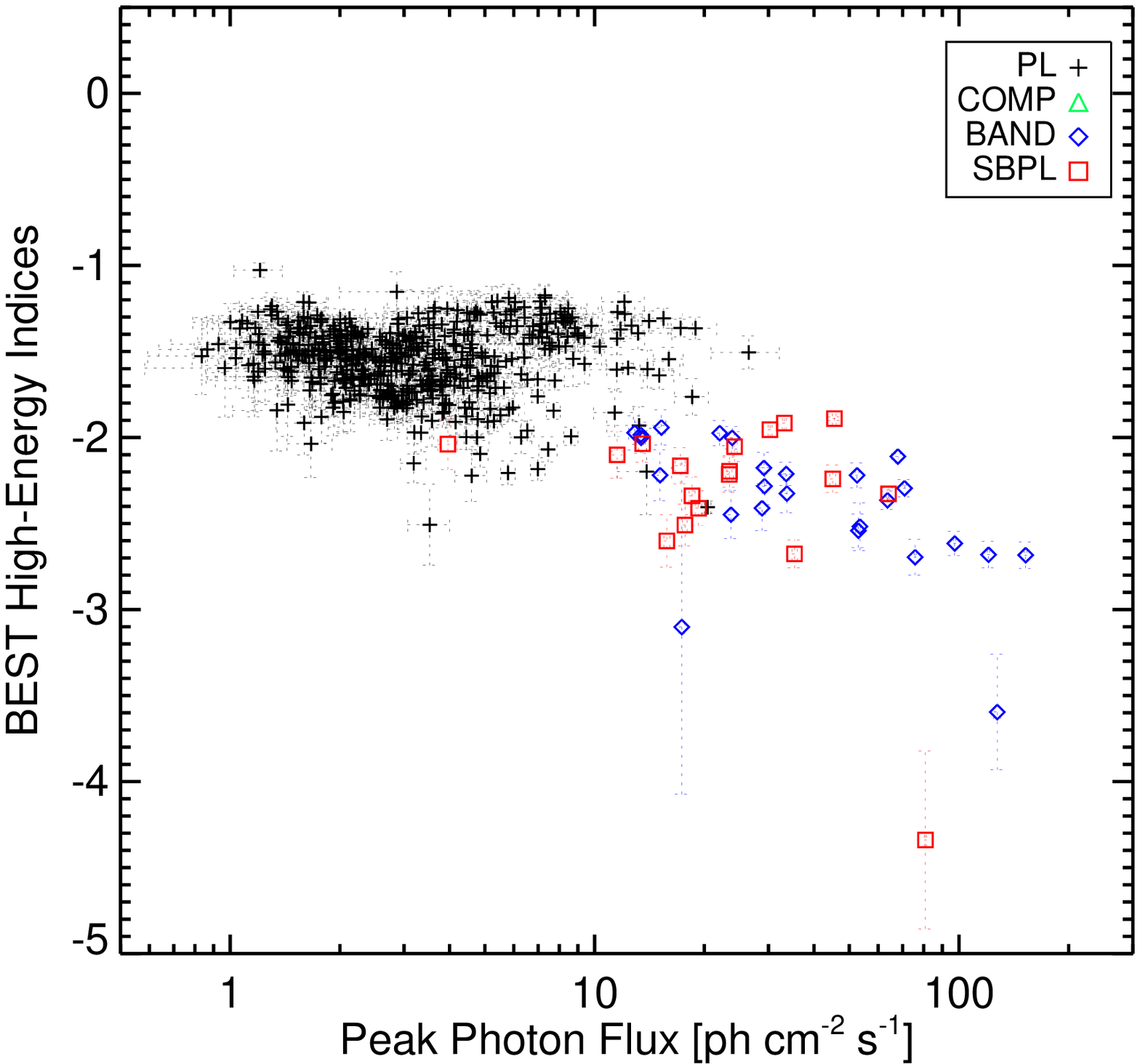}}\\
		\end{minipage}
		\begin{minipage}[t]{1\textwidth}
		\centering
		\subfigure[]{\label{fluxepeak}\includegraphics[scale=0.45]{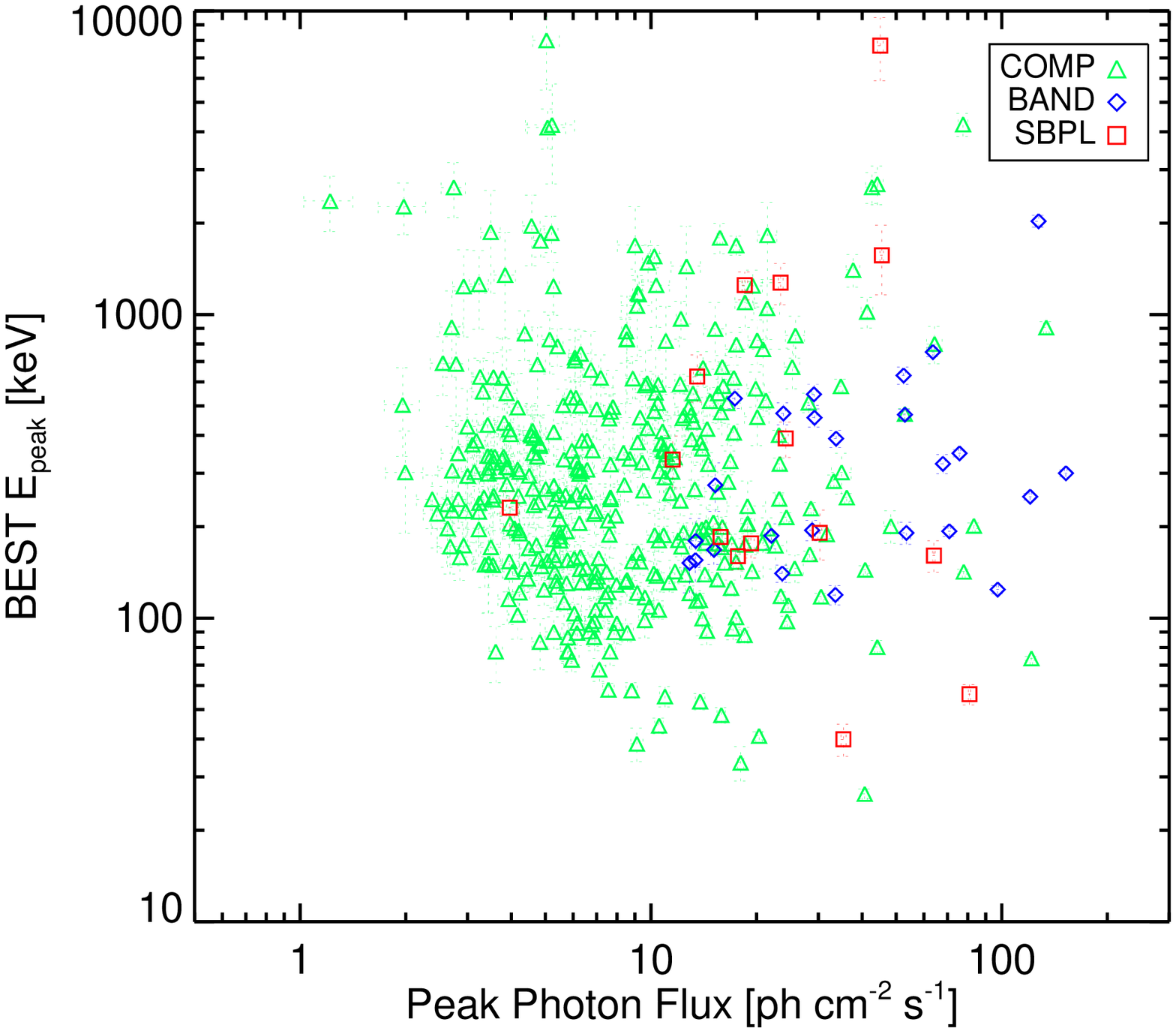}}
		\end{minipage}
	\end{center}
\caption{BEST spectral parameters as a function of the model peak photon flux of the \emph{P} spectral fits.  Note that the \pwrlw index is shown in both \subref
{fluxalpha} and \subref{fluxbeta} for comparison. \label{fluxparms}}
\end{figure}

To aid in the study of the systematics of the parameter estimation, as well as the effect of statistics on the fitting 
process, we investigate the behavior parameter values as a function of accumulation time and the photon fluence for the \emph{F} BEST spectra and peak photon flux for the \emph{P} BEST spectra, respectively.  These distributions are shown in Figures \ref{acctimeparms}, \ref{fluenceparms} and \ref{fluxparms}, respectively.  
As the photon fluence is correlated with the duration of a burst (see again Figure~\ref{timepfluence}) any correlation of a spectral parameter with the accumulation time will also be correlated with photon fluence.  

When fitting the time-integrated spectrum of a burst, we find the low- and high-energy indices trend towards steeper values for bursts with longer accumulation times. The simple \pwrlw index trends from shallow values of $\sim-1.3$ to a steeper value of $\sim-2$.  
The low-energy index for a spectrum with curvature tends to exhibit an unusually shallow value of $\sim-0.4$ for 
short GRBs or extremely low fluence spectra, and steepens to $\sim-1.5$ for longer GRBs or higher fluence spectra.  Similarly, the high-energy index trends from $\sim-1.2$ at short durations or low fluence to $\sim-2.7$ at long durations or high fluence, although this is complicated by unusually steep and poorly constrained indices that indicate that an exponential cutoff may result in a more reliable spectral fit.  When inspecting the \epeak as a function of accumulation time the resulting plot is reminiscent of the hardness/duration correlation with two distinct regions for the long and short GRBs. As for the correlation between \epeak and photon fluence a trend is much less apparent. If a burst is assumed to have significant spectral evolution, then obviously the \epeak will change values through the time history of the burst, typically following the traditional hard-to-soft energy evolution.  For this reason, spectra that integrate over increasingly more time will tend to suppress the highest energy of \epeak within the burst, so a general decrease in \epeak is expected with longer integration times.  However, the photon 
fluence convolves the integration time with the photon flux. This means that an intense burst with a short duration and high \epeak may have on the order 
the same fluence as a much longer but less intense burst with a smaller \epeak.  This causes significant broadening 
to the decreasing trend as shown in Figure \ref{fluenceepeak}.  

The distribution of parameters as a function of the peak photon 
flux, however, is much less clear.  The distributions shown in Figure \ref{fluxparms} are more susceptible to uncertainty because 
of smaller statistics involved in the study of the \emph{P} section of the GRB, except in some cases where the peak 
photon flux is on the order of the photon fluence.  Ignoring the regions where the parameters are poorly constrained, another 
trend emerges from the low-energy indices; they appear to become slightly more shallow as the photon flux increases.  The 
high-energy indices, however, appear to be unaffected by the photon flux, although a trend of steeper indices with higher photon fluxes seems apparent.  Finally, no obvious trend is visible for \epeak as a function of the peak photon flux (Spearman rank correlation coefficient of -0.05 with a chance probability of $P=0.9$). However, when inspecting \epeak as a function of peak energy flux a correlation becomes apparent (Spearman rank correlation coefficient of of $\sim 1$ with $P=8\times10^{-08}$). The lack and existence of a correlation between \epeak and the photon and energy flux, respectively is expected  because it is the energy of collected photons and not their number that drives the hardness (and thus \epeak) of a GRB.  

The spectral results, including the best fit spectral parameters and the photon model, are stored in files following the FITS standard and will be hosted as a public data archive on HEASARC\footnote{\url{http://heasarc.gsfc.nasa.gov/W3Browse/fermi/fermigbrst.html}}. The values returned by default in a Browse search of the catalog are the most recent values, which will be the results presented here upon acceptance of this paper. However, the values and results obtained by \citet{goldstein12} will still be available by ftp. The keyword \textit{scatalog} indicates which catalog a file belongs to. Both the differences between the methodologies and how to retrieve the older files will be fully documented at HEASARC.

\section{Summary \& Discussion}\label{summary}
After four successful years, GBM continues to be a prime instrument for GRB observations. The second \gbm spectral catalog includes 943 GRBs detected by \gbm during 4 years of operation. We presented the spectral properties of these bursts both from a time--integrated, and a peak flux analysis. Four photon model fits were applied to each selection, resulting in more than 7500 spectral fits.  We have described subsets of the full results in the form of data cuts based on parameter uncertainty, as well as employing model comparison techniques to select the most statistically preferred model for each GRB.  The analysis of each GRB was performed as objectively as possible, in an attempt to minimize biased systematic errors inherent in subjective analysis.  The methods we have described treat all bursts equally, and we have presented the ensemble of observed spectral properties of GBM GRBs. 

Although most of the resulting parameter distributions are similar to the ones found by other GRB missions, some contain important differences. For example, thanks to the two BGO detectors, burst data extend into the high--energy domain which, in the past, has been accessible only by the \emph{Solar Maximum Mission} Gamma Ray Spectrometer \citep{harris98} and the PHEBUS instrument onboard the Granat spacecraft \citep{barat06}. While there does not seem to be a new population of bursts with high \epeak values in their time-integrated spectra, 40 GRBs of the BEST fluence spectra and 37 of the BEST \emph{P} spectrum have \epeak values above 1~MeV. In order to verify whether the lack of GRBs with high \epeak values in their time-integrated spectra is intrinsic or an instrumental bias (decreasing effective are in the BGOs) requires  simulations which go beyond the scope of this catalog. We emphasize that there are indeed GRBs with very high \epeak values in their time-resolved spectra, such as e.g. GRB~110721A \citep{axelsson12} which has the highest observer-frame \epeak value ($15\pm2$~MeV) known to date. In a forthcoming catalog of time-resolved spectra of \gbm GRBs \citep{yu14} this will be investigated in more detail.

 Another interesting result of the parameter distributions is the $\Delta S$ parameter, the difference between the low- and high-energy spectral indices.  This is an important quantity because current models for the GRB prompt emission mechanism can be broken into two categories: magnetic \citep[e.g.][]{Lee} or driven by internal/external shocks \citep[e.g.][]{Rees}.  Thus, by comparing $\Delta S$ to the predictions by the synchrotron shock model, it is possible to obtain useful insights into the emission mechanisms of GRBs. As our results for the $\Delta S$ distribution are compatible and in line with the results obtained by \citet{Preece} and \citet{Kaneko06} it can concluded that the predictions of the SSM model, in its simplest form, are not reconcilable neither with the observations made by BATSE nor those performed by \gbm.
   
Thanks to the rapid localizations of the \emph{Swift} satellite, necessary to identify a redshift for any given burst, and the broad coverage of the \gbm we could study many spectral properties in the rest-frame of the GRB. This results in the largest \eprest, \eiso and \Liso sample to date which have been analyzed coherently and consistently with an unified methodology, thus minimizing systematic effects. This may be helpful in assessing whether or not well known correlations in the GRB literature are indeed real or due to statistical fluctuations and systematical issues.

 The differences between various GBM spectral catalogs \citep[e.g.][]{Nava} were extensively discussed in \citet{goldstein12}. Small discrepancies between \citet{goldstein12}'s first GBM spectral catalog and the one presented here (e.g. different BEST models for the same GRB, fraction of bursts best fit with a given model) are mainly due to enhanced and modified criteria in determining the BEST and GOOD sample, updated analysis software and response matrices and the usage of different statistical criteria, i.e. C-Stat instead of $\chi^2$ to determine the preferred model. These differences are representative of divergent methodologies and samples; therefore the reader should take care to understand how the values in the catalogs are derived. Certainly, there are avenues of investigation that require more detailed work and analysis or perhaps a different methodology.  This catalog should be treated as a starting point for future research on interesting bursts and ideas.  As has been the case in previous GRB spectral catalogs, we hope this catalog will be of great assistance and importance to the search for the physical properties of GRBs and other related studies.

\section{Acknowledgments}
We thank the reviewer for his/her comments, which significantly contributed to improving the quality of the publication. 
The GBM project is supported by the German Bundesministeriums 
f\"ur Wirtschaft und Technologie (BMWi)  via the Deutsches Zentrum f\"ur LuftÐ und Raumfahrt (DLR) under the contract 
numbers 50~QV~0301 and 50~OG~0502.


\clearpage




\clearpage

\clearpage

\clearpage


\clearpage

\begin{figure}
	\begin{center}
		\subfigure[]{\label{fluxepeak}\includegraphics[scale=0.45]{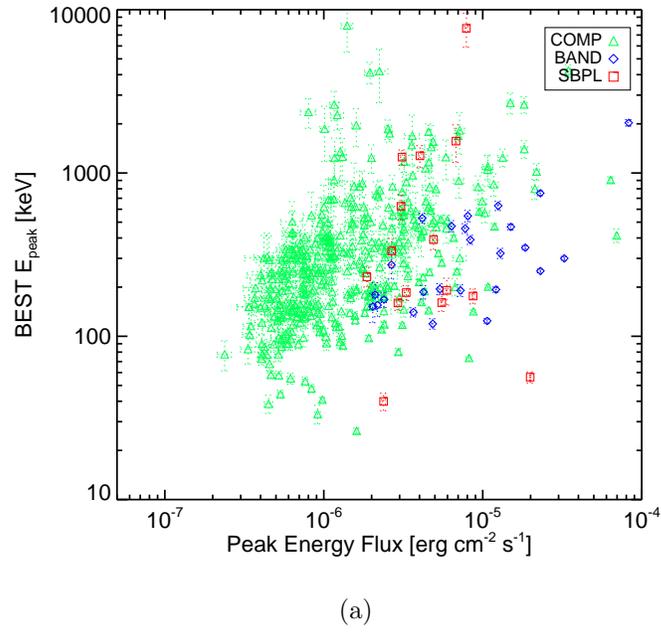}}
	\end{center}
\caption{BEST \epeak as a function of the model energy flux of the \emph{P} spectral fits.
\label{parmsvsenflux}}
\end{figure}

\clearpage

\clearpage

{\renewcommand{\arraystretch}{1.2}
\begin{deluxetable}{ c  c  c  c }
\tablecolumns{4}
\tablewidth{0pt}
\tabletypesize{\scriptsize}
\tablecaption{Determination of $\Delta$C-Stat$_{\rm crit}$ \label{CstatFluence}}
\startdata
	\hline
	\bf Burst & \bf Fluence & \bf $\Delta$C-Stat$_{\rm crit}$ & \bf $\Delta$C-Stat$_{\rm crit}$ \\ 
	 & \bf  [erg] & \bf \pwrlw vs. \comp & \bf \comp vs. \band \\ \hline
	
	\multicolumn{4}{ c }{} \\  
	
	GRB~120608.489 & 5.2e-07 & 8.55 & 13.35 \\ 
	GRB~110227.240 & 1.9e-06 & 7.75 & 9.56 \\ 
	GRB~120129.580 & 5.8e-05 & 8.25 & 10.55 \\
	GRB~120526.303  &1.3E-04 & 9.75 & 13.85 \\	\hline
	\multicolumn{4}{ c }{} \\  
	average &$-$&8.58&11.83\\ 
\enddata
\end{deluxetable}

\begin{deluxetable}{c  c  c  c  c }
\tablecolumns{5}
\tablewidth{0pt}
\tabletypesize{\scriptsize}
\tablecaption{BEST and GOOD GRB models.\label{BestTable}}
\startdata
	\hline
 & \bf \pwrlw & \bf \sbpl & \bf \band & \bf \comp \\ \hline
 
	\multicolumn{5}{ c }{\bf Fluence Spectra} \\ \hline 
BEST & 282 (29.9\%) & 62 (6.6\%) & 81 (8.6\%) & 516 (54.7\%) \\  
GOOD & 941 (99.7\%) & 392 (41.5\%) & 342 (36.2\%) & 684 (72.5\%) \\ \hline 
 
\multicolumn{4}{ c }{\bf Peak Flux Spectra} \\ \hline
BEST & 514 (54.4\%) & 18 (1.9\%) & 25 (2.6\%) & 375 (39.7\%) \\  
GOOD & 932 (98.7\%) & 196 (20.8\%) & 153 (16.2\%) & 430 (45.6\%) \\ \hline 
 
\enddata
\end{deluxetable}

\begin{deluxetable}{ c  c  c  c  c  c  c  c}
\tablecolumns{7}
\tablewidth{0pt}
\tabletypesize{\scriptsize}
\tablecaption{The Median Parameter Values and the 68\% C.L. of the Distributions of the GOOD Sample \label{ParamTable}}
\startdata
	\hline
	\multirow{2}{*}{\bf Model} & \bf Low-Energy & \bf High-Energy & $\bf E_{peak}$ & $\bf
	 E_{break}$ & \bf Photon Flux & \bf Energy Flux & \bf $\Delta S$\\

 	& \bf Index & \bf Index & \bf (keV) & \bf (keV) & \bf (ph $\bf s^{-1} \ cm^{-2}$) & \bf ($\bf 10^{-7} \ erg \ s^{-1} \ cm^{-2}$) & \\ \hline

	\multicolumn{7}{ c }{\bf Fluence Spectra} \\ \hline 

		\pwrlw & $-1.54_{-0.20}^{+0.17}$ & - & - & - & $2.52_{-1.11}^{+3.91}$ & $3.41_{-1.48}^{+7.11}$ & -  \\
		\comp & $-0.96_{-0.30}^{+0.37}$ & - & $211_{-109}^{+333}$ & - & $2.53_{-1.10}^{+4.22}$ & $3.19_{-1.66}^{+8.55}$ & - \\
 	\sbpl & $-0.98_{-0.23}^{+0.28}$ & $-2.35_{-0.52}^{+0.35}$ & $166_{-82}^{+293}$ & $112_{-55}^{+151}$ & $3.29_{-1.54}^{+4.71}$ & $4.54_{-2.40}^{+9.51}$ & $1.38_{-0.37}^{+0.63}$ \\
 	\band & $-0.86_{-0.25}^{+0.33}$ & $-2.29_{-0.39}^{+0.30}$ & $174_{-73}^{+286}$ & $118_{-41}^{+168}$ & $3.16_{-1.55}^{+4.85}$ & $4.64_{-2.50}^{+7.96}$ & $1.43_{-0.39}^{+0.53}$ \\
 	BEST & $-1.08_{-0.44}^{+0.43}$ & $-2.14_{-0.37}^{+0.27}$ & $196_{-100}^{+336}$ & $103_{-63}^{+129}$ & $2.38_{-1.05}^{+3.68}$ & $3.03_{-1.40}^{+7.41}$ & $1.26_{-0.32}^{+0.50}$\\
	\hline
		\multicolumn{7}{ c }{\bf Peak Flux Spectra} \\ \hline 
		\pwrlw & $-1.50_{-0.20}^{+0.16}$ & - & - & - & $4.77_{-2.69}^{+9.55}$ & $7.29_{-4.25}^{+16.25}$ & -\\
		\comp & $-0.75_{-0.28}^{+0.38}$ & - & $270_{-138}^{+370}$ & - & $8.49_{-4.57}^{+12.79}$ & $13.19_{-7.39}^{+32.57}$ & -\\
 	\sbpl & $-0.79_{-0.23}^{+0.27}$ & $-2.49_{-0.64}^{+0.33}$ & $202_{-105}^{+422}$ & $141_{-63}^{+239}$ & $14.46_{-7.74}^{+18.70}$ & $24.44_{-13.77}^{+48.07}$ & $1.76_{-0.46}^{+0.68}$\\
 	\band & $-0.64_{-0.28}^{+0.32}$ & $-2.41_{-0.46}^{+0.33}$ & $239_{-116}^{+327}$ & $153_{-54}^{+240}$ & $14.62_{-7.81}^{+18.40}$ & $24.69_{-13.02}^{+45.58}$ & $1.78_{-0.47}^{+0.57}$\\
 	BEST & $-1.32_{-0.33}^{+0.74}$ & $-2.24_{-0.38}^{+0.26}$ & $261_{-130}^{+364}$ & $133_{-39}^{+349}$ & $4.57_{-2.49}^{+8.82}$ & $6.49_{-3.46}^{+17.52}$ & $1.64_{-0.36}^{+0.59}$\\
	\enddata
\end{deluxetable}

\begin{deluxetable}{ c  c  c  c  c  c  c }
\tablecolumns{7}
\tablewidth{0pt}
\tabletypesize{\scriptsize}
\tablecaption{The Median Parameter Values and the 68\% C.L. of the BEST Model Fits \label{BestComparison}}
\startdata
	\hline
	\multirow{2}{*}{\bf Dataset} & \bf Low-Energy & \bf High-Energy & $\bf E_{peak}$ & $\bf
	 E_{break}$ & \bf Photon Flux & \bf Energy Flux \\
 	& \bf Index & \bf Index & \bf (keV) & \bf (keV) & \bf (ph $\bf s^{-1} \ cm^{-2}$) & \bf ($\bf 10^{-7} \ erg \ s^{-1} \ cm^{-2}$) \\ 
	\hline

	\multicolumn{7}{ c }{\bf Fluence} \\ \hline
	This Catalog BEST& $-1.08_{-0.44}^{+0.43}$ & $-2.14_{-0.37}^{+0.27}$ & $196_{-100}^{+336}$ & $103_{-63}^{+129}$ & $2.38_{-1.05}^{+3.68}$ & $3.03_{-1.40}^{+7.41}$ \\

	Goldstein et al. (2012) & $-1.05^{+0.44}_{-0.45}$ & $-2.25^{+0.34}_{-0.73}$ & $205^{+359}_{-121}$ & $123^{+240}_{-80.4}$ & $2.92^{+3.96}_{-1.31}$ & $4.03^{+9.38}_{-2.13}$\\

	\citet{Nava} & $-0.99^{+0.48}_{-0.49}$ & $-2.33^{+0.17}_{-0.38}$ & $174^{+320}_{-100}$ & $-$ & $-$ & $2.26^{+9.48}_{-1.34}$\\ 

	\citet{Kaneko06} & $-1.14^{+0.20}_{-0.22}$ & $-2.33^{+0.24}_{-0.26}$ & $251^{+122}_{-68}$ & $204^{+76}_{-56}$ & $-$ & $-$\\ \hline
	\multicolumn{7}{ c }{\bf Peak Flux Spectra} \\ \hline 
	This Catalog BEST & $-1.32_{-0.33}^{+0.74}$ & $-2.24_{-0.38}^{+0.26}$ & $261_{-130}^{+364}$ & $133_{-39}^{+349}$ & $4.57_{-2.49}^{+8.82}$ & $6.49_{-3.46}^{+17.52}$ \\

	Goldstein et al. (2012) & $-1.12^{+0.61}_{-0.50}$ & $-2.27^{+0.44}_{-0.50}$ & $223^{+352}_{-126}$ & $172^{+254}_{-100}$ & $5.39^{+10.18}_{-2.87}$ & $8.35^{+22.61}_{-4.98}$ \\

	\citet{Nava} & {$\left(-0.56^{+0.40}_{-0.37}\right)$\tablenotemark{a}}& $-2.39^{+0.23}_{-0.62}$ & $225^{+391}_{-122}$ & $-$ & $-$ & $13.5^{+79.8}_{-10.1}$\\

	\citet{Kaneko06} & $-1.02^{+0.26}_{-0.28}$ & $-2.33^{+0.26}_{-0.31}$ & $281^{+139}_{-99}$ & $205^{+72}_{-55}$ & 
	$-$ & $-$\\ \hline

	\hline 

\enddata

\tablenotetext{a}{$\textnormal{L}$ow-energy index of the peak flux spectra with curved function only.}
\end{deluxetable}


\begin{thebibliography}{0}%
\makeatletter
\providecommand \@ifxundefined [1]{%
 \@ifx{#1\undefined}
}%
\providecommand \@ifnum [1]{%
 \ifnum #1\expandafter \@firstoftwo
 \else \expandafter \@secondoftwo
 \fi
}%
\providecommand \@ifx [1]{%
 \ifx #1\expandafter \@firstoftwo
 \else \expandafter \@secondoftwo
 \fi
}%
\providecommand \natexlab [1]{#1}%
\providecommand \enquote  [1]{``#1''}%
\providecommand \bibnamefont  [1]{#1}%
\providecommand \bibfnamefont [1]{#1}%
\providecommand \citenamefont [1]{#1}%
\providecommand \href@noop [0]{\@secondoftwo}%
\providecommand \href [0]{\begingroup \@sanitize@url \@href}%
\providecommand \@href[1]{\@@startlink{#1}\@@href}%
\providecommand \@@href[1]{\endgroup#1\@@endlink}%
\providecommand \@sanitize@url [0]{\catcode `\\12\catcode `\$12\catcode
  `\&12\catcode `\#12\catcode `\^12\catcode `\_12\catcode `\%12\relax}%
\providecommand \@@startlink[1]{}%
\providecommand \@@endlink[0]{}%
\providecommand \url  [0]{\begingroup\@sanitize@url \@url }%
\providecommand \@url [1]{\endgroup\@href {#1}{\urlprefix }}%
\providecommand \urlprefix  [0]{URL }%
\providecommand \Eprint [0]{\href }%
\providecommand \doibase [0]{http://dx.doi.org/}%
\providecommand \selectlanguage [0]{\@gobble}%
\providecommand \bibinfo  [0]{\@secondoftwo}%
\providecommand \bibfield  [0]{\@secondoftwo}%
\providecommand \translation [1]{[#1]}%
\providecommand \BibitemOpen [0]{}%
\providecommand \bibitemStop [0]{}%
\providecommand \bibitemNoStop [0]{.\EOS\space}%
\providecommand \EOS [0]{\spacefactor3000\relax}%
\providecommand \BibitemShut  [1]{\csname bibitem#1\endcsname}%
\let\auto@bib@innerbib\@empty
\end{thebibliography}%


\begin{thebibliography}{}
\bibitem[Ade~et~al.(2013)]{cosmo} Ade, P. A. R., Aghanim, N., Armitage-Caplan, C. et al. 2013,  arXiv: 1303.50761
\bibitem[Abdo~et~al.(2009)]{abdo09} Abdo, A~A., Ackermann, M., Ajello, M. et al. 2009,  \apj, 706, 138
\bibitem[Ackermann~et~al.(2010)]{ackermann10} Ackermann, M., Asano., K., Atwood, W. B., et al. 2010,  \apj, 716, 1178
\bibitem[Ackermann~et~al.(2011)]{ackermann11} Ackermann, M., Ajello, M., Asano, K., et al. 2011,  \apj, 729, 114
\bibitem[Ackermann~et~al.(2012a)]{sfl12} Ackermann, M., Ajello, M., Allafort, A., et al. 2012,  \apj, 745, 144
\bibitem[Ackermann~et~al.(2012b)]{kocevskibeta} Ackermann, M., Ajello, M., Baldini, L., et al. 2012,  \apj, 754, 121
\bibitem[Axelsson~et~al.(2012)]{axelsson12} Axelsson, M., Baldini, L., Barbiellini, G., et al. 2012,  \apj, 757, 31
\bibitem[Atwood~et~al.(2009)]{Atwood} Atwood, W.~B., Abdo, A. A., Ackermann, M., et al. 2009,  \apj, 697, 1071
\bibitem[Band~et~al.(1993)]{Band93} Band, D.~L., Matteson, J., Ford, L.,  et al. 1993, \apj, 413, 281
\bibitem[Barat~et~al.(1993)]{barat06} Barat, C., Skinner, G. K., \& Lestrade, J. P. 2006, AdSpR, 38, 1329
\bibitem[Bissaldi~et~al.(2009)]{Bissaldi} Bissaldi, E., von Kienlin, A., Lichti G., et al. 2009, \emph{Exp. Astron.}, 24, 47
\bibitem[Briggs~et~al.(1999)]{Briggs} Briggs, M.~S., Band, D. L., Kippen R. M., et al. 1999, \apj, 524, 82
\bibitem[Briggs~et~al.(2013)]{briggs11} Briggs, M.~S., et al. 2013, J. Geophys. Res. Space Physics, 50205
\bibitem[Case~et~al.(2011)]{Case2011} Case, G.~L., Cherry, M. L., Wilson-Hodge C. A., et al.\ 2011, \apj, 729, 105 
\bibitem[Cash(1979)]{Cash}Cash, W. 1979, \apj, 228, 939
\bibitem[Collazzi~et~al.(2011)]{collazzi}Collazzi, A.~C., et al. 2011, \apj, 729, 89
\bibitem[Crider~et~al.(1999)]{crider99}Crider, A.~C., Liang, E.~P., Preece, R.~D., et al. 1999, \apj, 519, 206
\bibitem[Geng \& Huang (2013)]{geng} Geng, J. J. \& Huang, Y. F., 2013, \apj, 764, 75
\bibitem[Ghirlanda~et~al.(2004)]{Ghirlanda} Ghirlanda, G., Ghisellini, G., \& Lazzati, D. 2004, \apj, 616, 331
\bibitem[Goldstein~et~al.(2012)]{goldstein12} Goldstein, A., Burgess, J. M., Preece, R. D., et al. 2012, \apjs, 199, 19
\bibitem[Gruber~et~al.(2011a)]{grubersfl11} Gruber, D., Lachowicz, P., Bissaldi, E., et al. 2011, \aap, 533, 61
\bibitem[Gruber~et~al.(2011b)]{restframe} Gruber, D., Greiner, J., von Kienlin, A., et al. 2011, \aap, 531, 20
\bibitem[Gruber~et~al.(2012)]{untrig} Gruber, D., et al. 2012, PoS, GRB 2012, 036
\bibitem[Guiriec~et~al.(2010)]{Guiriec} Guiriec, S., Briggs, M. S., Connaughton, V., et al. 2010, \apj, 725, 225
\bibitem[Guiriec~et~al.(2011)]{guiriec11} Guiriec, S., Connaughton, V., Briggs, M. S., et al. 2011, \apj, 727, 33
\bibitem[Guiriec~et~al.(2013)]{guiriec13} Guiriec, S., Daigne, F., Hasco\"{e}t, R., et al. 2013, \apj, 770, 32
\bibitem[Harmon~et~al.(2002)]{Harmon2002} Harmon, B.~A., Fishman, G. J., Wilson, C. A., et al. 2002, \apjs, 138, 149 
\bibitem[Harris \& Share (1998)]{harris98} Harris, M. J. \& Share, G. H. 1998, \apj, 494, 724
\bibitem[Hoover~et~al.(2008)]{Hoover2008} Hoover, A.~S., et al. 2008, AIP Conf. Proc. 1000, Gamma-Ray Bursts 2007: Proceedings of the Santa Fe Conference, ed. M. Galassi, D. Palmer, \& E. Fenimore,565
\bibitem[Hoover~et~al.(2010)]{Hoover2010} Hoover, A., Kippen, M., \& Wallace, M. 2010, General Response Simulation System (GRESS) Software User's Guide, Document GRESS-SUG-001-21 (Los Alamos, NM: LANL), http://public.lanl.gov/mkippen/gress/
\bibitem[Israel~et~al.(2008)]{Israel} Israel, G.~L., Romano, P., Mangano, V., et al. 2008, \apj, 685, 2
\bibitem[Kaneko~et~al.(2006)]{Kaneko06} Kaneko, Y., Preece, R. D., Briggs, M. S., et al. 2006, \apjs, 166, 298
\bibitem[Kouveliotou et al.(1993)]{Kouveliotou} Kouveliotou, C., Meegan, C. A., Fishman, G. J., et al. 1993,  \apj, 413, L101
\bibitem[Lebrun~et~al.(2003)]{Lebrun} Lebrun, F., Leray, J. P., Lavocat, P., et al. 2003, \aap , 411, L141
\bibitem[Lee~et~al.(2000)]{Lee} Lee, H.~K., Wijers, R.~A.~M.~J, \& Brown, G.~E. 2000, \physrep, 325, 83
\bibitem[Lin~et~al.(2011)]{Lin} Lin, L., Kouveliotou, C., Baring, M. G., et al. 2011, \apj, 739, 87
\bibitem[Lloyd~et~al.(2000)]{Lloyd00} Lloyd, N.~M., Petrosian, V., \& Mallozzi, R.~S. 2000, \apj, 534, 227 
\bibitem[Lu~et~al.(2012)]{lu12} Lu, R.-J., Wei, J.-J., Liang, E.-W., et al. 2012, \apj, 756, 112 
\bibitem[Mallozzi~et~al.(1995)]{Mallozzi95} Mallozzi, R.~S., Paciesas, W.~S., Pendleton, G.~N., Briggs, M.~S., Preece, R.~D., Meegan, C.~A., \& Fishman, G.~J. 1995, \apj, 454, 597
\bibitem[Meegan~et~al.(2009)]{Meegan} Meegan, C.~A., Lichti, G., Bhat, P. N., et al. 2009, \apj, 702, 791
\bibitem[Mitrofanov~et~al.(1999)]{Mitrofanov} Mitrofanov, I.~G., Anfimov, D. S., Litvak, M. L., et al. 1999, \apj, 522, 1069
\bibitem[Nava~et~al.(2011)]{Nava} Nava, L., Ghirlanda, G., Ghisellini, G., \& Celotti, A. 2011, \aap, 530, A21 
\bibitem[Paciesas~et~al.(2012)]{paciesas12} Paciesas, W.~S., Meegan, C. A., von Kienlin, A., et al. 2012, \apjs, 199, 18
\bibitem[Planck~Collaboration(2005)]{planck} Planck Collaboration. 2005, ESA publication ESA-SCI(2005)/01, arXiv:astroph/0604069
\bibitem[Preece~et~al.(1998a)]{Preece98} Preece, R.~D., Pendleton, G. N., Briggs, M. S., et al. 1998, \apj, 496, 849
\bibitem[Preece~et~al.(1998b)]{LineOfDeath} Preece, R.~D., Briggs, M. S., Mallozzi, R. S., et al. 1998, \apj, 506, 23
\bibitem[Preece~et~al.(2002)]{Preece} Preece, R.~D., Briggs, M. S., Giblin, T. W., et al. 2002, \apj, 581, 1248
\bibitem[Rees~\&~Meszaros(1992)]{Rees} Rees, M.~J. \& Meszaros, P. 1992, \mnras, 258, 41P
\bibitem[Ryde(1999)]{Ryde} Ryde, F., 1999, \emph{Astro. Lett. and Comm.}, 39, 281
\bibitem[Sari~et~al.(1998)]{Sari} Sari, R., Piran, T., \& Narayan, R. 1998, \apj, 497, L17
\bibitem[Sz{\'e}csi~et~al.(2012)]{szecsi} Sz{\'e}csi, D., Bagoly, Z., K{\'o}bori, J. et al. 2013, \aap, 557, 8
\bibitem[Vedrenne~et~al.(2003)]{Vedrenne} Vedrenne, G., Roques, J.-P., Sch\"{o}nfelder, V., et al. 2003, \aap, 411, L63
\bibitem[von~Kienlin~et~al.(2009)]{Andreas} von Kienlin, A., Briggs, M. S., Connoughton, V., et al. 2009,  AIP Conf. Proc. 1133, Gamma-ray Bursts: Sixth Huntsville Symposium, ed. C.~A.~Meegan, C.~Kouveliotou, \& N.~Gehrels, 446
\bibitem[von~Kienlin~et~al.(2012)]{kienlin12} von Kienlin, A., Gruber, D., Kouveliotou, C., et al. 2012, \apj, 755, 150
\bibitem[von~Kienlin~et~al.(submitted)]{kienlin13} von Kienlin, A., Meegan, C. A., Paciesas, W. S., et al. 2013, \apjs, submitted
\bibitem[Yu~et~al.(in prep.)]{yu14} Yu, H.~F. , et al. in preparation
\end{thebibliography}
\end{document}